\documentclass[onecolumn]{emulateapj}
\usepackage{subfigure,epsfig,graphicx}
\usepackage{natbib}
\usepackage{rotating}

\newcommand{\Msun}{M_{\odot}}

\newcommand{\Mstar}{M_{\star}}

\newcommand{\Fig}[1]{Fig \ref{#1}}

\newcommand{\beq}{\begin{eqnarray}}
\newcommand{\eeq}{\end{eqnarray}}

\def\lsim{\mathrel{\rlap{\lower4pt\hbox{\hskip1pt$\sim$}} \raise1pt\hbox{$<$}}}                
\def\gsim{\mathrel{\rlap{\lower4pt\hbox{\hskip1pt$\sim$}}\raise1pt\hbox{$>$}}}                
\begin{document}

\title{An Analysis of Jitter and Transit Timing Variations in the HAT-P-13 System}
\shorttitle{Jitter and TTVs in the HAT-P-13 system}
\shortauthors{Payne\,\&\,Ford}

\author{Matthew J. Payne and Eric B. Ford}
\address{Department of Astronomy, University of
  Florida, 211 Bryant Space Science Center, P.O. Box 112055,
  Gainesville, FL 32611-2055}
\email{matthewjohnpayne@gmail.com}

\begin{abstract}
If the two planets in the HAT-P-13 system are coplanar, the orbital states provide a probe of the internal planetary structure. Previous analyses of radial velocity and transit timing data of the system suggested that the observational constraints on the orbital states were rather small. We reanalyze the available data, treating the jitter as an unknown MCMC parameter, and find that a wide range of jitter values are plausible, hence the system parameters are less well constrained than previously suggested. For slightly increased levels of jitter ($\sim 4.5\,m\,s^{-1}$) the eccentricity of the inner planet can be in the range $0<e_{inner}<0.07$, the period and eccentricity of the outer planet can be $440<P_{outer}<470$ days and $0.55<e_{outer}<0.85$ respectively, while the relative pericenter alignment, $\eta$, of the planets can take essentially any value $-180^{\circ}<\eta<+180^{\circ}$. It is therefore difficult to determine whether $e_{inner}$ and $\eta$ have evolved to a fixed-point state or a limit cycle, or to use $e_{inner}$ to probe the internal planetary structure. We perform various transit timing variation (TTV) analyses, demonstrating that current constraints merely restrict $e_{outer}<0.85$, and rule out relative planetary inclinations within $\sim 2^{\circ}$ of $i_{rel}=90^{\circ}$, but that future observations could significantly tighten the restriction on both these parameters. We demonstrate that TTV profiles can readily distinguish the theoretically favored inclinations of $i_{rel}=0^{\circ}\,\&\,45^{\circ}$, provided that sufficiently precise and frequent transit timing observations of HAT-P-13b can be made close to the pericenter passage of HAT-P-13c. We note the relatively high probability that HAT-P-13c transits and suggest observational dates and strategies.
\end{abstract}


\maketitle


\noindent{\it celestial mechanics; methods: numerical; methods: data analysis; methods: observational; planetary systems}:


\section{Introduction}\label{INTRO}
The HAT-P-13 system \citep{Bakos09} was the first extrasolar planetary system to be discovered in which both a transiting planet and an additional confirmed companion were known to coexist. Since this initial discovery, further transit-plus-companion systems have been discovered: CoRoT-7 \citep{Queloz09} and HAT-P-7 \citep{Narita10} as well as the recent multi-transit systems from Kepler \citep{Steffen10, Holman10}, while a number of other transiting systems display RV trends symptomatic of outer companions \citep[E.g. HAT-P-11, ][]{Bakos10}.

Such systems are of interest for a number of reasons. The first relates to the observable effects which arise from interactions between the two planets. The gravitational interaction between multiple planets causes the planetary orbits to be perturbed away from Keplerian ellipses. When one of the planets is transiting, these perturbations mean that the duration of, and the period between, successive transits will not be strictly constant. It has been calculated that observations of such transit timing variations (TTVs) and transit duration variations (TDVs) would allow the (inferred) detection of terrestrial-mass planets in Hot-Jupiter systems \citep{Holman05,Agol05}, trojan planets \citep{Ford07b} and exoplanet moons \citet{Kipping09a,Kipping09b}. 

In addition, accurate measurements of transiting systems can allow us to observationally determine a huge range of system parameters \citep{Winn09b}, one such parameter being the tidal Love number of the planet (if the system architecture is convenient -  \citet{Wu02,Mardling07,Ragozzine09}). In an investigation by \citet{Batygin09}, it is demonstrated for the coplanar case that the HAT-P-13 system is indeed such a system, and a relationship is found between the Love number and the eccentricity of the inner planet. 

However, the determination of the Love number depends on the assumption that the inner planet tends towards a quasi-fixed point in ($e_{inner}$,$\eta$) space, where $e_{inner}$ is the eccentricity of the inner planet and $\eta = \varpi_{outer} - \varpi_{inner}$ is the difference in the alignment of the longitude of pericenters of the outer and inner planets. The recent work by \citet{Mardling10} looked at the evolution of a general \emph{non}-coplanar, two-planet system, in which the angular momentum of the outer planet dominates the angular momentum budget of the system, and revealed that in such a system the inner planet does \emph{not} tend to a fixed point, but instead tends to a \emph{limit cycle}, with $e_{inner}$ and $\eta$ constantly sampling along a closed trajectory.

The approach to the limit cycle is found to be strongly dependent on the relative inclination between the two planets, $i_{rel}$:
The average value of $e_{inner}$ around the cycle \emph{decreases} and the limit cycle amplitude \emph{increases} with increasing $i_{rel}$;
Limit cycle behaviour only exists for $0^{\circ}< i_{rel} <33^{\circ}$, $46^{\circ}< i_{rel} <54^{\circ}$, $126^{\circ}< i_{rel} <134^{\circ}$ and $147^{\circ}< i_{rel} <180^{\circ}$
For the regions $33^{\circ}< i_{rel} <46^{\circ}$ and $134^{\circ}< i_{rel} <147^{\circ}$, $\eta$ circulates and no limit cycle exists.
For the region $54^{\circ}< i_{rel} <126^{\circ}$, the effects of Kozai oscillations combined with tidal dissipation act to move the relative inclinations back towards $i_{rel} = 54$ or $i_{rel} = 126$ for prograde or retrograde orbits respectively, meaning that the system simply cannot exist with $54^{\circ}< i_{rel} <126^{\circ}$ for a tidal dissipation factor, $Q < 10^6$.
For the range of possible $e_{inner}$ and $R_{inner}$ (the radius of the inner planet) reported by \citet{Batygin09}, certain relative inclinations can be ruled out, suggesting that (for the prograde case), the system either has $i\lsim 10^{\circ}$, or $i\sim 45^{\circ}$.

It is apparent from the analyses in both \citet{Mardling10} and \citet{Batygin09} that the conclusions one can derive regarding the possible state of the system rather sensitively depend on (i) the measured eccentricity of the inner planet, (ii) the relative pericenter of the two planets  and (iii) the essentially unknown relative inclination between the two planets. It is our aim in this paper to try and understand in more detail what the current observational constraints on these quantities are.

We start by re-evaluating the published system parameters for HAT-P-13 from \citet{Bakos09} (henceforth B09) as well as those from the expanded analysis by \citet{Winn10} (henceforth W10), performing a Markov chain Monte-Carlo (MCMC) investigation that concentrates on the effect of assumptions about jitter and how to include it within a MCMC analysis. We use jitter as a mathematical parameter to quantify the magnitude of unmodeled variations in the  radial velocity observations.  Jitter can be due to undetected planets, stellar activity or unrecognized noise in the instrument or data analysis pipeline.  In the case of a complete model for the planetary system and ideal measurements, the jitter would reduce to the ``stellar jitter''. Stellar jitter is the noise introduced into radial velocity (RV) measurements by unknown changes in the surface of the observed star, driven by sun spots, bulk flows and other inhomogeneities on the stellar surface. Several investigations have examined this phenomenon, trying to correlate the magnitude of the jitter with observable stellar parameters \citep{Saar97,Saar98,Wright05}. They find that stars of a given type can have a wide range of jitter values, with stars similar to HAT-P-13 having jitters in the range $2\,-\,15\,m\,s^{-1}$ (see \citet{Wright05} and section \ref{Jitter} for further details).

In the papers of B09 and W10, subsequent to the MCMC analysis of the fitted planetary parameters, jitter levels of $\sigma_j = 3.0\,m\,s^{-1}$ and $\sigma_j = 3.4\,m\,s^{-1}$ respectively were required in order to give reduced $\chi^2$ values of 1 (see \S\ref{MCMC} for further discussion).  We wish to understand the effects of changing this methodology, and in particular to find out to what degree including the jitter as an MCMC model parameter loosens the constraints on quantities such as the eccentricity of the inner planet and the pericenter alignment of the two planets.

In a rigorous Bayesian analysis, the jitter should be treated as a model parameter simultaneously with the mass and orbital parameters. In this paper we show that correlations between the jitter and orbital parameters can lead to erroneous inferences if the jitter is held fixed during the modeling process. Thus, one should include the jitter in MCMC analyses alongside all of the other parameters that one is attempting to model. In this manner, one can arrive at a consistent statistical interpretation of oneÕs knowledge of the observables in the system.  In a Bayesian analysis, one must explicitly state assumptions for the prior distributions. While un-modeled observations (e.g., stellar color or temperature) can influence the choice of prior (e.g., F stars are more likely to have a jitter exceeding 10 m/s), the choice of prior for the jitter must not be influenced by the radial velocities themselves.  E.g., choosing a prior for the jitter based on the residuals to a fit results in Òdouble-countingÓ the radial velocity data and can result in misleadingly small uncertainties.

Further to this MCMC investigation, we go on to add an analysis of the TTVs for HAT-P-13. We do this to try and ascertain whether (i) combining the MCMC analysis with the TTV analysis can further restrict the range of parameter space available to the observed system quantities (eccentricities, alignments, etc), and (ii) we wish to understand whether the TTVs can provide some insight into the relative inclination of the planets (Relative planetary inclinations have previously been shown to be important in determining the expected TTVs in some systems, \citealt{Nesvorny09,Payne10a}).

We proceed in this paper as follows: In section \ref{Method} we outline the numerical methods we use to conduct our MCMC and TTV analyses; In section \ref{MCMC} we look at the effects of jitter in an MCMC analysis of the orbital elements of the planets in the HAT-P-13 system, and combine this with TTV constraints to understand whether this allows a more nuanced determination of the system parameters; In section \ref{FURTHER_TTVS} we look more generally at the TTVs in the HAT-P-13 system, focusing in detail on the effects that relative planetary inclinations may have on the expected TTVs; After this, we move on in section \ref{Projections} to consider the potential for future observations (both transit and RV) to better constrain the planetary orbits; and finally, in section \ref{Conclusion} we present a summary and discussion of our conclusions.


\section{Methodology}\label{Method}
\subsection{Radial Velocity \& Transit Observation Analysis}\label{Method:MCMC}
We analyze the radial velocity and transit observations using a Bayesian framework following \citet{Ford05b,Ford06}.  We assume priors that are uniform in log of orbital period, eccentricity, argument of pericenter, mean anomaly at epoch, and the velocity zero-point.  For the velocity amplitude ($K$) and jitter ($\sigma_j$), we adopt a prior of the form $p(x)=(x+x_o)^{-1}[log(1+x/x_o)]^{-1}$, with $K_o=\sigma_{j,o}=1$m/s, i.e. high values are penalized. For a discussion of priors, see \citet{Ford07c}.  We adopt a likelihood that is the product of two terms corresponding to the radial velocity and transit observations.  The likelihood for radial velocity terms assumes that each radial velocity observation ($v_i$) is independent and normally distributed about the true radial velocity with a variance of $\sigma_i^2+\sigma_j^2$, where $\sigma_i$ is the published measurement uncertainty.  

Instead of modeling each photometric observation, we account for the transit observations by including a likelihood term that is the product of three Bayesian penalties based on the orbital period, transit duration and ingress time of HAT-P-13b, assuming Gaussian distributions for each measurement with standard deviations taken from  the published uncertainties as derived by B09.  We use MCMC to calculate a sample from the posterior distribution \citep{Ford06}.  We calculate multiple Markov chains, each with $\sim~2\times 10^8$ states.  We discard the first half of the chains and calculate Gelman-Rubin test statistics for each model parameter and several ancillary variables.  We find no indications of non-convergence. Thus, we randomly choose a subsample ($10,000$ samples, large enough to give a statistically valid and accurate outcome, but small enough to be computationally tractable) from the posterior distribution for further investigation.

\subsection{Transit Timing Variations}\label{Method:TTV}
To investigate the TTV signature in the HAT-P-13 system, we employ the same basic method as that used in \citet{Payne10a} and \citet{Veras10}. We consider a fiducial system consisting of two planets: a transiting hot-Jupiter planet and an outer, non-transiting planet which perturbs the transit times of the inner planet. Given an initial specification of the planetary masses and orbital elements, we evolve each system forward in time for $\sim 3.5$ years, corresponding to several hundred transits by the inner planet in a system such as HAT-P-13 (assuming that the inner planet remains transiting throughout the integration). This 3.5 year integration time was also used in the studies of \citet{Payne10a} and \citet{Veras10} whose investigations attempted to illuminate potential Kepler missions observations, where the Kepler mission is expected to run for at least 3.5 years. To facilitate any comparison with the method and results of these papers, we chose to also maintain this 3.5 year simulation timescale.

The n-body integrations are performed using a conservative Bulirsch-Stoer integrator, derived from that of {\sc Mercury} (Chambers 1999).  We use a barycentric coordinate system and limit the time steps to no more than 0.04 times the orbital period.  After each time step, we test whether the star-planet separation projected onto the sky ($\Delta$) passed through a local minimum and the planet in question is closer to the observer than the star. Each time these conditions are met, we find the nearby time that minimizes $\Delta$ via Newton-Raphson iteration and increment an index $i$.  If the minimum $\Delta$ is less than the stellar radius, then we record the mid-time of the transit, $t_{i}$.  In calculating the observed transit time one needs to account for the light travel time, $\delta~t_{ltt}(i) \simeq -({\bf r}_p(t_{i}) \cdot {\bf \hat{r}}_{los})/c$, where ${\bf r}_p(t_i)$ is the barycentric vector of the planet at time $t_i$, ${\bf r}_{los}$ is a unit vector pointing to the observer, and $c$ is the speed of light.  The observable transit time variations are calculated as $\delta~t(i) = t_{i}+\delta~t_{llt}(i)- i\,P-t_0$, where the constants $P$ and $t_0$ are determined by linear least squares minimization of $\sum_i (\delta~t(i))^2$. We neglect any motion of the stellar center between the time of light emission and the time of transit.

It should be noted that the TTV investigations in this paper split into two main strands: (i) an investigation in \S\ref{MCMC_PLUS_TTV} of the TTVs that would arise in the systems that come out of our RV MCMC analysis, and (ii) the structured investigation of inclination effects in TTVs for HAT-P-13 carried out in \S\ref{FURTHER_TTVS}. The introduction to the respective results sections provides more detail on the precise manner in which each TTV investigation was conducted.


\section{MCMC Analysis of the Radial Velocity Observations}\label{MCMC}
\subsection{Jitter}\label{Jitter}
%
\begin{figure}
\centering
\begin{tabular}{ccc}
\includegraphics[trim = 0mm 0mm 0mm 0mm, clip, angle=-90, width=0.33\textwidth]{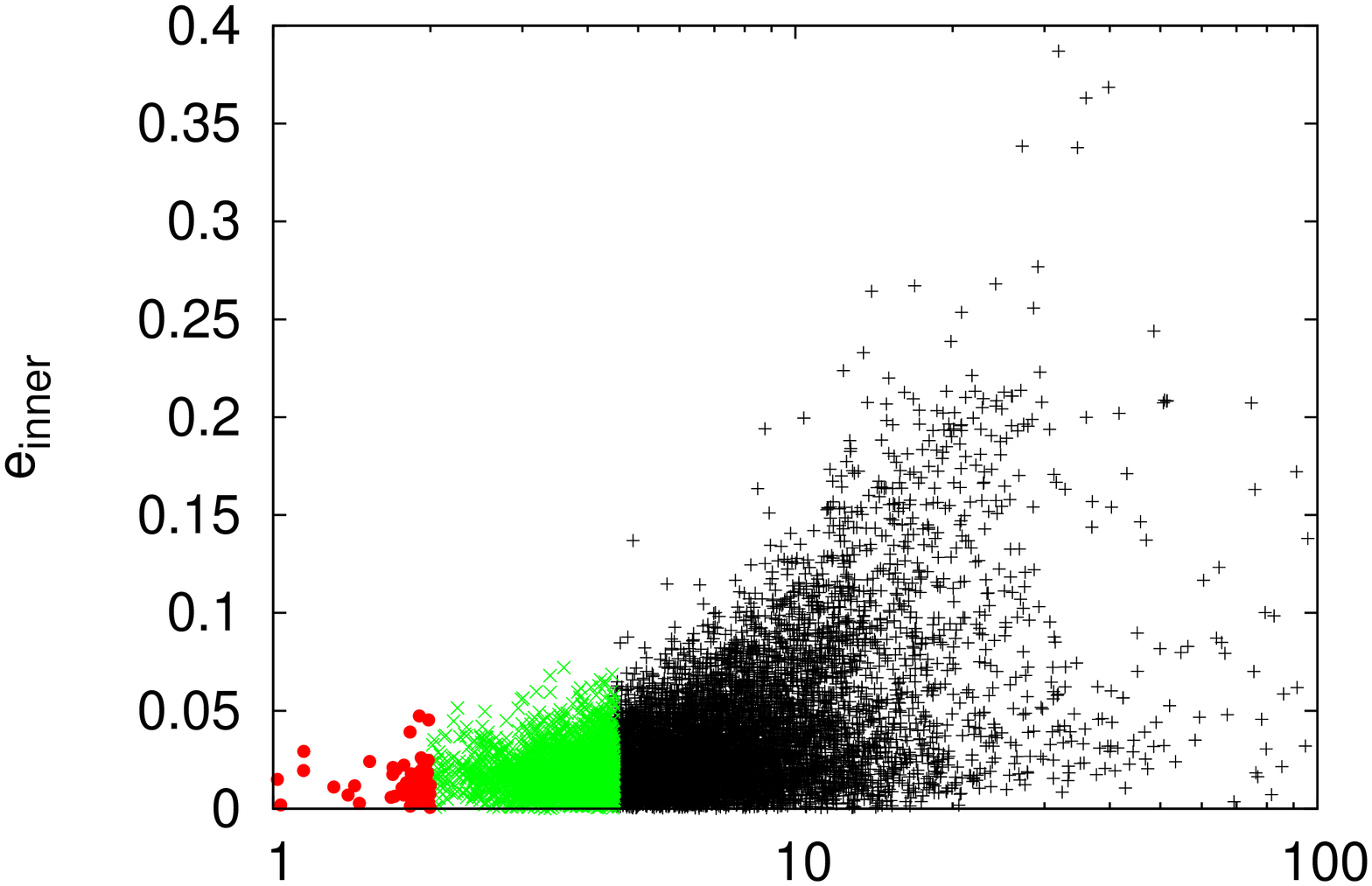}&
\includegraphics[trim = 0mm 0mm 0mm 0mm, clip, angle=-90, width=0.33\textwidth]{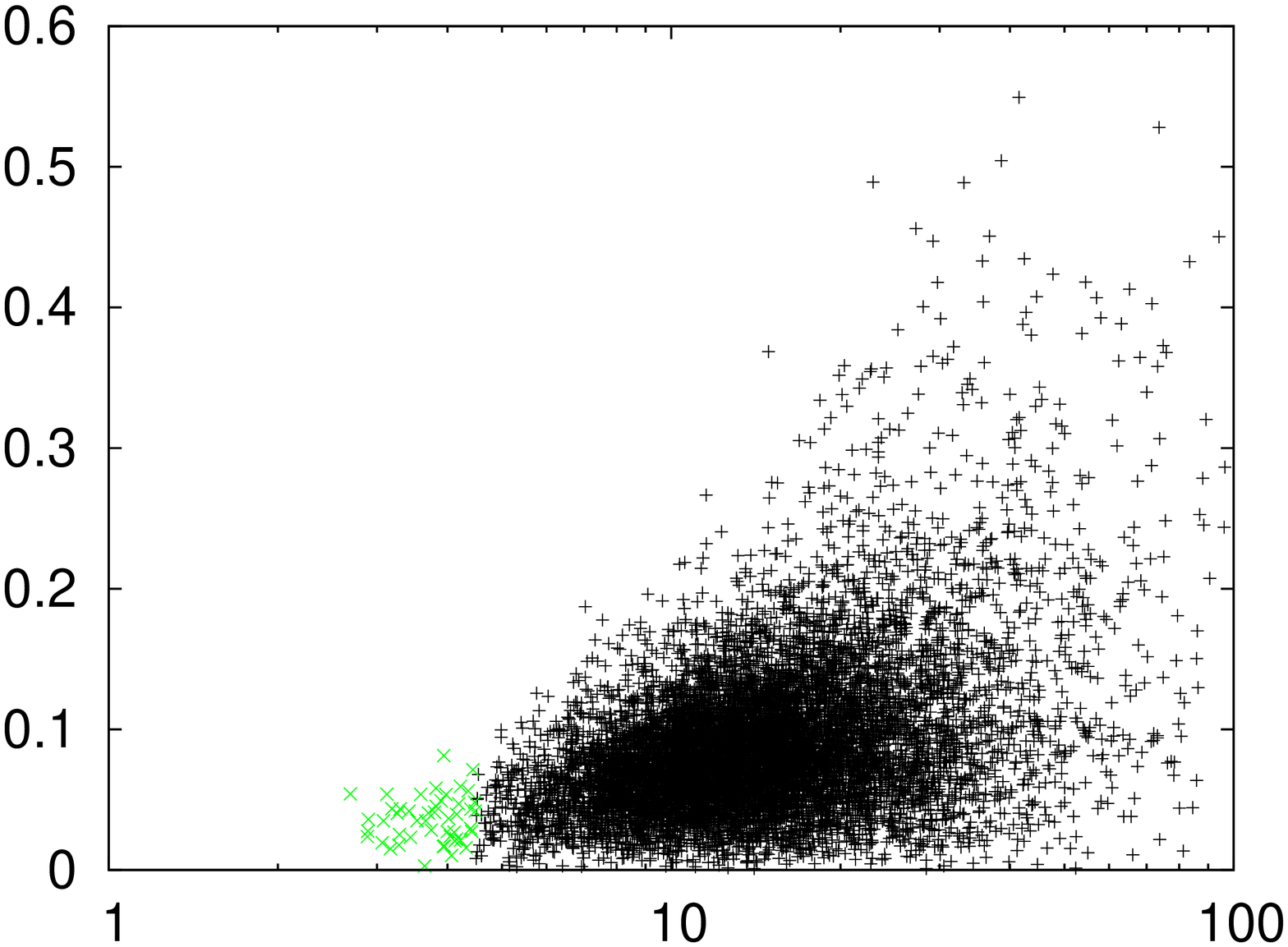}&
\includegraphics[trim = 0mm 0mm 0mm 0mm, clip, angle=-90, width=0.33\textwidth]{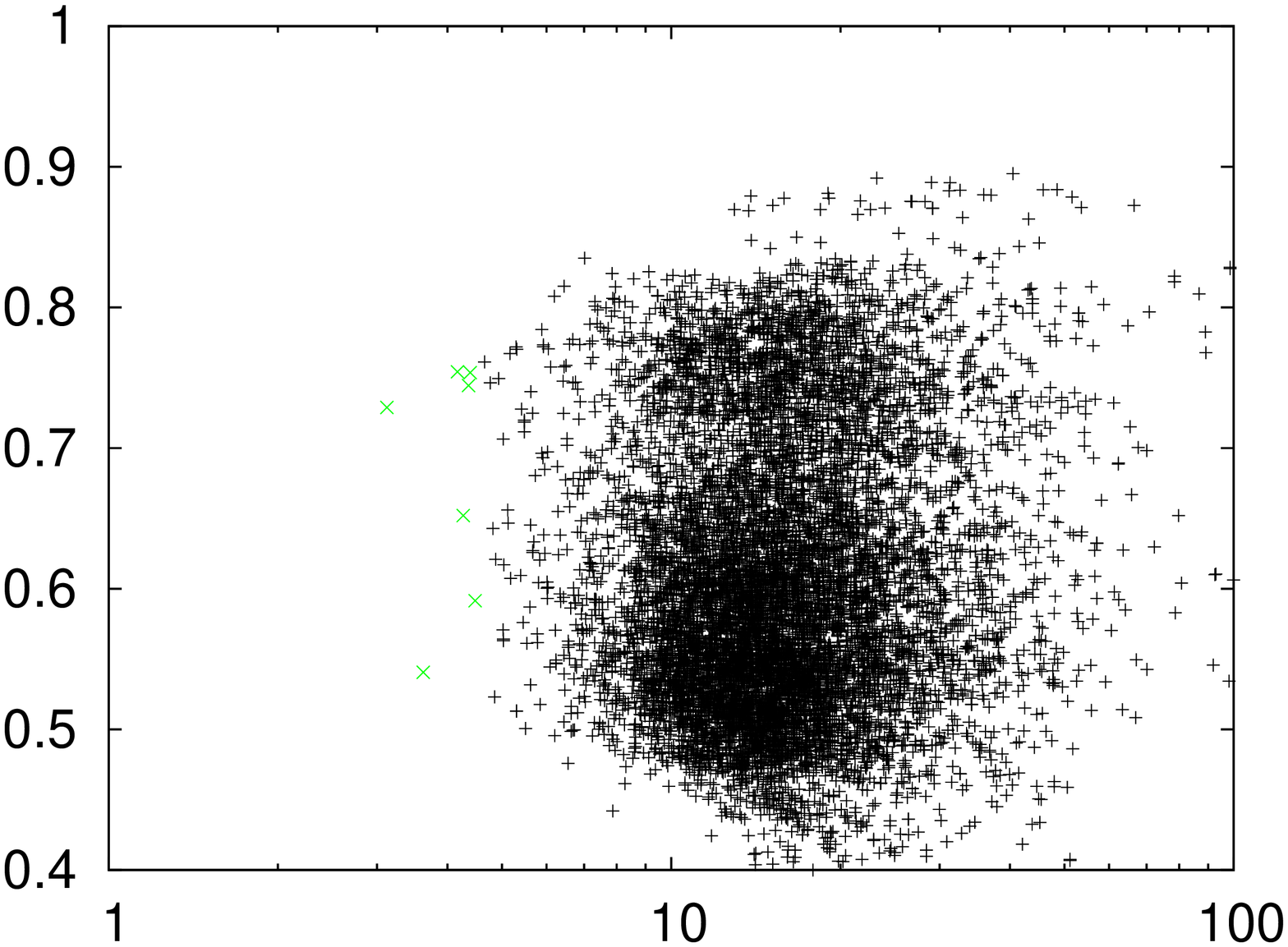}\\
\includegraphics[trim = 0mm 0mm 0mm 0mm, clip, angle=-90, width=0.33\textwidth]{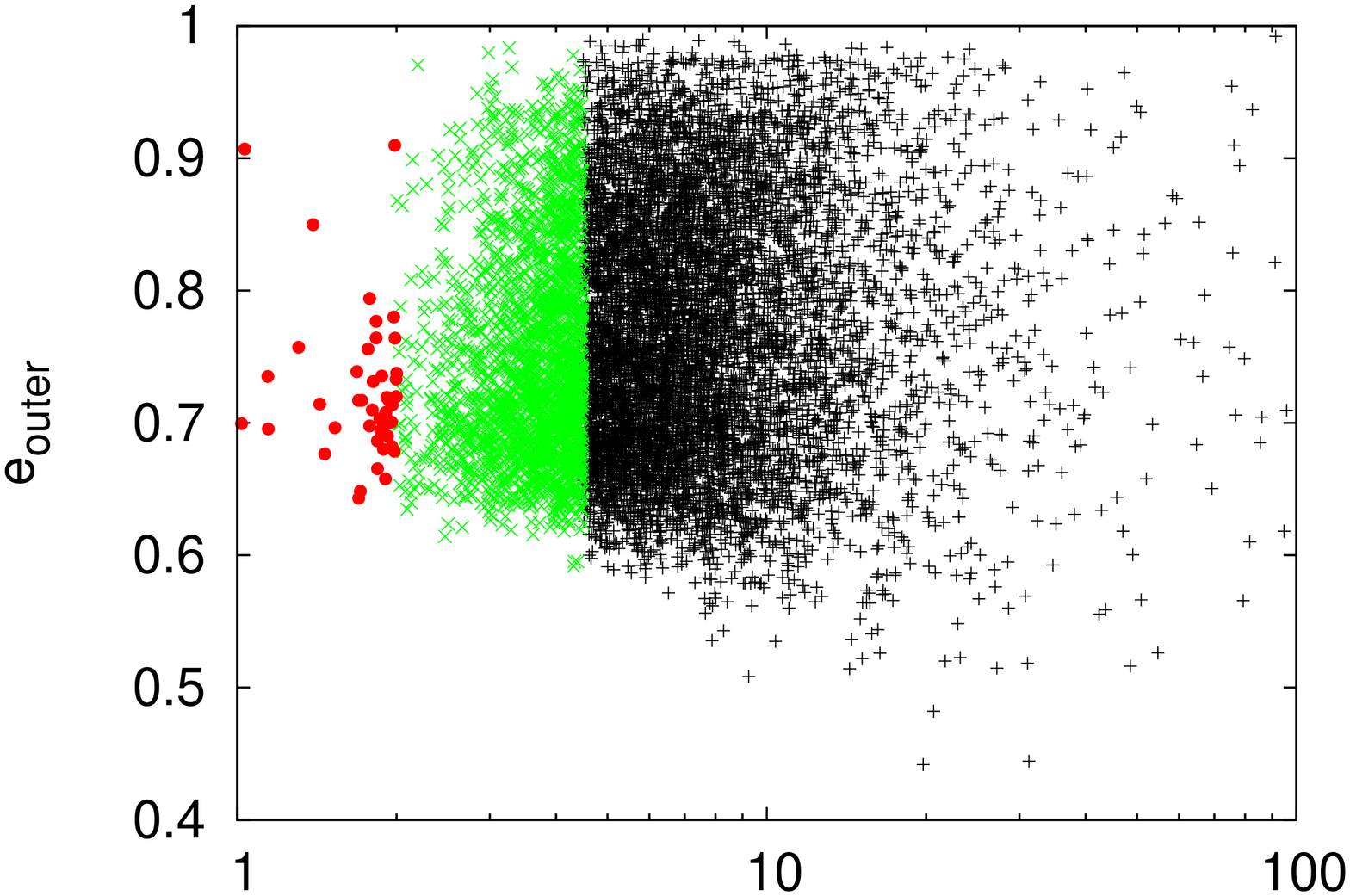}&
\includegraphics[trim = 0mm 0mm 0mm 0mm, clip, angle=-90, width=0.33\textwidth]{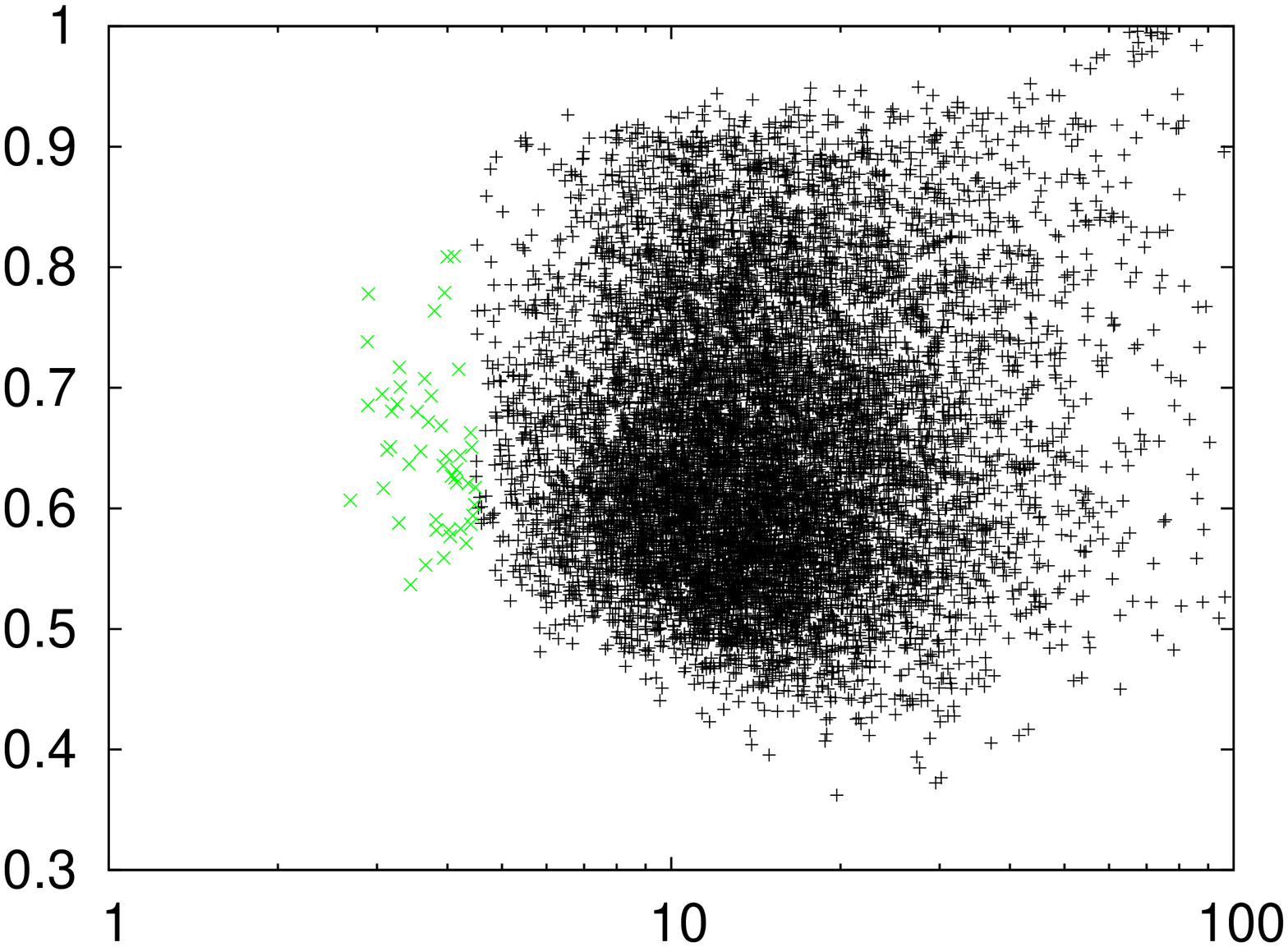}&
\includegraphics[trim = 0mm 0mm 0mm 0mm, clip, angle=-90, width=0.33\textwidth]{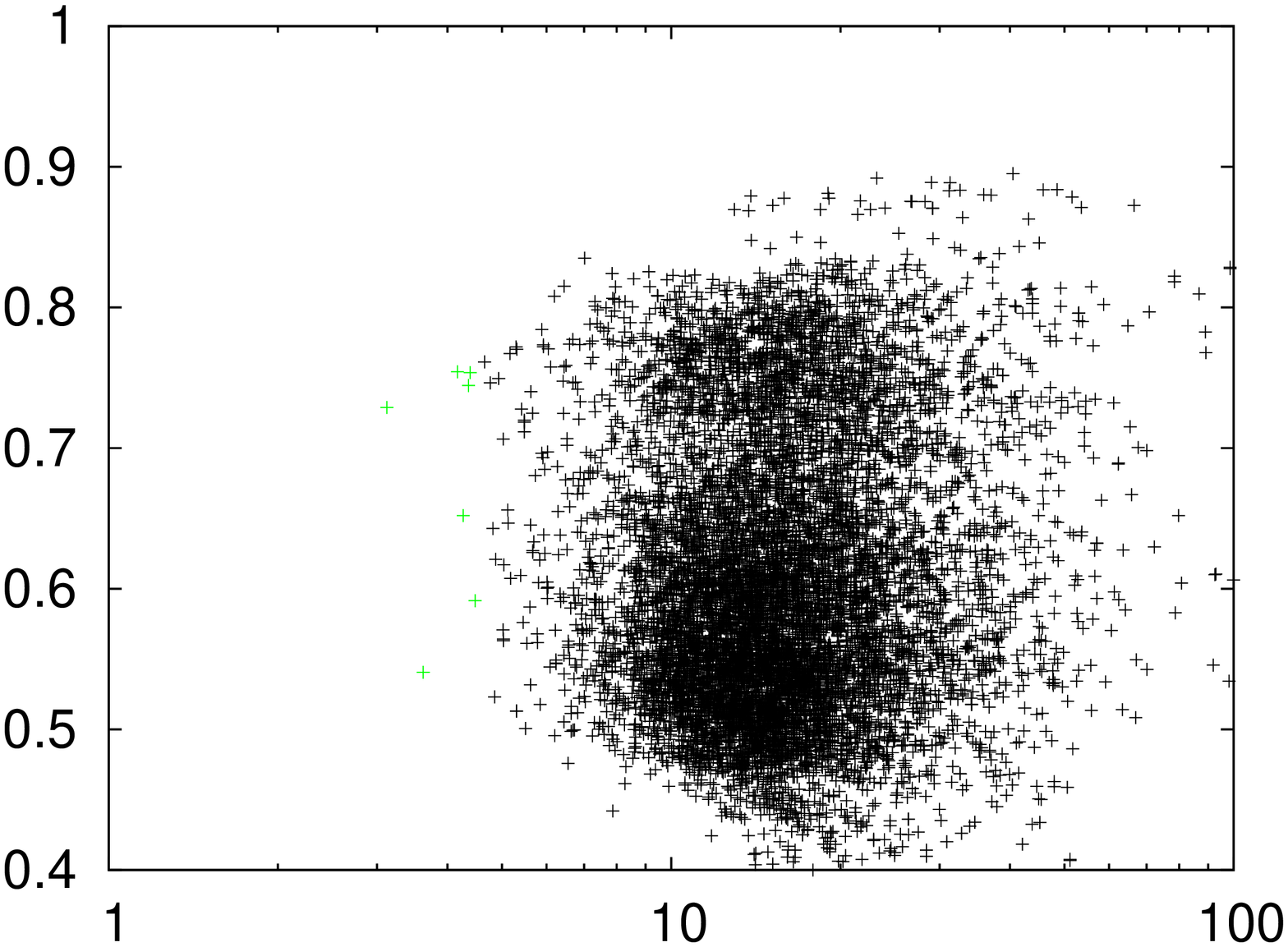}\\
\includegraphics[trim = 0mm 0mm 0mm 0mm, clip, angle=-90, width=0.33\textwidth]{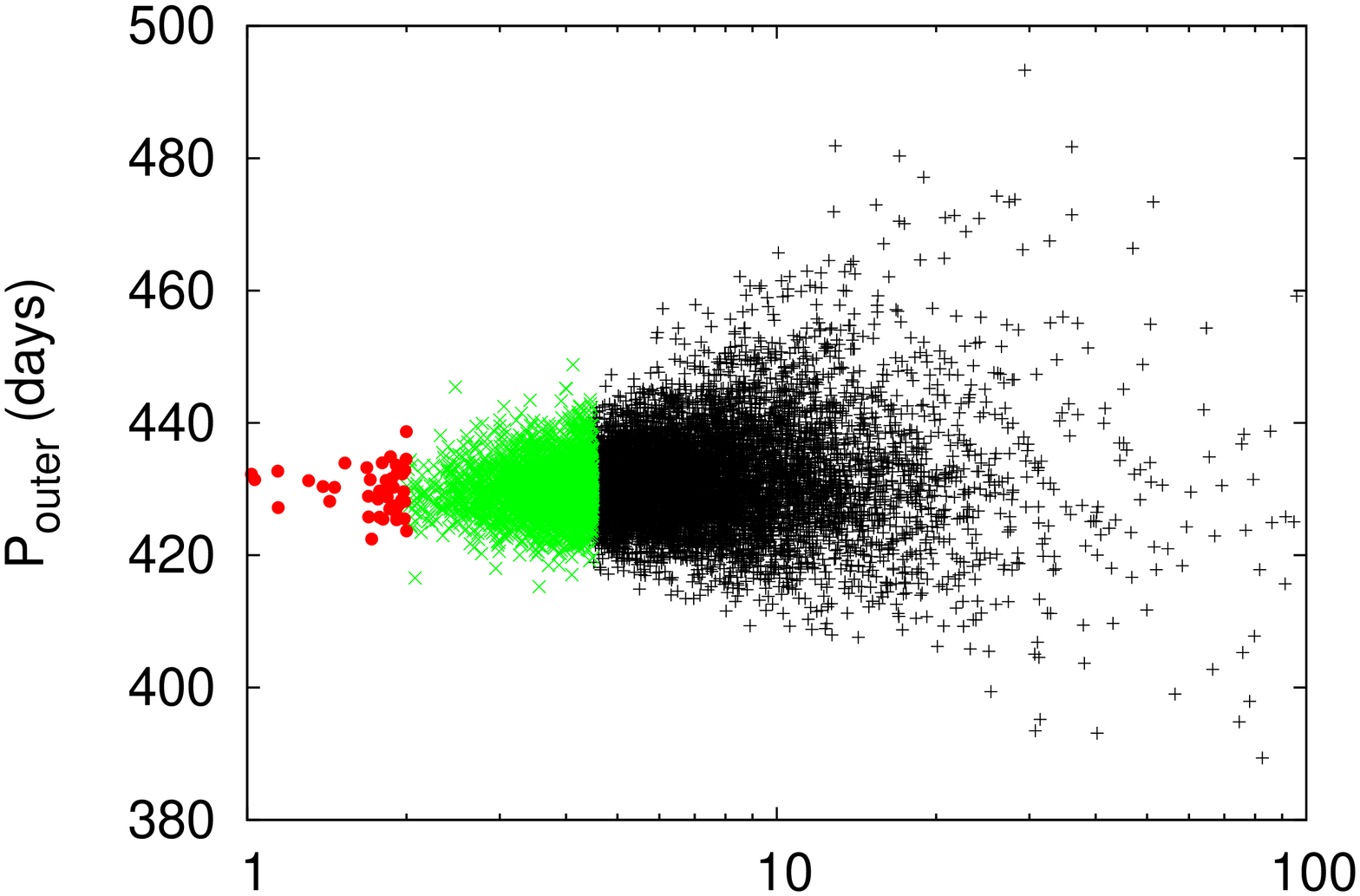}&
\includegraphics[trim = 0mm 0mm 0mm 0mm, clip, angle=-90, width=0.33\textwidth]{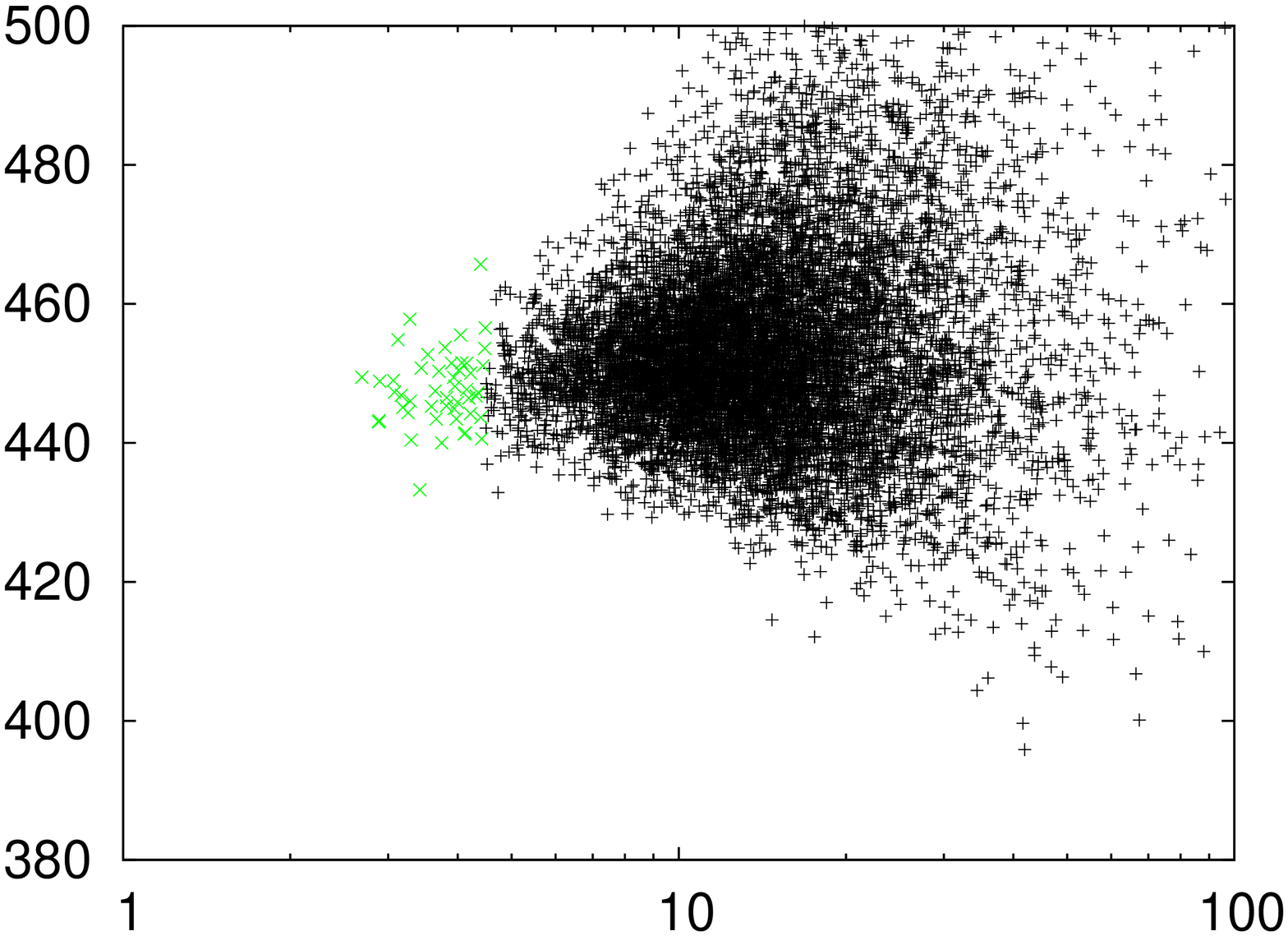}&
\includegraphics[trim = 0mm 0mm 0mm 0mm, clip, angle=-90, width=0.33\textwidth]{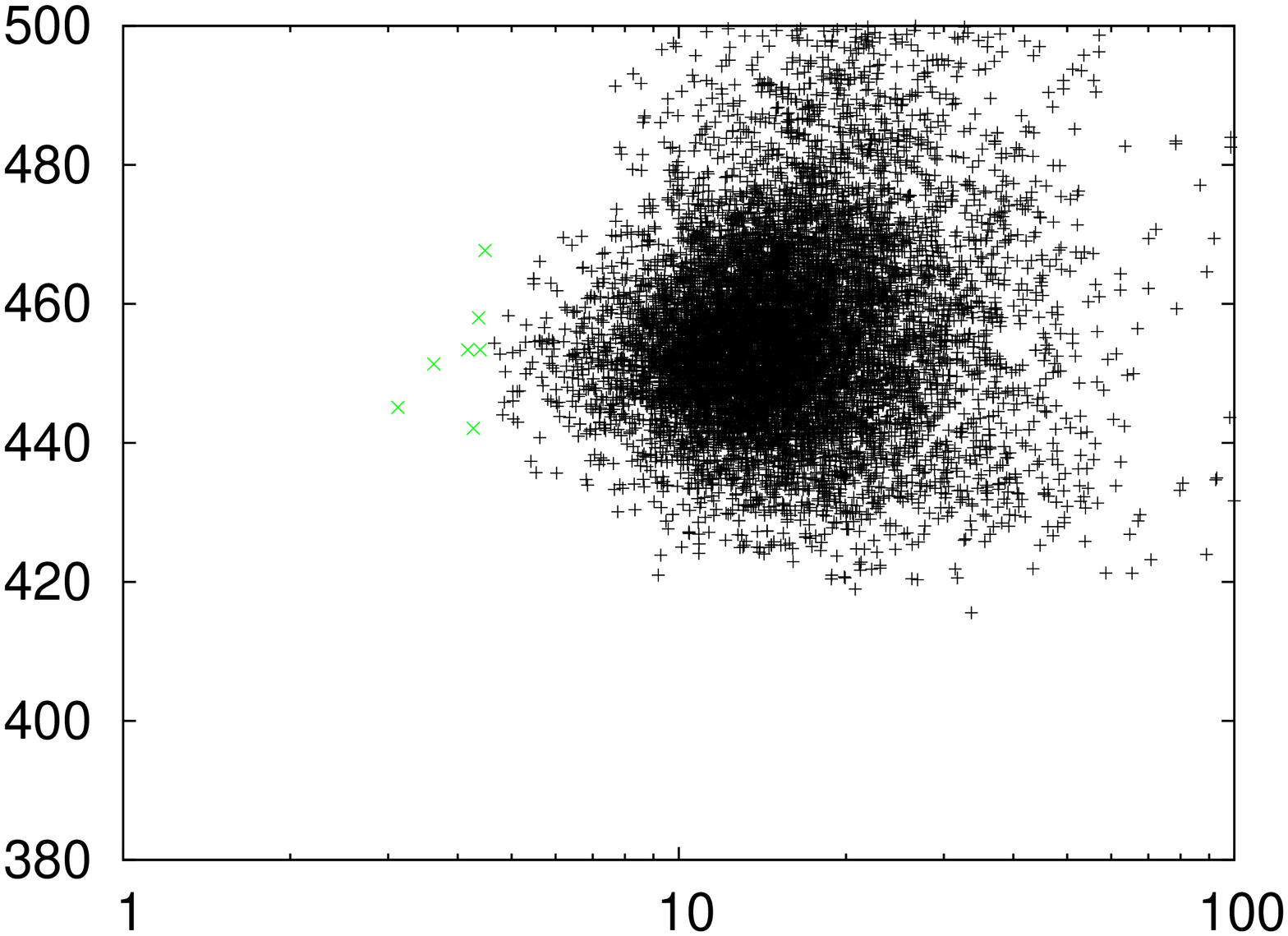}\\
\includegraphics[trim = 0mm 0mm 0mm 0mm, clip, angle=-90, width=0.33\textwidth]{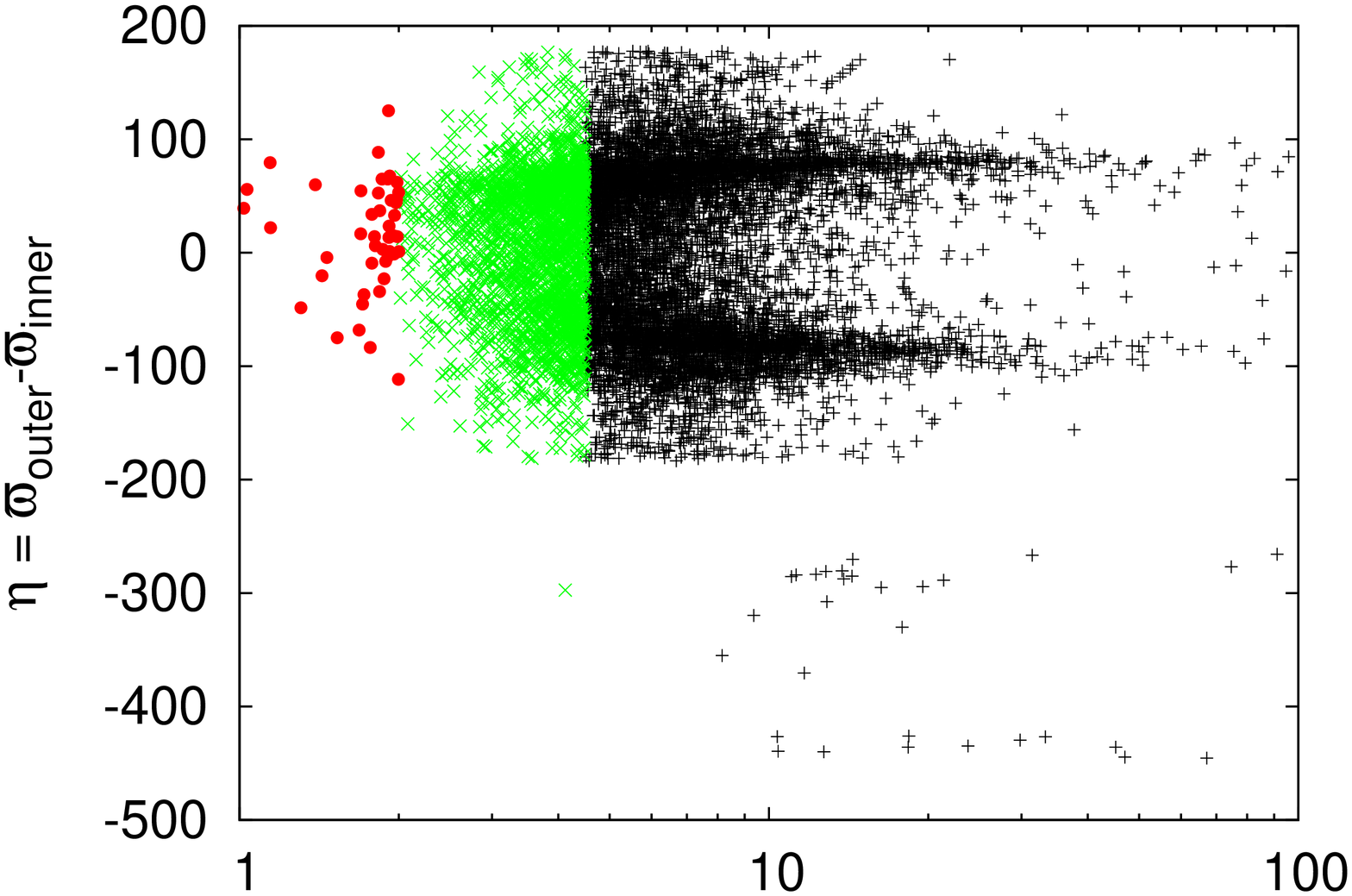}&
\includegraphics[trim = 0mm 0mm 0mm 0mm, clip, angle=-90, width=0.33\textwidth]{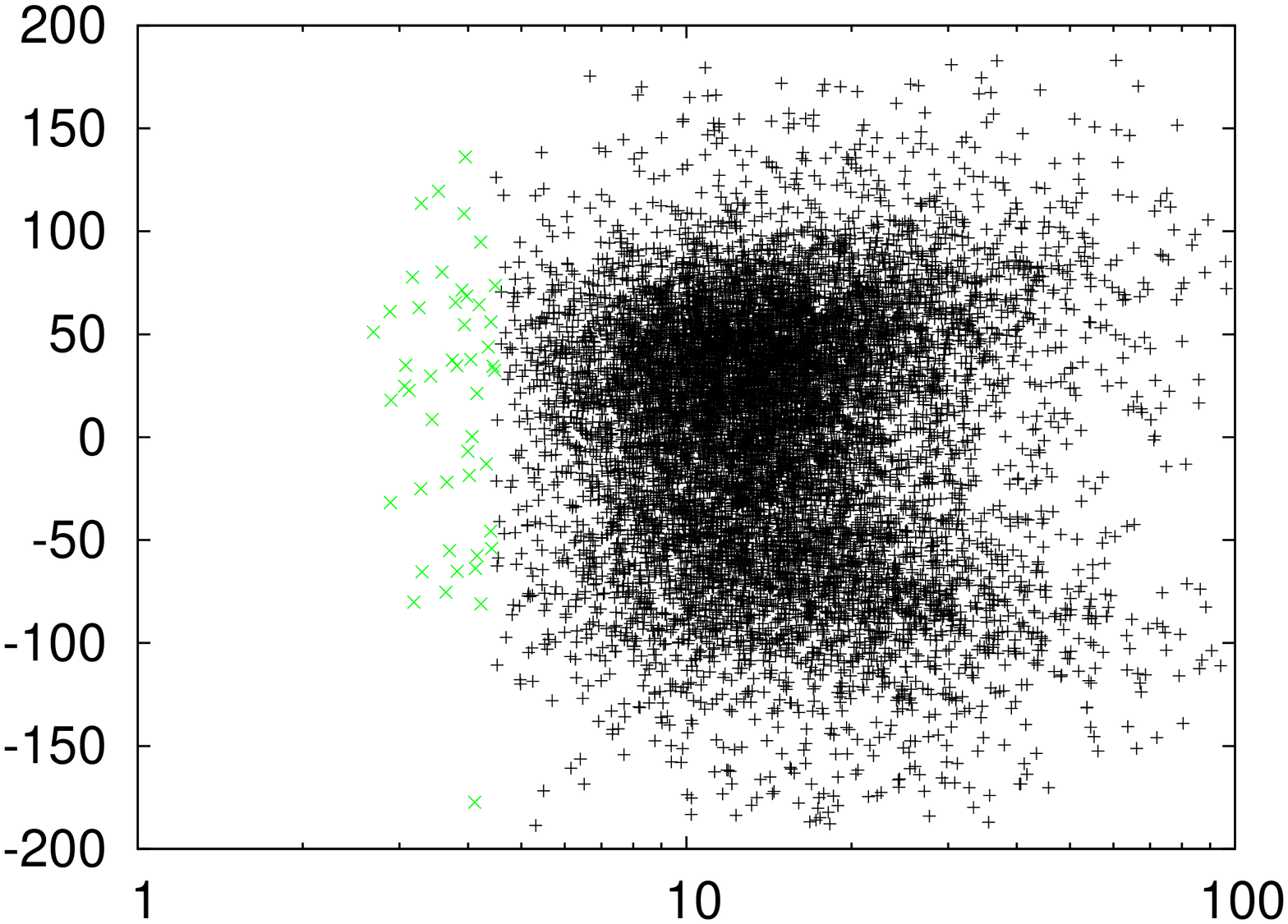}&
\includegraphics[trim = 0mm 0mm 0mm 0mm, clip, angle=-90, width=0.33\textwidth]{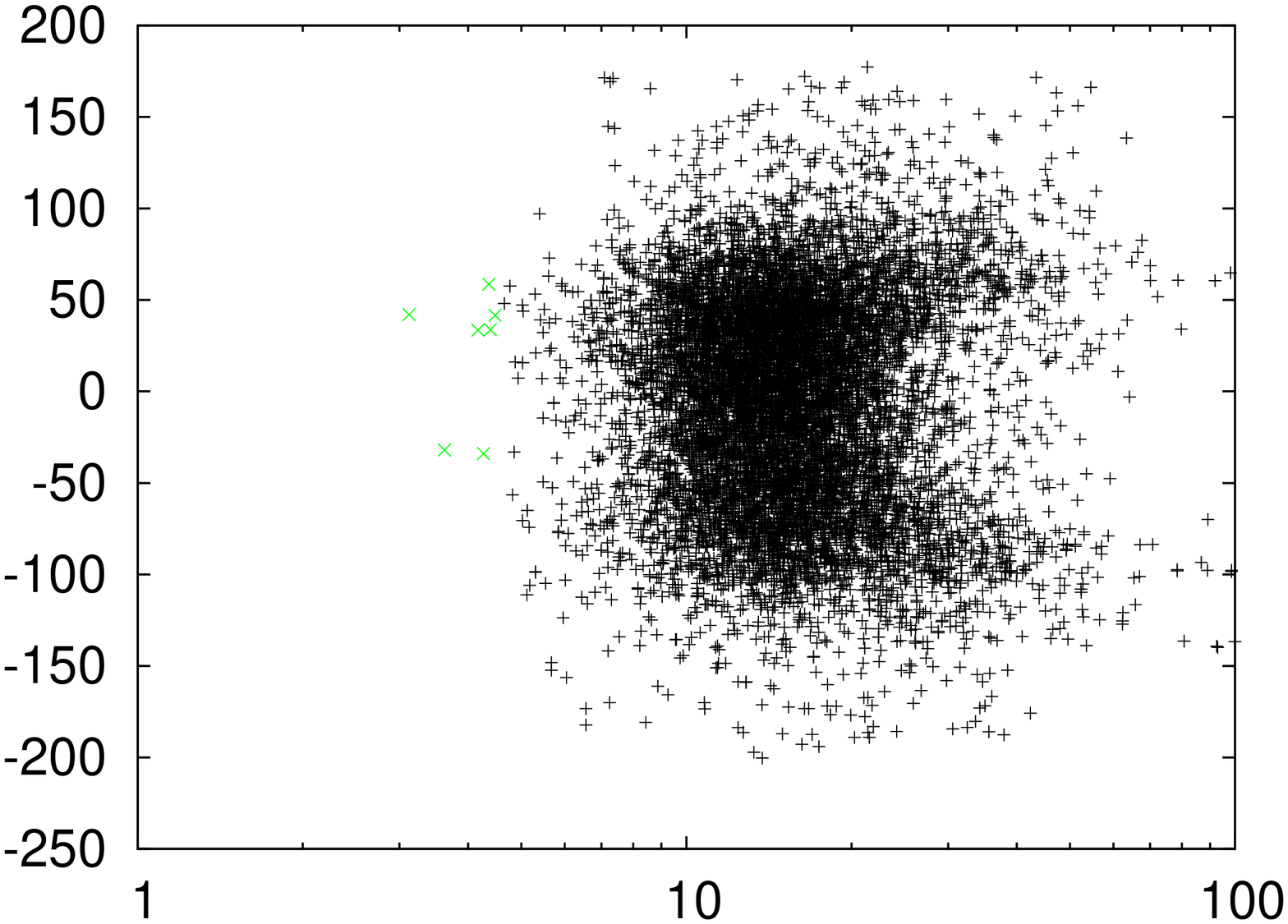}\\
\includegraphics[trim = 0mm 0mm 0mm 0mm, clip, angle=-90, width=0.33\textwidth]{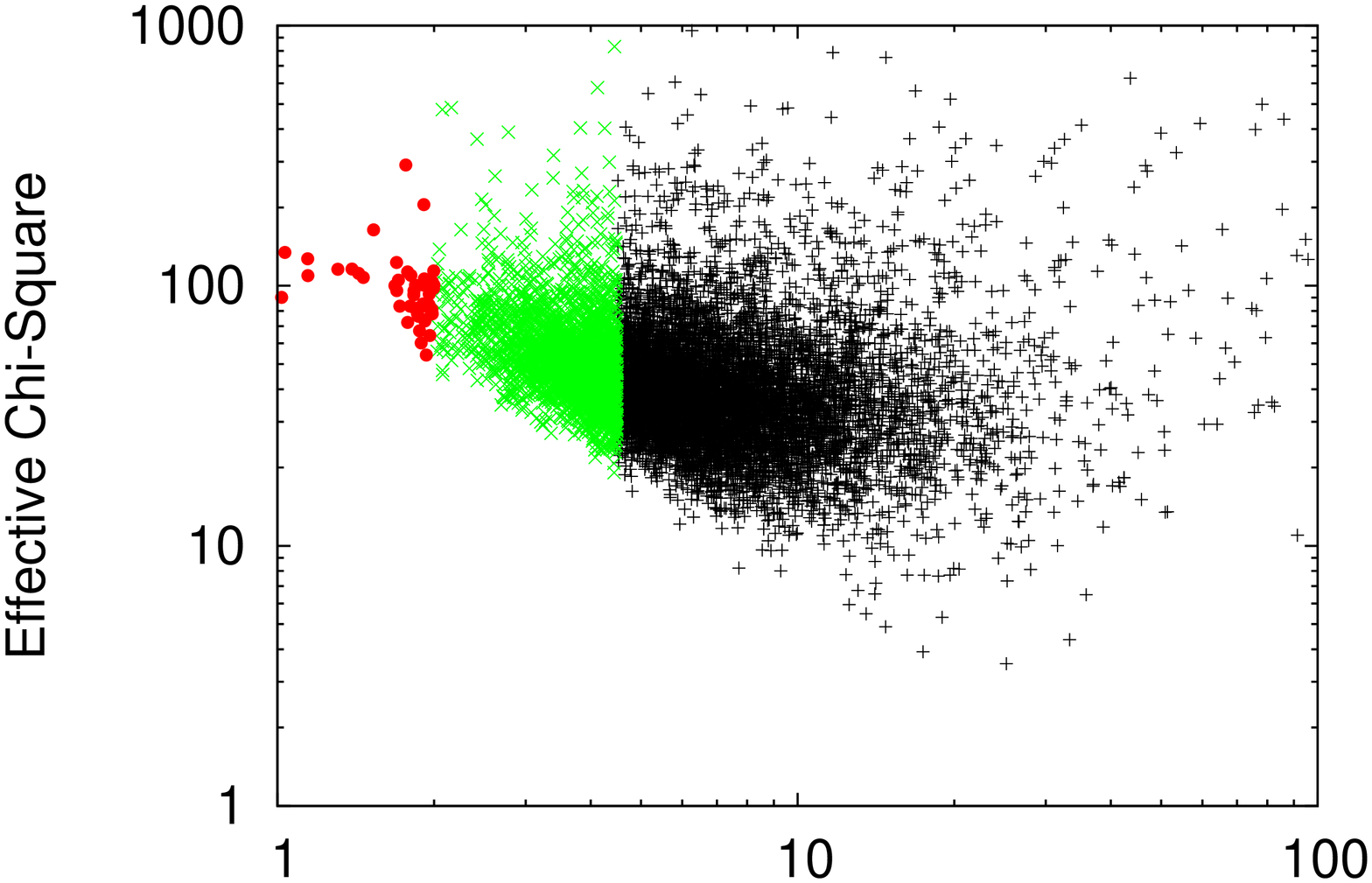}&
\includegraphics[trim = 0mm 0mm 0mm 0mm, clip, angle=-90, width=0.33\textwidth]{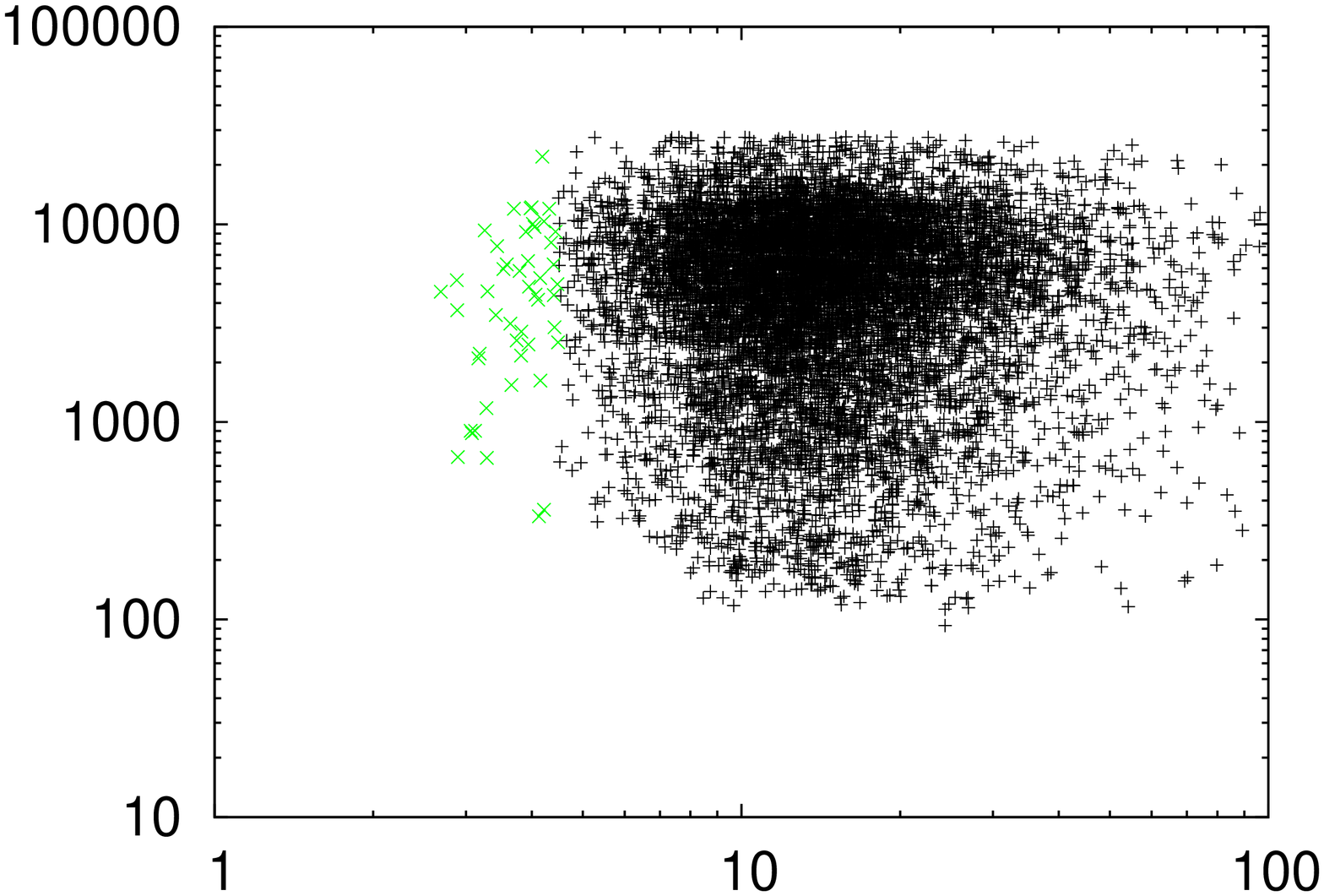}&
\includegraphics[trim = 0mm 0mm 0mm 0mm, clip, angle=-90, width=0.33\textwidth]{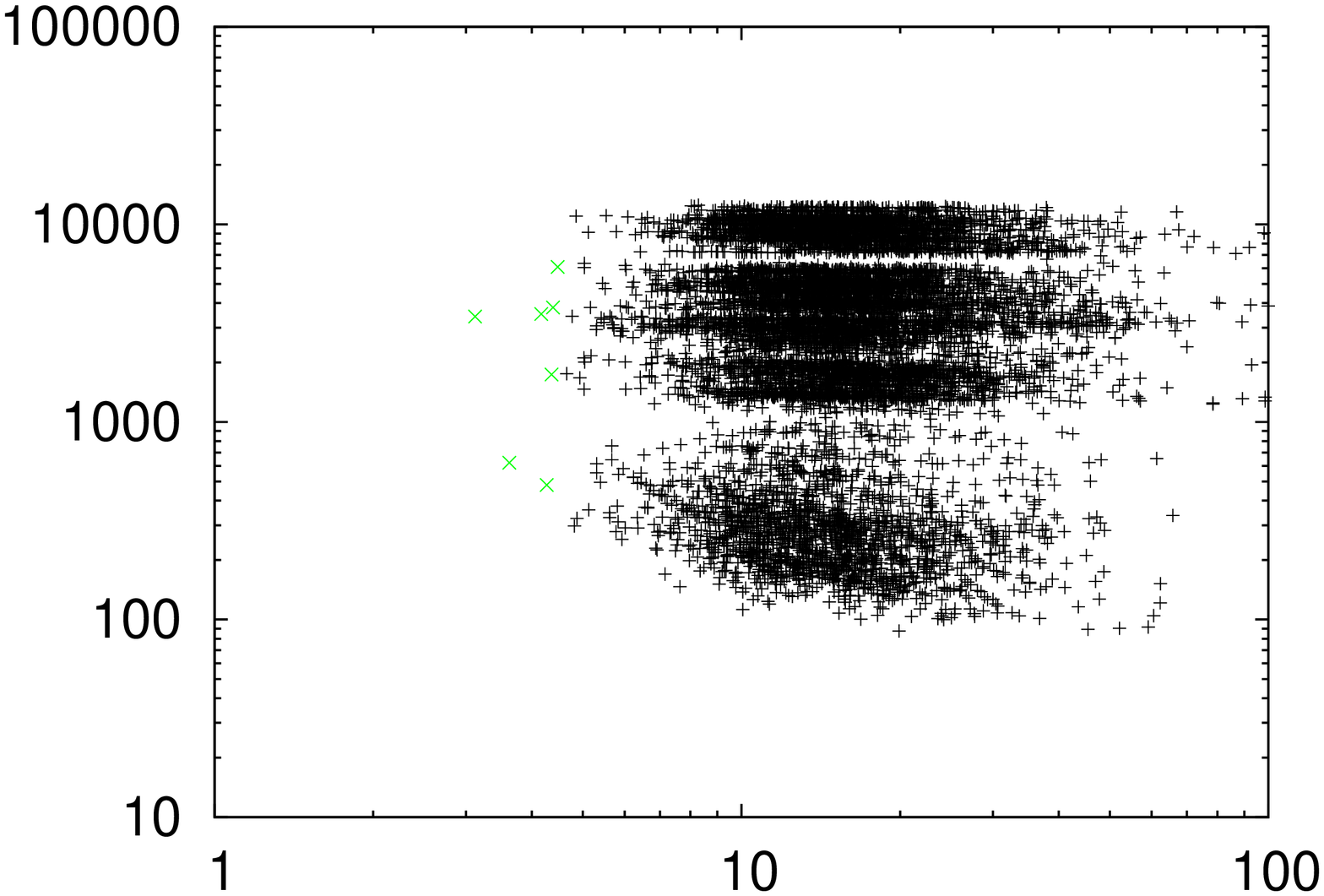}\\
%
  \put(220,0){\centering \bf \large Jitter [$m\,s^{-1}$]}
\end{tabular}
\caption{Sensitivity of selected system parameter determinations to assumptions regarding jitter.
The top row has $e_{outer}$ vs. Jitter, the second row has $P_{outer}$ vs. Jitter, the third row has $\eta = \varpi_{outer} - \varpi_{inner}$ vs. Jitter, and the bottom row has Goodness-of-fit vs. Jitter.
In the left-hand column we present results using only the subset of data known at the time of publication by B09, assuming a 2-planet fit. In the center and right-hand columns we present results obtained using the full data set of W10, with the central column assuming 2-planets + a linear trend, while the right-hand column assumes a 3-planet fit.
Systems with $\sigma_j< 2.0\,m\,s^{-1}$ are plotted using a red filled circle (gray in the print version), those with $2.0 < \sigma_j< 4.5\,m\,s^{-1}$ are plotted using a green cross (gray in the print version), and those with $\sigma_j > 4.5\,m\,s^{-1}$ are plotted using black $+$ symbols.
If one regards the jitter as being unknown and allow a range of values to be sampled by the MCMC routine, we can see that in this sample of plots for various system parameters, there are a much larger range of parameters which can plausibly fit the data than if one assumes $\sigma_j = 3.0\,m\,s^{-1}$.
The Effective-chi-square measure plotted here in the final row illustrates the rather large range of fit ``qualities'' that the different MCMC runs result in and that the 2-planet fit to B09 has rather poor fits at low jitter levels.
}
  \label{FIG:Jitter}
\end{figure}
%
In the B09  discovery paper, upon performing an RV fit (in conjunction with an analysis of the observed transit timing constraints) they found that a jitter of $3.0\,m\,s^{-1}$ was required in order to give a reduced $\chi^2$ figure of 1.0. 
Similarly W10 found a jitter of $3.4\,m\,s^{-1}$ was required in order to give a reduced $\chi^2$ figure of 1.0.
In the work of \citet{Wright05}, it was found that from a sample of $\sim 30$ stars similar to HAT-P-13 (mass, $\Mstar = 1.22\Msun$) that the distribution of jitters was such that the 20th, 50th and 80th percentiles were found to be 2.6, 4.0 and $6.2\,m\,s^{-1}$ respectively, with an upper bound of $\sim 15\,m\,s^{-1}$. It is thus plausible that the actual jitter value for HAT-P-13 is substantially different from the value of $\sim 3.0\,m\,s^{-1}$ used in B09  and W10. As discussed in the introduction, we feel that it is important that the jitter be placed on an equal footing and analyzed in the same manner as all of the other parameters that one can constrain using the RV data.  As such, we include the jitter as a prior in our MCMC analysis (see Section \ref{Method:MCMC} for details). We illustrate in \Fig{FIG:Jitter} the sensitivity of a number of the key parameters to the value of the jitter that is used in the MCMC analysis. 

To illustrate the general effect of including jitter as an integral part the MCMC analysis, we look at two different data cuts for HAT-P-13: The first set (S1) uses only the RV data that was included in the discovery paper of B09, while the second set (S2) uses the larger data set used in the later paper of W10. We have analyzed these data sets in such a manner as to afford direct comparison with the original papers: we analyze S1 assuming a two-planet solution (as was done in B09), while we analyze S2 by allowing for the possibility of an additional ``slope'' in the RV data (see Eqn 1 of W10) Additionally, for S2 we also perform a full 3-planet analysis. We plot some of the results of these MCMC analyses in \Fig{FIG:Jitter}. 

In general, it can be observed from the various plots in  \Fig{FIG:Jitter} that the MCMC routine investigates a wide range of jitter levels, and that (as one might expect), the larger the level of jitter it tries, the wider the range of planetary parameters it can accommodate. However, we see that this process is not completely unlimited, as there are few points in any of the plots which have jitters above $30\,m\,s^{-1}$, and the majority of them are significantly below this level. 

In more detail, we see that from the left-hand column of  \Fig{FIG:Jitter} (in which the S1 results are plotted), that if we restrict ourselves to examining jitter ranges $2.0 < \sigma_j < 3.0\,m\,s^{-1}$ (see the green crosses (gray in the print version) in \Fig{FIG:Jitter} which label regions with $\sigma_j = 3.0\,m\,s^{-1}\, \pm 50\%$), the eccentricity of the inner planet, $e_{inner}$, is constrained to be $0 < e_{inner} < 0.07$ (best fit value of $e_{inner}=0.017^{+0.013}_{-0.009}$, where the uncertainty figures cover $68\%$ of the data, i.e. they equate to 1-sigma error bars), whilst the difference in the alignment of the longitude of pericenters, $\eta = \varpi_{outer} - \varpi_{inner}$, can take on essentially the full range of available values: $ -180^{\circ} < \eta < + 180^{\circ}$ (best fit value of $\eta = 15.9^{50.6+}_{-83.7}$). If we allow a higher range of possible jitter, then the available parameter space expands, such that if we take all of the MCMC results with $0 < \sigma_j < 10\,m\,s^{-1}$, the eccentricity of the inner planet could take any value in the range $0 < e_{inner} < 0.15$ (best fit value of $e_{inner}=0.021^{+0.023}_{-0.012}$), while the eccentricity, $e_{outer}$, and period, $P_{outer}$, of the outer planet also expand to encompass ranges of $0.6<e_{outer}<0.95$ (best fit value of $e_{outer}=0.75^{+0.12}_{-0.08}$) and $410<P_{outer}<450$ days (best fit value of $P_{outer} = 430.2^{+6.3}_{-4.9}$) respectively.  Interestingly, we note that at such high values of $\sigma_j$, the alignment of the longitude of pericenters starts to favor orthogonal values, $\eta \approx \pm 90^{\circ}$.

The middle column in \Fig{FIG:Jitter} details the results of the 2-planet + slope analysis of S2, i.e. analogous to the analysis of W10. We see that there are now very few systems in the analysis which have jitter levels below $5\,m\,s^{-1}$. If we restrict our attention to the very few systems which have $\sigma_j = 3.0\,m\,s^{-1}\, \pm 50\%$), the eccentricity of the inner planet, $e_{inner}$, is constrained to be $0 < e_{inner} < 0.09$ (best fit value of $e_{inner}=0.038^{+0.022}_{-0.018}$, a value very much larger than that found in S1 (or indeed, in the analysis of W10).  If we allow a higher range of possible jitter, then the available parameter space expands, such that if we take all of the MCMC results with $0 < \sigma_j < 10\,m\,s^{-1}$, the eccentricity of the inner planet could take any value in the range $0 < e_{inner} < 0.2$ (best fit value of $e_{inner}=0.06^{+0.032}_{-0.026}$), while the eccentricity, $e_{outer}$, and period, $P_{outer}$, of the outer planet also expand to encompass ranges of $0.45<e_{outer}<0.9$ (best fit value of $e_{outer}=0.64^{+0.13}_{-0.08}$) and $430<P_{outer}<490$ days (best fit value of $P_{outer} = 450.6^{+8.4}_{-7.1}$) respectively. 

Finally, in the third column of  \Fig{FIG:Jitter}, we display the results from our 3-planet analysis of S2. The results pertaining to $e_{inner}$, $e_{outer}$, $P_{outer}$ and $\eta$ as displayed are qualitatively very similar to those of the 2-planet + slope analysis, as find that for $\sigma_j = 3.0\,m\,s^{-1}\, \pm 50\%$ the best fit values are  $e_{inner}=0.087^{+0.06}_{-0.039}$,  $e_{outer}=0.61^{+0.15}_{-0.09}$ and  $P_{outer} = 455.0^{+18.0}_{-13.7}$.  The fitted parameters for the third planet in this analysis are extremely ill-constrained. Given the qualitative similarity in the two sets of results for the S2 system for the inner planets and the very poor constraints that can be placed on any third planet, we chose to concentrate the rest of the analysis on the 2-planet + slope analysis to ensure ease of comparison with the published results of W10.

The goodness-of-fit parameters plotted in the final row are an effective-chi-square measure, used here to illustrate the rather large range of fit ``qualities'' that the different MCMC runs result in. We can see that in the 2-planet analysis of B09 in the left-hand column, it is particularly noticeable that the low jitter systems tend to have a worse fit (higher effective chi-square measure) than do the systems with higher jitter. The analyses of the W10 data show a wide range of effective chi-square measures right across the jitter range sampled by the MCMC routines. 

We provide two tables at the end of the subsequent section (Table \ref{TAB:PARAMETERS} and Table \ref{TAB:PARAMETERS2}) in which we list some key statistics for the fitted values of the various system parameters arising from our respective analyses of S1 and S2. We give the median value for each parameter, and then also give as uncertainty figures the spread in parameter values which are required to cover $68\%$ of the data in the sample, i.e. corresponding approximately to the 1-sigma deviation figure as quoted in B09. 

We discuss further some of the implications of these jitter figures in \S \ref{MCMCSummary}.

\subsection{Combining RV and TTV Data}\label{MCMC_PLUS_TTV}
%
\begin{figure*}
\centering
\begin{tabular}{ccc}
    	\psfig{figure=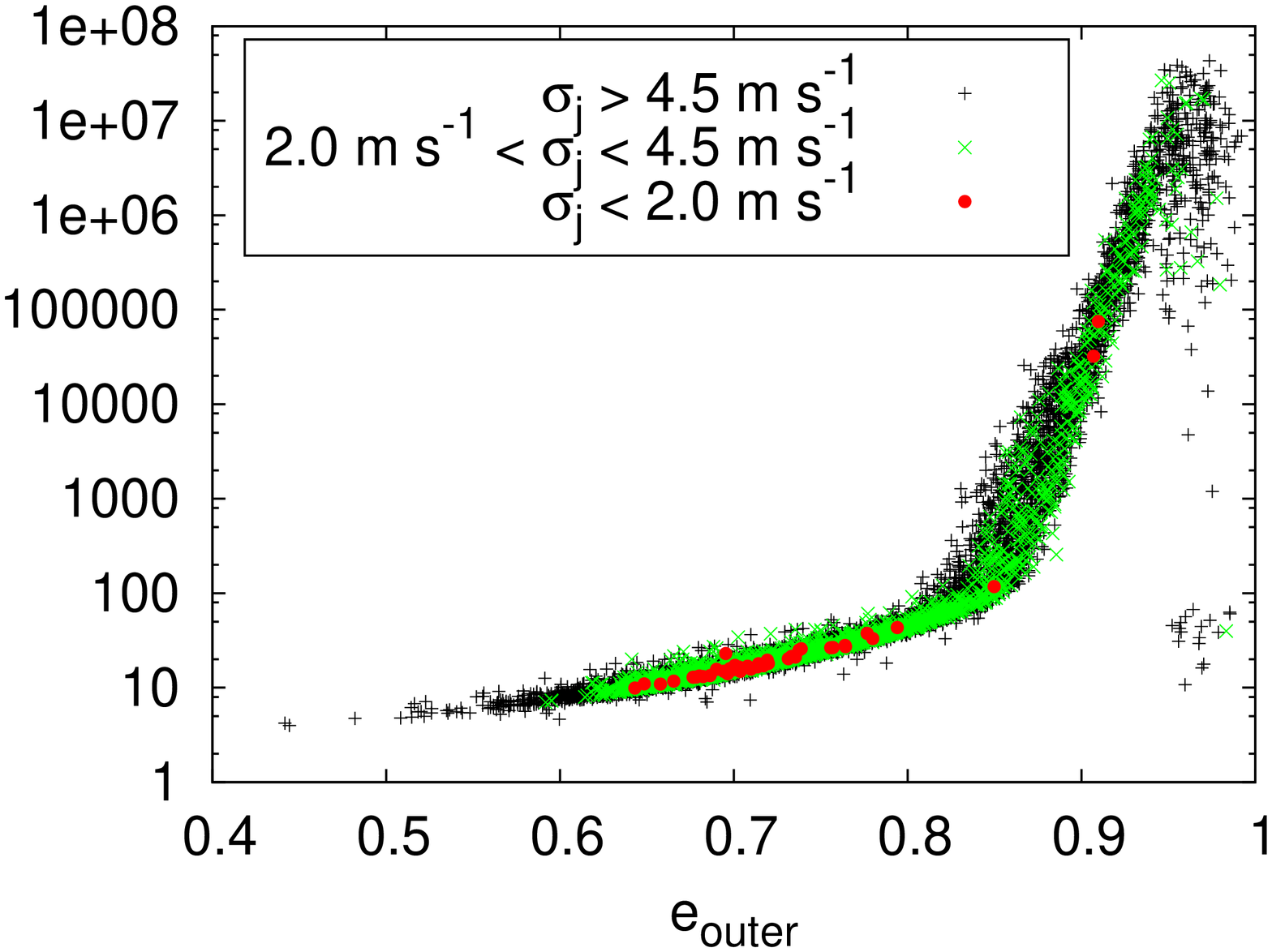,angle=-90,width=0.37\textwidth}&
    	\psfig{figure=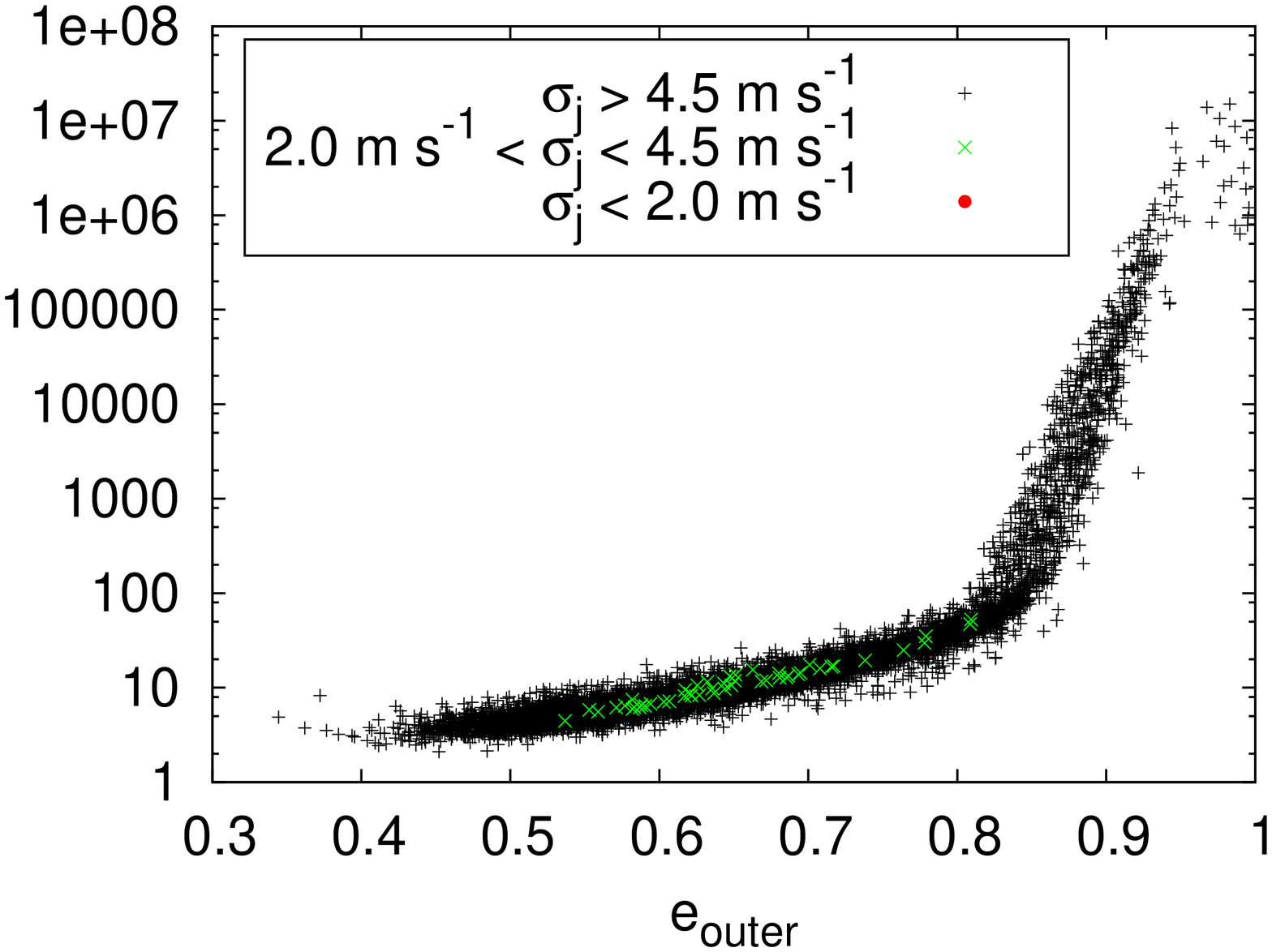,angle=-90,width=0.37\textwidth}\\
    	\psfig{figure=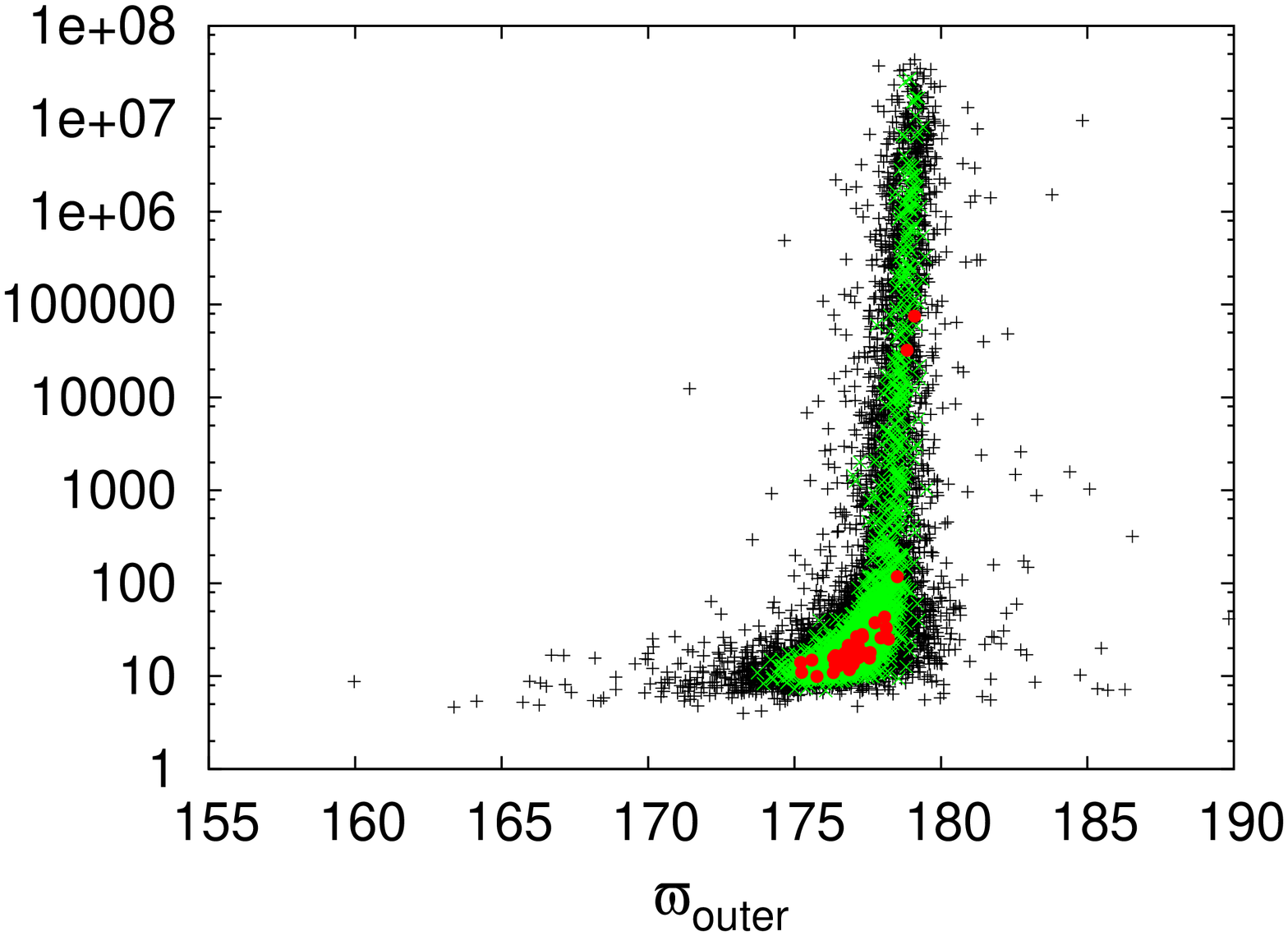,angle=-90,width=0.37\textwidth}&
    	\psfig{figure=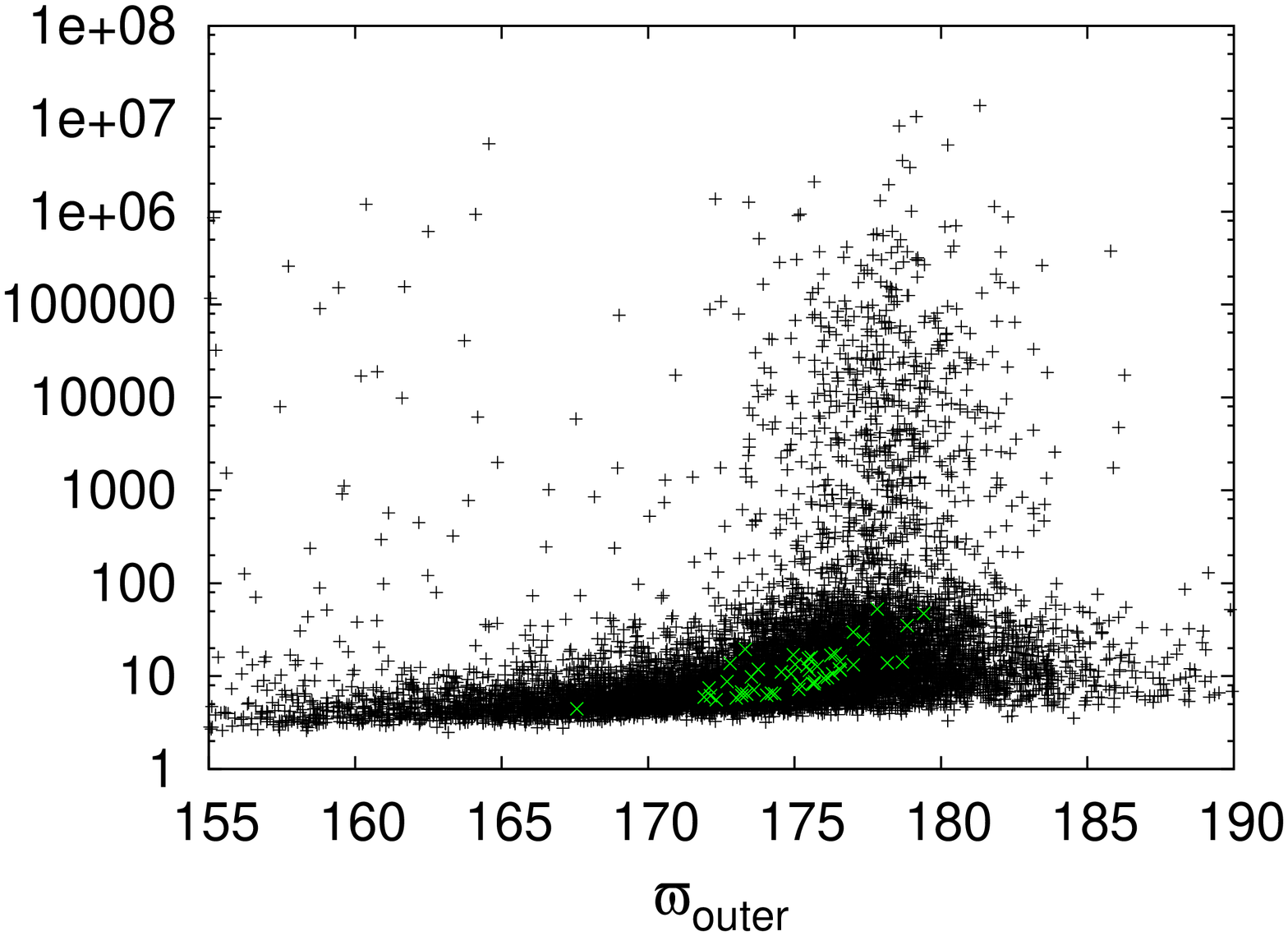,angle=-90,width=0.37\textwidth}\\
    	\psfig{figure=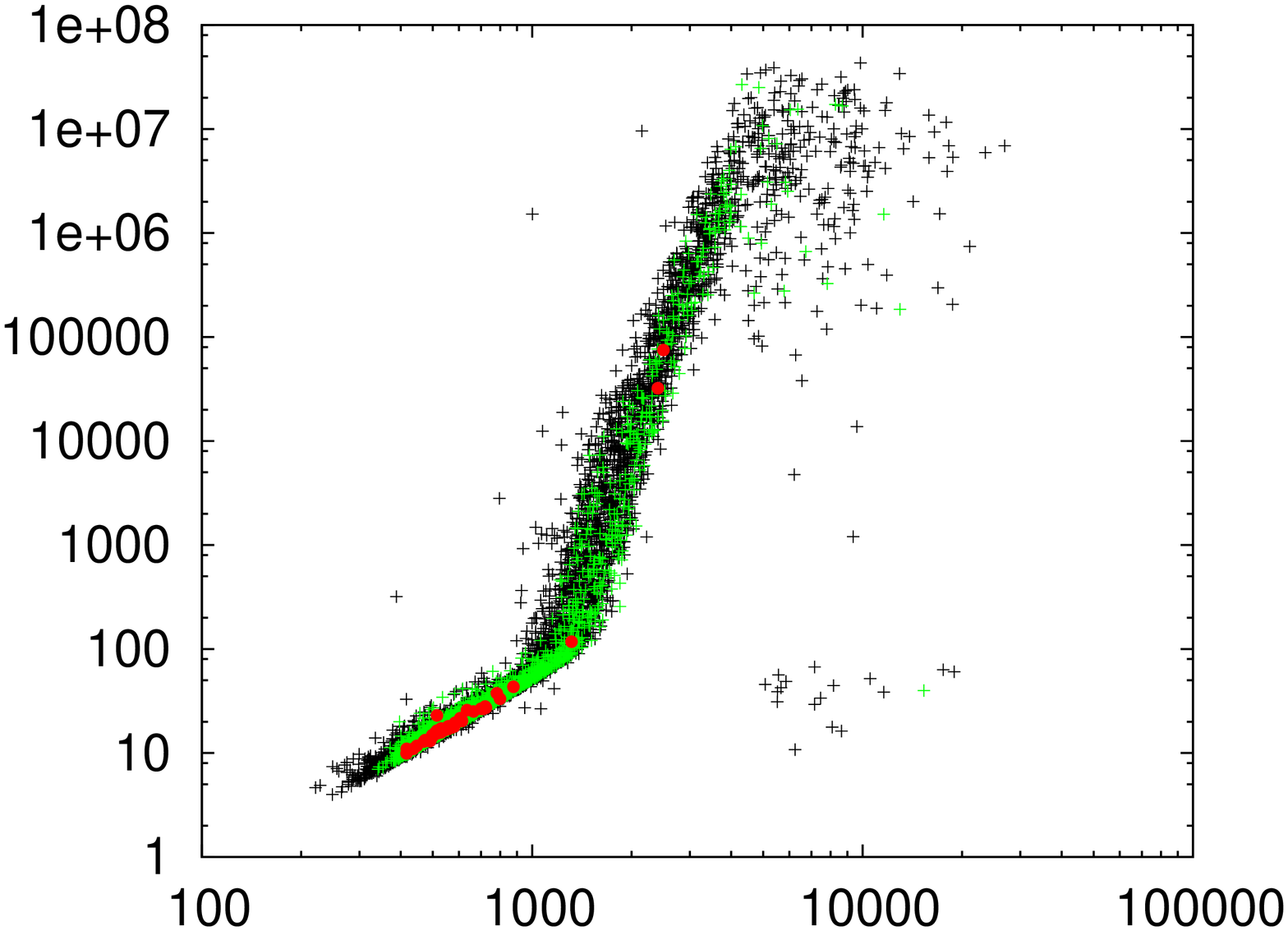,angle=-90,width=0.37\textwidth}&
    	\psfig{figure=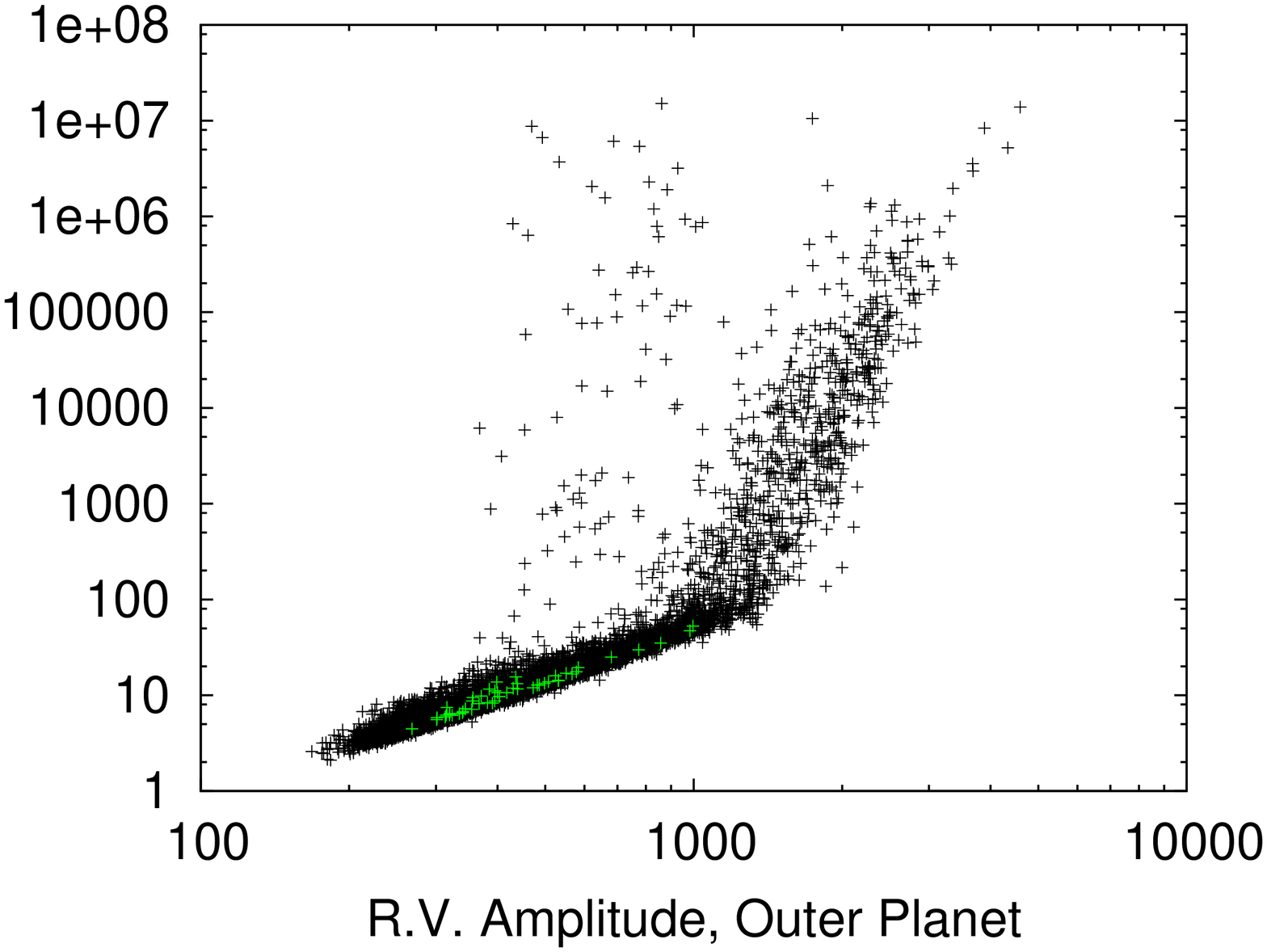,angle=-90,width=0.37\textwidth}\\
    	\psfig{figure=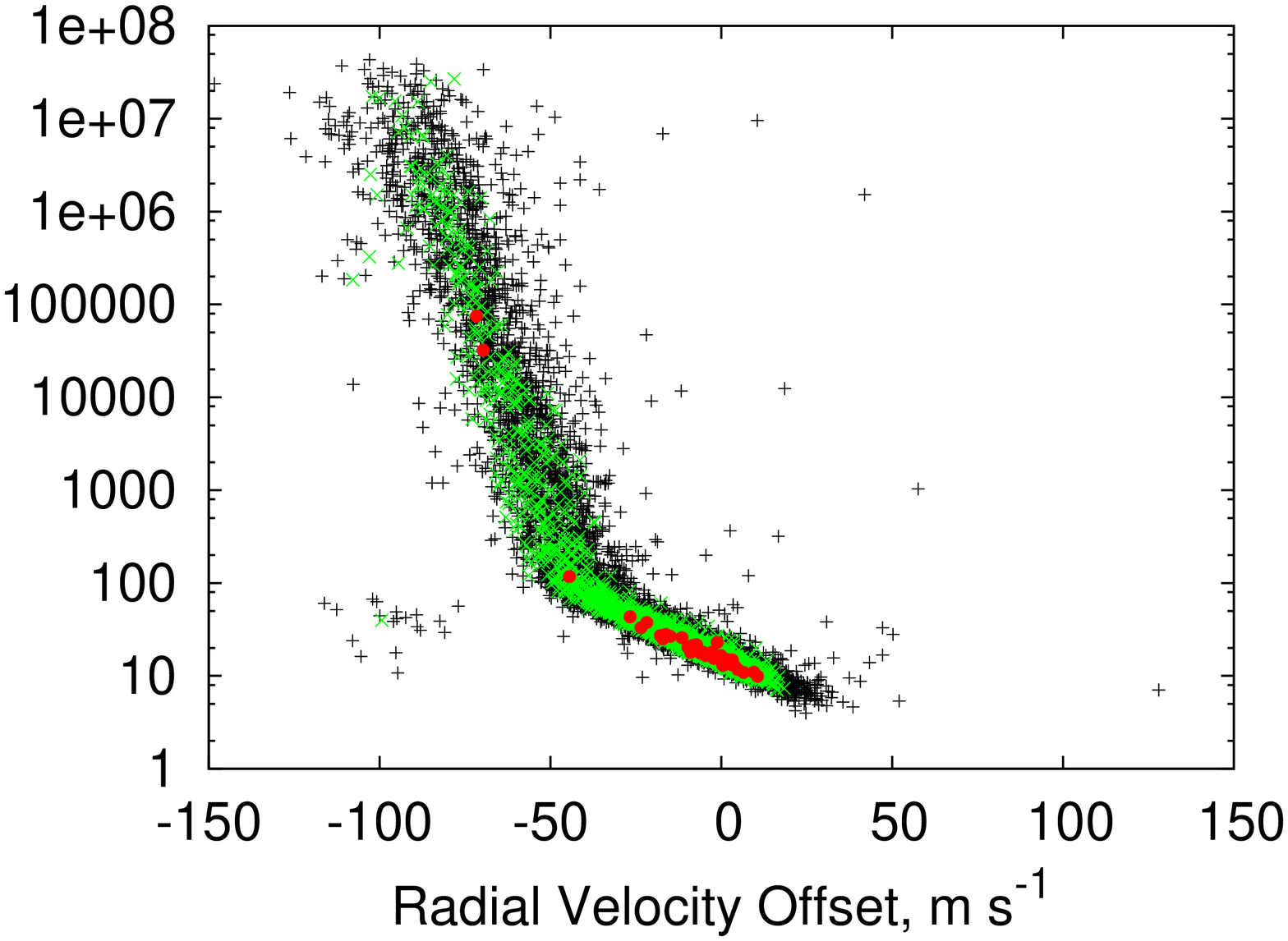,angle=-90,width=0.37\textwidth}&
    	\psfig{figure=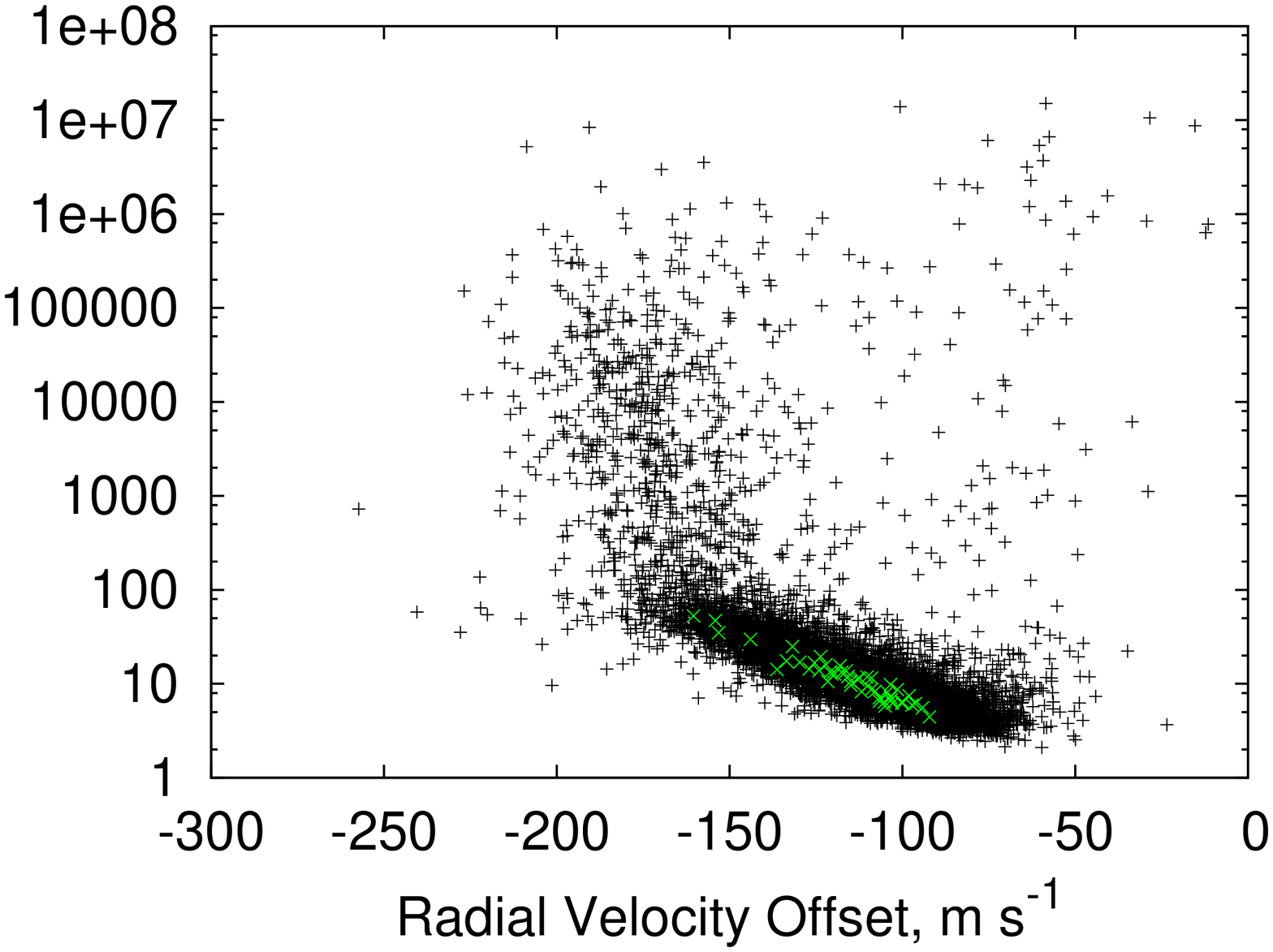,angle=-90,width=0.37\textwidth}\\
\end{tabular}
  \put(-425,-100){{\begin{sideways} \centering {\bf \large RMS TTV Amplitude, [s]} \end{sideways}}}
\caption{Understanding the constraints given by combining the TTVs and the Jitter: I.
TTV Amplitude vs. System Parameters. In the various plots above we show the RMS TTV amplitude as a function of various system parameters (planetary eccentricity, etc). As is Fig. 1, systems with $\sigma_j< 2.0\,m\,s^{-1}$ are plotted using a red circle (gray in the print version), those with $2.0 < \sigma_j< 4.5\,m\,s^{-1}$ are plotted using a green $\times$ (gray in the print version), and those with $\sigma_j > 4.5\,m\,s^{-1}$ are plotted using black $+$ symbols. 
In the left-hand column we present results using only the subset of data known at the time of publication by B09, assuming a 2-planet fit. In the right-hand column we present results obtained using the full data set of W10, assuming 2-planets + a linear trend.
We can see that imposing a cut such that the TTV amplitude is $< 100s$ helps to eliminate a sizeable proportion of the overall solutions, but note that the \emph{majority} of the systems with jitter $< 3.5\,m\,s^{-1}$ tend to naturally fall into areas which have TTV amplitudes $<100s$.
The quantities plotted above are the \emph{only} variables which have a potentially significant correlation with the RMS TTV amplitude.
We note that comparing the W10 data to the B09 data, once again the TTVs can be used to (approximately) constrain the eccentricity of the outer planet, but now is even worse at constraining the other parameters due to the much larger scatter.
}
\label{FIG:TTV1}
\end{figure*}
%
\begin{figure*}
\centering
\begin{tabular}{ccc}
  \psfig{figure=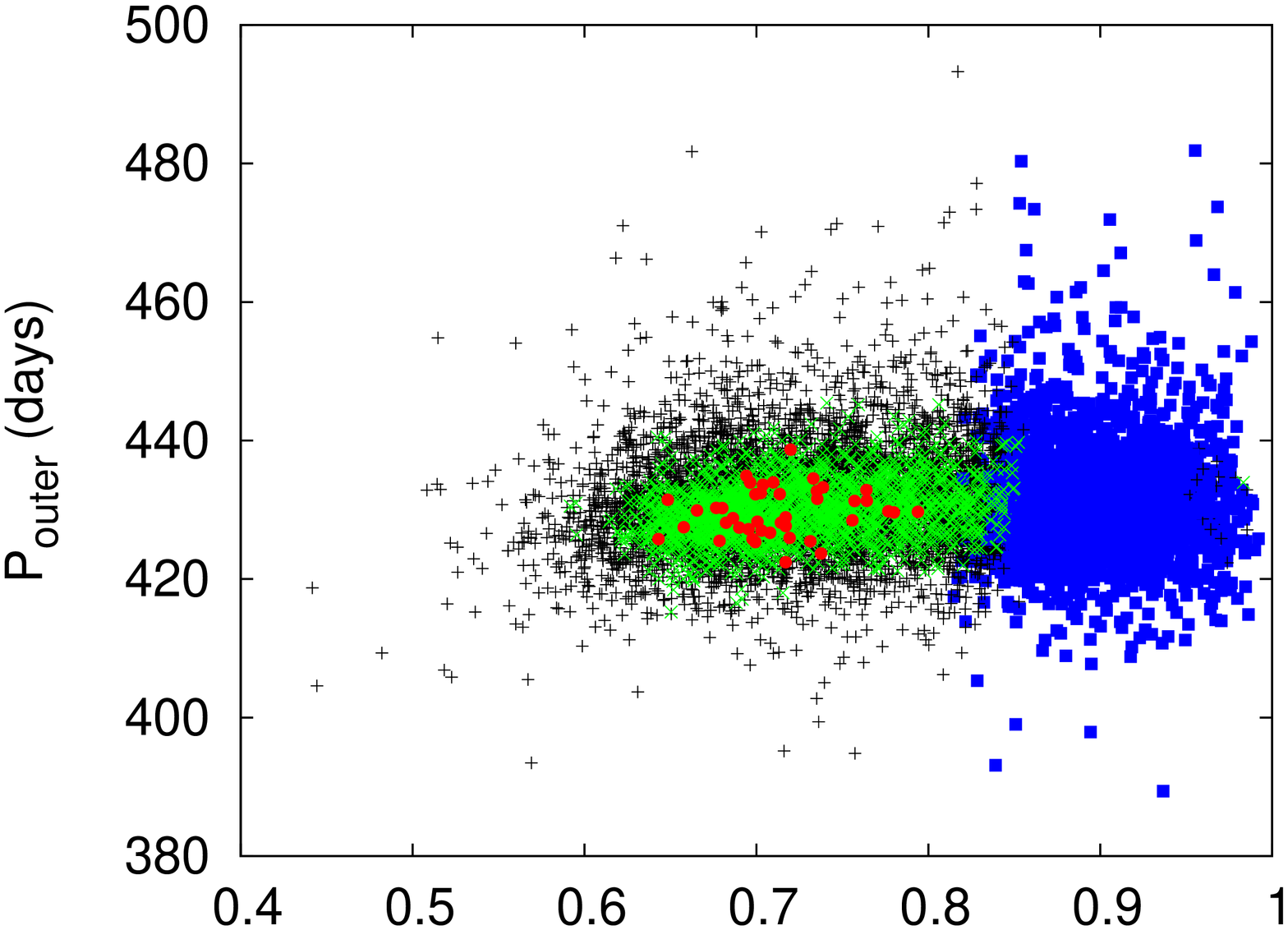,angle=-90,width=0.43\textwidth}&
  \psfig{figure=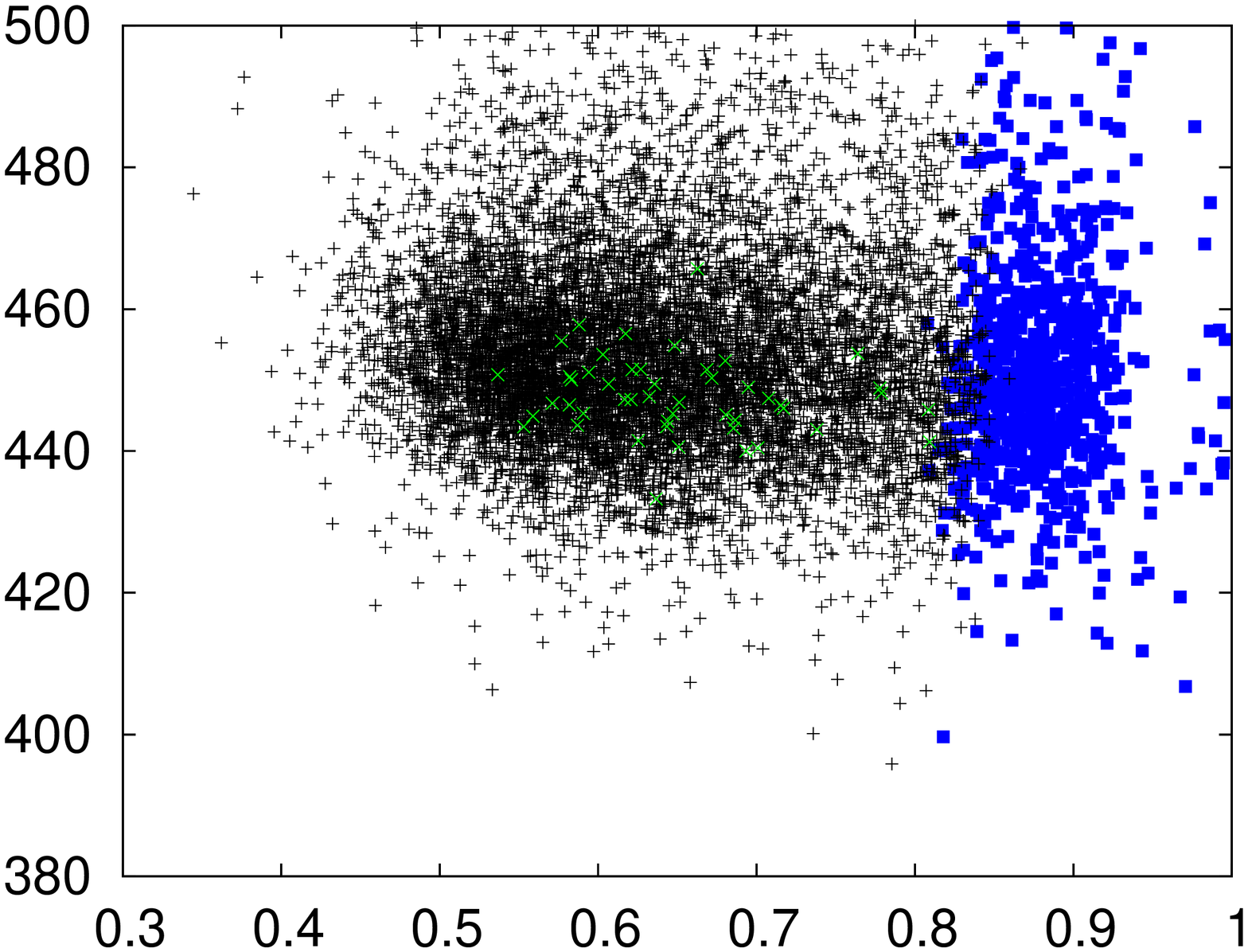,angle=-90,width=0.43\textwidth}\\
  \psfig{figure=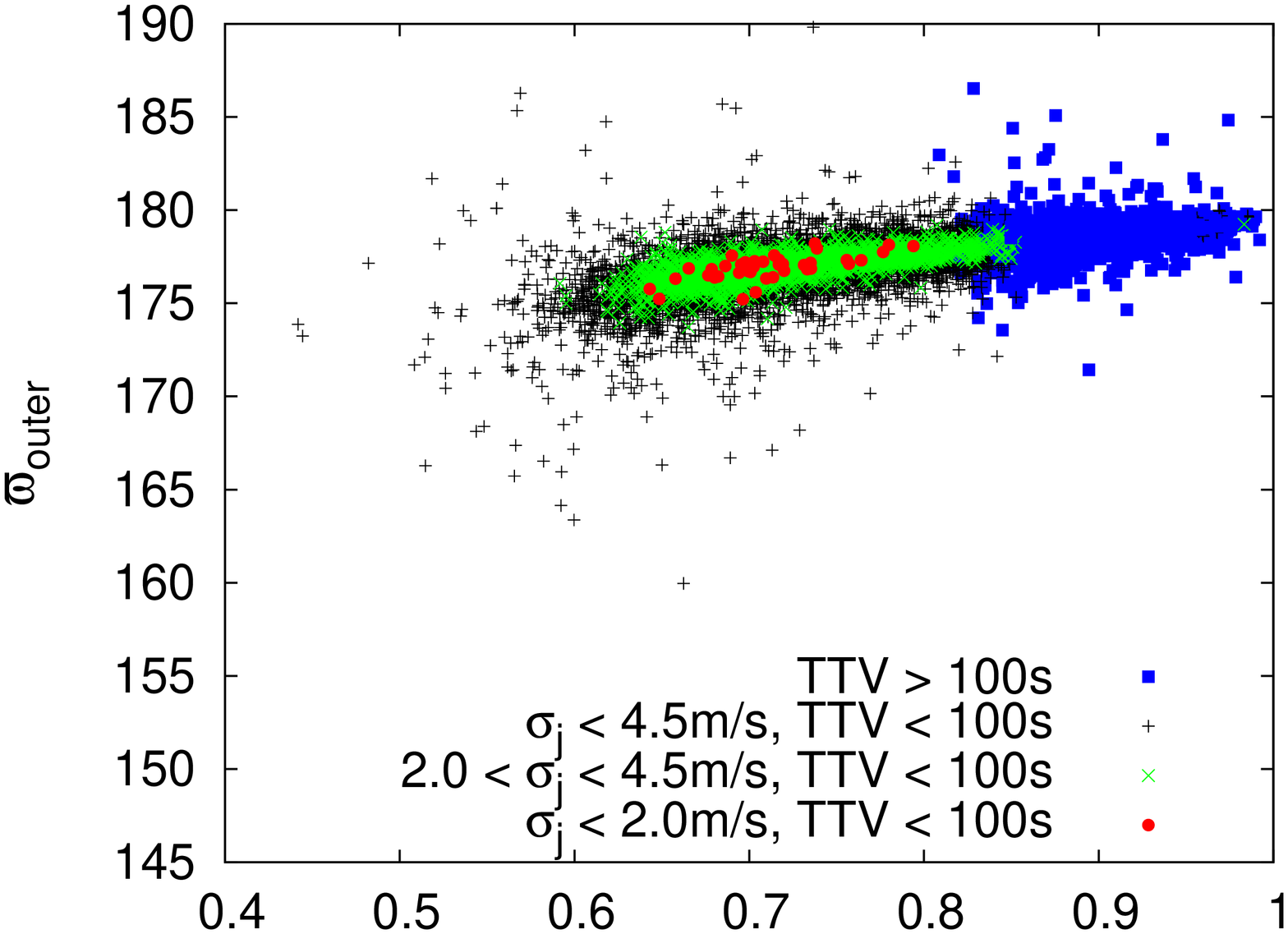,angle=-90,width=0.43\textwidth}&
  \psfig{figure=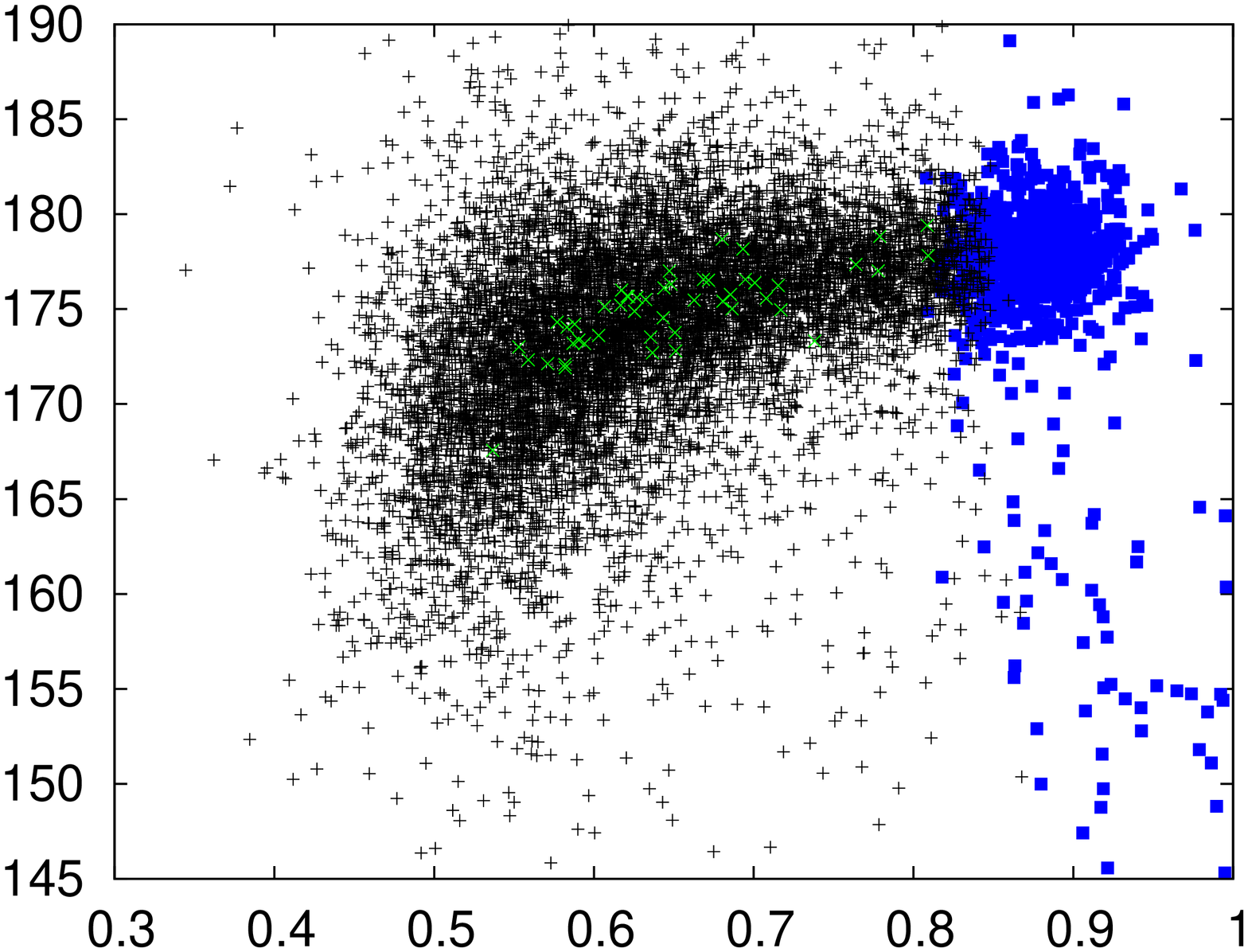,angle=-90,width=0.43\textwidth}\\
  \psfig{figure=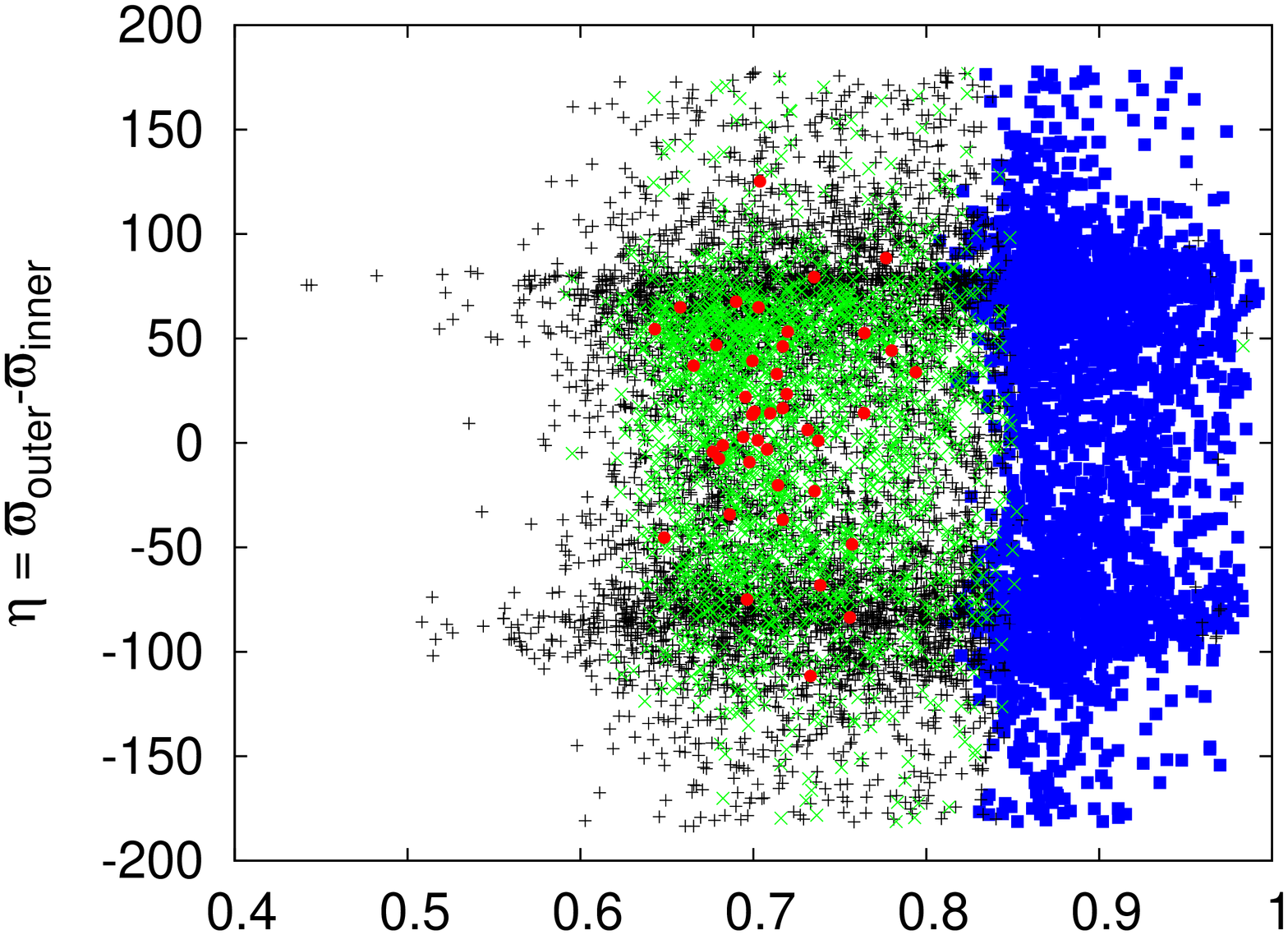,angle=-90,width=0.43\textwidth}&
  \psfig{figure=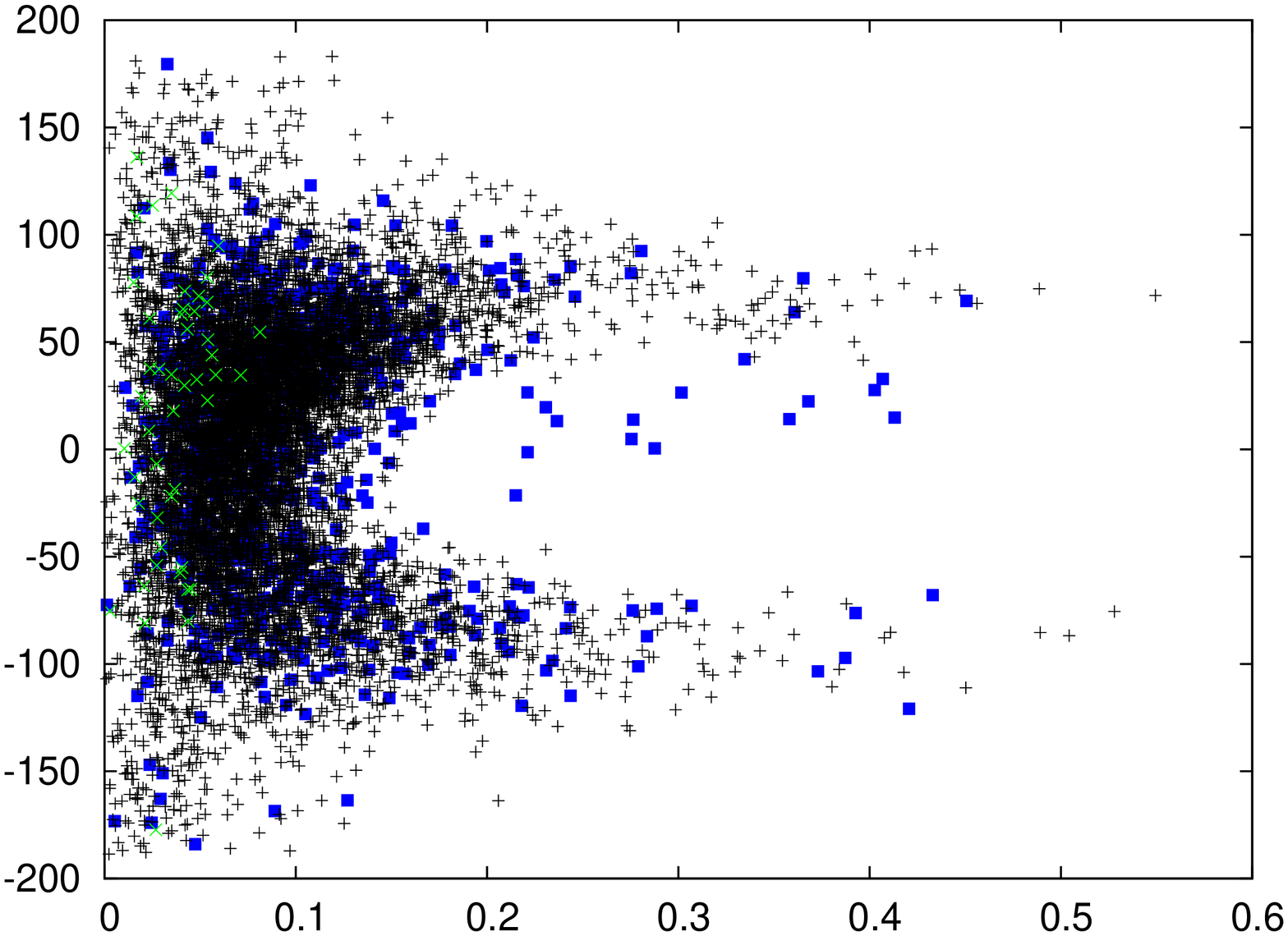,angle=-90,width=0.43\textwidth}\\
%
  \put(200,0){\centering \bf \large $e_{outer}$}
\end{tabular}
\label{FIG:TTV2}
\caption{Understanding the constraints given by combining the TTVs and the Jitter: II.
We plot a variety of the possible system configurations arising from the MCMC analysis, and then constrain the possibilities by indicating (i) Using blue squares (gray in the print version) the systems which have TTV amplitudes bigger than 100s, and then (ii) In other colors indicate the range of jitter assumed for the systems: systems with $\sigma_j< 2.0\,m\,s^{-1}$ are plotted using a red circle (gray in the print version), those with $2.0 < \sigma_j< 4.5\,m\,s^{-1}$ are plotted using a green cross (gray in the print version), and those with $\sigma_j > 4.5\,m\,s^{-1}$ are plotted using black addition symbols. 
As in Fig. 1, in the left-hand column we present results using only the subset of data known at the time of publication by B09, assuming a 2-planet fit. In the right-hand column we present results obtained using the full data set of W10, assuming 2-planets + a linear trend.
We find that the eccentricity of the outer planet is strongly constrained to be $\lsim 0.85$, but that other parameters are largely unaffected / unconstrained.
}
  \label{FIG:TTV2}
\end{figure*}
%
In their discovery paper, B09 also perform an analysis of the transit timings to look for any evidence of TTVs. They find that the TTVs in the system are restricted to be $\lsim 100$ seconds. We wish to add-in this constraint directly to the MCMC analysis performed in section \ref{MCMC}. To do this, we take a 10,000-strong subsample of the MCMC systems (see section \ref{Method} for further discussion of methodology) and use these systems as the basis for a TTV investigation. 

We take the fitted parameters for each of the subsample systems, and use these to set up the planetary masses and orbits in an n-body simulation. Given that the relative inclination of the planets is unconstrained in the RV analysis, we performed the simulations assuming coplanarity. This simulation is then run in the manner described in section \ref{Method:TTV} and a TTV analysis performed on the inner planet. This allows us to directly investigate whether any particular range of the MCMC subsample can be excluded by considering the observational TTV constraints. Extremely high TTV signals generally arise due to close approaches between objects in the simulation, giving a strong indiction that such systems are unstable (see Veras et al 2010 and Payne et al. 2010 for further discussion of such high TTV signals, and the likely Hill and/or Lagrange stability.). However, irrespective of whether any such systems are absolutely stable or not, we can use the fact that they have TTV amplitudes $> 100s$ to remove them from the analysis.

We show a sample of the results from the TTV analysis in \Fig{FIG:TTV1}. On the left-hand side we plot results for a 2-planet analysis of the S1 data set, while on the right-hand side we plot results for the 2-planet + slope analysis of the S2 data set. The majority of the observable parameters do \emph{not} show any obvious correlation with the RMS TTV amplitude (we provide some examples of such plots in an appendix). However, the overall RV offset, as well as the RV Amplitude, $K_{outer}$ , longitude of pericenter, $\varpi_{outer}$, and the eccentricity, $e_{outer}$, of the outer planet all have such a correlation. We could thus hope to use this upper TTV amplitude cut-off of $100s$ to limit the allowed range of these variables. Whilst this is certainly possible on a gross scale, we note that when we color-code the results according to the assumed jitter (as done in \Fig{FIG:TTV1}), the majority of the low jitter systems tend to fall into the low-TTV area of parameter space. This means that using the TTV amplitude constraints is probably only of some marginal help in restricting the range of plausible system parameters to be considered.

In \Fig{FIG:TTV2}, we go on to plot a selection of the key orbital parameters, using colored symbols to show the regions of parameter space which are favored or ruled out by the combined jitter and TTV amplitude analyses discussed above. As in  \Fig{FIG:TTV1}, on the left-hand side we plot results for 2-planet analysis of the S1 data set, while on the right-hand side we plot results for the 2-planet + slope analysis of the S2 data set. These suggest that for the inner planet, even for a fairly tightly constrained jitter of $2.0 - 4.5 m\,s^{-1}$, there can be substantial variation in the eccentricity, $e_{inner}$, such that it can take any value $0 < e_{inner} < 0.07$;

For the outer planet, much greater variations are possible:
(i) Perhaps the strongest result we find from using an observed constraint on the TTV amplitude of $\lsim 100$ seconds, is in constraining the eccentricity of the outer planet to be $e_{outer} < 0.85$, irrespective of assumed jitter values, giving an approximate overall range of $0.5 < e_{outer} < 0.85$ (best fit value of $e_{outer} = 0.017^{+0.013}_{-0.009}$, assuming jitter is in the range 2.0 to 4.5 $m\,s^{-1}$). This is clearly comprehensible in the light of the above argument concerning stability: extremely high eccentricity values for the outer planet can easily lead to close approaches and/or orbit-crossing. These  results stand for the analyses of both S1 and S2.
(ii) While the longitude of pericenter of the outer planet ($ \varpi_{outer} $) is strongly constrained to be $\sim 180$ (best fit value of $\varpi_{outer}=177.0^{+0.7}_{-0.9}$ for S1, $\varpi_{outer}=171.2^{+6.3}_{-7.3}$ for S2),  the longitude of pericentre of the inner planet can take almost any value, meaning that there is no strong preference to indicate that the planetary pericentres are aligned ($-180^{\circ}<\eta<+180^{\circ}$, best fit value for S1 of $\eta =16.0 ^{+50.2}_{-82.0}$, while for S2 the best fit value has  $\eta =3.2 ^{+64.4}_{-82.5}$).
(iii) In the S1 analysis, the outer planet can have periods $420 < P_{outer} < 440$ days (best fit value of $P_{outer} = 429.7^{+4.3}_{-3.6}$), while in the S2 analysis, the period is pushed to higher values $430 < P_{outer} < 490$ days (best fit value of $P_{outer} = 455.0^{+18.0}_{-13.7}$).

Note that Table \ref{TAB:PARAMETERS} also contains a section in which the figures are analyzed for systems with TTVs less than 100 seconds only. As an example of the data contained therein, it can be seen that the lower allowed range on the outer planet's eccentricity from the TTV analysis manifests itself in the table as a reduction in both the upper uncertainty and the median value of the fitted outer eccentricity.

\begin{table}
\centering
\caption{We provide a table summarizing the results of our MCMC analysis for the HAT-P-13 system \emph{using just the data from the work of \citet{Bakos09}}. The different columns give the figures which result from different jitter assumptions, with the top half of the table showing results for \emph{all} data, whereas the bottom half shows the data for just those systems found to have TTV $<100$ seconds. The figures given are the \emph{median} values from the 10,000 selected MCMC realization, with the \emph{$\pm$} figures giving the spread required in order to encompass $68\%$ of the data, i.e. giving a figure which approximates the 1-sigma result tabulated in the discovery paper of \citet{Bakos09}.}
\begin{tabular}{llcccccc}
\hline
TTV  & 
Parameter  & 
$\sigma_j^{\rm b} < 3.0$ & 
$\sigma_j^{\rm b} < 4.5$ & 
$\sigma_j^{\rm b} < 10.0$ & 
All $\sigma_j^{\rm b}$ & 
$2.5 < \sigma_j^{\rm b} < 3.5$ & 
$2.0 < \sigma_j^{\rm b} < 4.5$ 
\\
\hline
\\
ALL & $P_{inner}$ (days)         &      $2.91626 ^{+0.00002}_{-0.00001}$         &      $2.91626 ^{+0.00002}_{-0.00002}$         &      $2.91626 ^{+0.00002}_{-0.00002}$         &      $2.91626 ^{+0.00002}_{-0.00002}$         &      $2.91626 ^{+0.00002}_{-0.00001}$      &       $2.91626 ^{+0.00002}_{-0.00002}$         \\
ALL & $e_{inner}$        &      $0.017 ^{+0.009}_{-0.009}$       &      $0.017 ^{+0.012}_{-0.009}$       &      $0.021 ^{+0.023}_{-0.012}$       &      $0.023 ^{+0.032}_{-0.013}$       &      $0.016 ^{+0.012}_{-0.009}$       &      $0.017 ^{+0.013}_{-0.009}$       \\
ALL & $\varpi_{inner}^{\rm c}$   &      $157.1 ^{+77.0}_{-37.9}$         &      $161.7 ^{+85.9}_{-49.8}$         &      $164.9 ^{+97.4}_{-63.2}$         &      $164.4 ^{+99.4}_{-65.1}$         &      $158.1 ^{+86.3}_{-42.3}$         &      $161.9 ^{+86.1}_{-50.1}$      \\
ALL & $T_{P,inner}^{\rm a}$      &      $4780.5 ^{+0.7}_{-0.3}$       &      $4780.6 ^{+0.7}_{-0.5}$       &      $4780.8 ^{+0.8}_{-0.7}$       &      $4780.9 ^{+0.7}_{-0.8}$       &      $4780.6 ^{+0.8}_{-0.4}$       &      $4780.6 ^{+0.7}_{-0.5}$       \\
ALL & $P_{outer}$ (days)         &      $429.4 ^{+3.8}_{-3.2}$   &      $429.7 ^{+4.4}_{-3.7}$   &      $430.2 ^{+6.3}_{-4.9}$   &      $430.3 ^{+7.0}_{-5.4}$   &      $429.9 ^{+4.2}_{-3.8}$   &      $429.7 ^{+4.4}_{-3.7}$   \\
ALL & $e_{outer}$        &      $0.71 ^{+0.10}_{-0.04}$  &      $0.73 ^{+0.12}_{-0.06}$  &      $0.75 ^{+0.12}_{-0.08}$  &      $0.75 ^{+0.12}_{-0.08}$  &      $0.72 ^{+0.12}_{-0.05}$  &      $0.73 ^{+0.12}_{-0.06}$  \\
ALL & $\varpi_{outer}^{\rm c}$   &      $177.1 ^{+0.9}_{-0.8}$   &      $177.2 ^{+1.0}_{-0.9}$   &      $177.4 ^{+1.1}_{-1.2}$   &      $177.4 ^{+1.2}_{-1.4}$   &      $177.1 ^{+1.0}_{-0.9}$   &      $177.2 ^{+1.0}_{-1.0}$   \\
ALL & $T_{P,outer}^{\rm a}$      &      $4890.1 ^{+0.6}_{-0.7}$       &      $4890.2 ^{+0.7}_{-0.8}$       &      $4890.1 ^{+0.9}_{-1.0}$       &      $4890.1 ^{+1.1}_{-1.2}$       &      $4890.1 ^{+0.7}_{-0.7}$       &      $4890.2 ^{+0.7}_{-0.8}$       \\
ALL & $\eta$   &      $20.6 ^{+38.4}_{-75.3}$  &      $15.6 ^{+50.7}_{-83.3}$  &      $12.3 ^{+64.8}_{-95.8}$  &      $12.2 ^{+67.1}_{-98.5}$  &      $17.7 ^{+44.9}_{-81.6}$  &      $15.9 ^{+50.6}_{-83.7}$        \\
\hline
\\
ALL & $T_{T,outer},Bakos^{\rm a}$        &      $4872.4 ^{+8.3}_{-4.0}$       &      $4874.0 ^{+9.2}_{-5.5}$       &      $4875.9 ^{+8.9}_{-7.6}$       &      $4876.1 ^{+9.1}_{-8.0}$       &      $4873.0 ^{+9.8}_{-4.3}$       &      $4874.1 ^{+9.2}_{-5.6}$    \\
ALL & $T_{T,outer},2010^{\rm a}$         &      $5302.6 ^{+9.4}_{-6.6}$       &      $5304.6 ^{+9.4}_{-7.9}$       &      $5306.7 ^{+10.2}_{-9.8}$      &      $5307.0 ^{+10.9}_{-10.4}$     &      $5303.9 ^{+9.9}_{-7.3}$       &      $5304.6 ^{+9.5}_{-7.9}$    \\
ALL & $T_{T,outer},2011^{\rm a}$         &      $5732.1 ^{+12.0}_{-9.2}$      &      $5734.3 ^{+12.1}_{-10.4}$     &      $5736.7 ^{+15.1}_{-12.8}$     &      $5737.1 ^{+16.5}_{-13.8}$     &      $5733.8 ^{+12.7}_{-10.2}$     &      $5734.3 ^{+12.2}_{-10.4}$  \\
ALL & $T_{T,outer},2012^{\rm a}$         &      $6161.3 ^{+14.0}_{-11.5}$     &      $6164.0 ^{+15.7}_{-13.2}$     &      $6166.6 ^{+21.2}_{-16.6}$     &      $6167.1 ^{+23.4}_{-18.0}$     &      $6164.1 ^{+15.6}_{-13.6}$     &      $6164.1 ^{+15.9}_{-13.2}$  \\
\hline
\hline
\\
$< 100s$ & $P_{inner}$ (days)    &      $2.91626 ^{+0.00002}_{-0.00001}$         &      $2.91626 ^{+0.00002}_{-0.00002}$         &      $2.91626 ^{+0.00002}_{-0.00002}$         &      $2.91626 ^{+0.00002}_{-0.00002}$         &      $2.91626 ^{+0.00002}_{-0.00001}$      &       $2.91626 ^{+0.00002}_{-0.00002}$         \\
$< 100s$ & $e_{inner}$   &      $0.017 ^{+0.009}_{-0.010}$       &      $0.017 ^{+0.013}_{-0.009}$       &      $0.021 ^{+0.023}_{-0.012}$       &      $0.023 ^{+0.030}_{-0.013}$       &      $0.016 ^{+0.012}_{-0.009}$       &      $0.017 ^{+0.013}_{-0.009}$       \\
$< 100s$ & $\varpi_{inner}^{\rm c}$      &      $155.8 ^{+79.9}_{-36.4}$         &      $161.2 ^{+85.0}_{-49.1}$         &      $165.3 ^{+96.0}_{-62.7}$         &      $165.5 ^{+97.6}_{-65.5}$         &      $156.8 ^{+85.9}_{-41.0}$         &      $161.0 ^{+86.2}_{-48.9}$       \\
$< 100s$ & $T_{P,inner}^{\rm a}$         &      $4780.5 ^{+0.7}_{-0.3}$       &      $4780.6 ^{+0.7}_{-0.4}$       &      $4780.8 ^{+0.8}_{-0.7}$       &      $4780.9 ^{+0.7}_{-0.8}$       &      $4780.5 ^{+0.8}_{-0.4}$       &      $4780.6 ^{+0.7}_{-0.5}$    \\
$< 100s$ & $P_{outer}$ (days)    &      $429.4 ^{+3.8}_{-3.2}$   &      $429.7 ^{+4.3}_{-3.6}$   &      $430.2 ^{+6.0}_{-4.8}$   &      $430.3 ^{+6.7}_{-5.2}$   &      $430.0 ^{+3.9}_{-3.9}$   &      $429.7 ^{+4.3}_{-3.6}$   \\
$< 100s$ & $e_{outer}$   &      $0.70 ^{+0.07}_{-0.04}$  &      $0.71 ^{+0.07}_{-0.05}$  &      $0.72 ^{+0.08}_{-0.06}$  &      $0.72 ^{+0.08}_{-0.06}$  &      $0.71 ^{+0.07}_{-0.04}$  &      $0.72 ^{+0.07}_{-0.05}$  \\
$< 100s$ & $\varpi_{outer}^{\rm c}$      &      $176.9 ^{+0.7}_{-0.7}$   &      $177.0 ^{+0.7}_{-0.8}$   &      $177.0 ^{+0.9}_{-1.1}$   &      $177.0 ^{+1.0}_{-1.3}$   &      $176.9 ^{+0.8}_{-0.7}$   &      $177.0 ^{+0.7}_{-0.9}$   \\
$< 100s$ & $T_{P,outer}^{\rm a}$         &      $4890.1 ^{+0.5}_{-0.7}$       &      $4890.1 ^{+0.7}_{-0.8}$       &      $4890.1 ^{+0.9}_{-1.0}$       &      $4890.0 ^{+1.0}_{-1.2}$       &      $4890.1 ^{+0.6}_{-0.7}$       &      $4890.1 ^{+0.7}_{-0.8}$    \\
$< 100s$ & $\eta$      &      $20.6 ^{+37.6}_{-75.3}$  &      $15.9 ^{+50.0}_{-82.3}$  &      $11.7 ^{+64.2}_{-94.6}$  &      $11.0 ^{+66.8}_{-96.9}$  &      $20.5 ^{+42.2}_{-84.3}$  &      $16.0 ^{+50.2}_{-82.9}$ \\
\hline
\\
$< 100s$ & $T_{T,outer},Bakos^{\rm a}$   &      $4871.7 ^{+6.0}_{-3.6}$       &      $4872.5 ^{+6.4}_{-4.5}$       &      $4873.1 ^{+6.7}_{-5.8}$       &      $4873.1 ^{+6.7}_{-6.1}$       &      $4872.0 ^{+6.5}_{-3.9}$       &      $4872.6 ^{+6.4}_{-4.5}$    \\
$< 100s$ & $T_{T,outer},2010^{\rm a}$    &      $5301.5 ^{+7.5}_{-5.8}$       &      $5302.5 ^{+8.2}_{-6.4}$       &      $5303.6 ^{+8.9}_{-8.0}$       &      $5303.7 ^{+9.4}_{-8.5}$       &      $5301.9 ^{+8.9}_{-5.9}$       &      $5302.5 ^{+8.2}_{-6.5}$    \\
$< 100s$ & $T_{T,outer},2011^{\rm a}$    &      $5730.5 ^{+10.5}_{-8.0}$      &      $5732.2 ^{+11.9}_{-9.4}$      &      $5733.5 ^{+14.1}_{-11.4}$     &      $5733.7 ^{+15.5}_{-12.4}$     &      $5732.0 ^{+12.4}_{-9.2}$      &      $5732.2 ^{+11.9}_{-9.4}$   \\
$< 100s$ & $T_{T,outer},2012^{\rm a}$    &      $6159.9 ^{+13.5}_{-10.7}$     &      $6161.7 ^{+16.1}_{-12.6}$     &      $6163.5 ^{+20.1}_{-15.7}$     &      $6163.8 ^{+21.8}_{-17.1}$     &      $6162.0 ^{+16.8}_{-13.0}$     &      $6161.7 ^{+16.1}_{-12.6}$  \\
\hline
\end{tabular}
\label{TAB:PARAMETERS}
We strove to make our data comparable to the discovery paper where at all possible: as such, it should be noted that the parameter fits given for $P$, $e$, $\varpi$ and $T_{peri}$ are based on a coplanar MCMC analysis, while the transit time calculations for the outer planet are made using the assumption that the outer planet is inclined at $90^{\circ}$ to the plane of the sky.\\
$^{\rm a}$ All transit and pericenter passage times are in Julian days - 2,450,000\\
$^{\rm b}$ All jitter constraints are in $m\,s^{-1}$\\
$^{\rm c}$ All pericenter alignments are in degrees
%
\end{table}
%
\begin{table}
\centering
\caption{We provide a table summarizing the results of our MCMC analysis for the HAT-P-13 system \emph{using the full set of data available from W10}. The different columns give the figures which result from different jitter levels, with the top half of the table showing results for \emph{all} data, whereas the bottom half shows the data for just those systems found to have TTV $<100$ seconds. The figures given are the \emph{median} values from the 10,000 selected MCMC realization, with the \emph{$\pm$} figures giving the spread required in order to encompass $68\%$ of the data, i.e. giving a figure which approximates to a 1-sigma result.}
\begin{tabular}{llcccccc}
\hline
TTV  & 
Parameter  & 
$\sigma_j^{\rm b} < 3.0$ & 
$\sigma_j^{\rm b} < 4.5$ & 
$\sigma_j^{\rm b} < 10.0$ & 
All $\sigma_j^{\rm b}$ & 
$2.5 < \sigma_j^{\rm b} < 3.5$ & 
$2.0 < \sigma_j^{\rm b} < 4.5$ 
\\
\hline
\\
ALL & $P_{inner}$ (days)         &      $2.91627 ^{+0.00004}_{-0.00006}$         &      $2.91627 ^{+0.00004}_{-0.00004}$         &      $2.91626 ^{+0.00004}_{-0.00004}$         &      $2.91626 ^{+0.00004}_{-0.00004}$         &      $2.91627 ^{+0.00005}_{-0.00004}$         &      $2.91627 ^{+0.00004}_{-0.00004}$         \\
ALL & $e_{inner}$        &      $0.028 ^{+0.014}_{-0.013}$       &      $0.038 ^{+0.022}_{-0.018}$       &      $0.060 ^{+0.032}_{-0.026}$       &      $0.079 ^{+0.053}_{-0.036}$       &      $0.032 ^{+0.016}_{-0.017}$       &      $0.038 ^{+0.022}_{-0.018}$       \\
ALL & $\varpi_{inner}$ (degrees)  &      $145.3 ^{+90.1}_{-35.1}$         &      $162.6 ^{+86.1}_{-50.6}$         &      $158.5 ^{+73.9}_{-40.6}$         &      $159.2 ^{+83.4}_{-46.6}$         &      $152.7 ^{+85.3}_{-44.0}$         &      $162.6 ^{+86.1}_{-50.6}$         \\
ALL & $T_{P,inner}$      &      $4780.4 ^{+0.7}_{-0.3}$       &      $4780.6 ^{+0.7}_{-0.4}$       &      $4780.5 ^{+0.7}_{-0.3}$       &      $4780.5 ^{+0.8}_{-0.4}$       &      $4780.5 ^{+0.8}_{-0.3}$       &      $4780.6 ^{+0.7}_{-0.4}$       \\
ALL & $P_{outer}$ (days)         &      $448.6 ^{+5.1}_{-4.3}$   &      $449.4 ^{+5.0}_{-4.8}$   &      $450.6 ^{+8.4}_{-7.1}$   &      $452.1 ^{+15.4}_{-10.7}$         &      $449.8 ^{+5.0}_{-4.3}$   &      $449.4 ^{+5.0}_{-4.8}$   \\
ALL & $e_{outer}$        &      $0.68 ^{+0.10}_{-0.06}$  &      $0.65 ^{+0.10}_{-0.06}$  &      $0.64 ^{+0.13}_{-0.08}$  &      $0.64 ^{+0.15}_{-0.09}$  &      $0.65 ^{+0.07}_{-0.06}$  &      $0.65 ^{+0.10}_{-0.06}$  \\
ALL & $\varpi_{outer}$ (degrees)  &      $175.5 ^{+2.0}_{-1.8}$   &      $175.5 ^{+2.0}_{-2.3}$   &      $175.1 ^{+2.9}_{-3.9}$   &      $174.9 ^{+4.1}_{-6.1}$   &      $175.3 ^{+1.9}_{-1.9}$   &      $175.5 ^{+2.0}_{-2.3}$   \\
ALL &  $\eta$  &      $30.5 ^{+36.9}_{-89.4}$  &      $12.4 ^{+52.2}_{-83.9}$  &      $15.9 ^{+42.3}_{-70.6}$  &      $14.3 ^{+49.4}_{-80.5}$  &      $23.3 ^{+45.3}_{-82.9}$  &      $12.4 ^{+52.2}_{-83.9}$  \\
\hline
\\
ALL & $T_{P,outer}$      &      $4889.1 ^{+1.6}_{-2.0}$       &      $4889.2 ^{+2.1}_{-2.5}$       &      $4888.5 ^{+3.2}_{-4.3}$       &      $4887.9 ^{+4.9}_{-6.7}$       &      $4888.9 ^{+2.3}_{-2.1}$       &      $4889.2 ^{+2.1}_{-2.5}$       \\
ALL & $T_{T,outer},Bakos$        &      $4868.2 ^{+9.8}_{-6.9}$       &      $4865.6 ^{+9.9}_{-7.7}$       &      $4863.4 ^{+13.6}_{-10.4}$     &      $4862.8 ^{+15.2}_{-13.2}$     &      $4864.8 ^{+7.3}_{-7.5}$       &      $4865.6 ^{+9.9}_{-7.7}$       \\
ALL & $T_{T,outer},2010$         &      $5314.9 ^{+14.2}_{-6.1}$      &      $5314.9 ^{+11.3}_{-7.4}$      &      $5314.1 ^{+14.8}_{-11.0}$     &      $5315.0 ^{+20.9}_{-15.3}$     &      $5314.2 ^{+9.7}_{-5.8}$       &      $5314.9 ^{+11.3}_{-7.4}$      \\
ALL & $T_{T,outer},2011$         &      $5766.4 ^{+12.1}_{-10.9}$     &      $5765.4 ^{+12.8}_{-11.7}$     &      $5764.8 ^{+20.1}_{-15.4}$     &      $5766.8 ^{+33.5}_{-22.6}$     &      $5764.0 ^{+12.2}_{-8.7}$      &      $5765.4 ^{+12.8}_{-11.7}$     \\
ALL & $T_{T,outer},2012$         &      $6215.8 ^{+15.8}_{-14.5}$     &      $6215.2 ^{+16.6}_{-15.6}$     &      $6215.5 ^{+27.3}_{-21.2}$     &      $6218.5 ^{+47.0}_{-32.2}$     &      $6214.7 ^{+17.6}_{-12.5}$     &      $6215.2 ^{+16.6}_{-15.6}$     \\
\hline
\hline
\\
$< 100s$ & $P_{inner}$ (days)    &      $2.91627 ^{+0.00004}_{-0.00007}$         &      $2.91627 ^{+0.00004}_{-0.00004}$         &      $2.91626 ^{+0.00004}_{-0.00004}$         &      $2.91626 ^{+0.00004}_{-0.00004}$         &      $2.91627 ^{+0.00005}_{-0.00004}$         &      $2.91627 ^{+0.00004}_{-0.00004}$         \\
$< 100s$ & $e_{inner}$   &      $0.030 ^{+0.012}_{-0.017}$       &      $0.038 ^{+0.022}_{-0.018}$       &      $0.060 ^{+0.032}_{-0.026}$       &      $0.078 ^{+0.052}_{-0.035}$       &      $0.032 ^{+0.015}_{-0.017}$       &      $0.038 ^{+0.022}_{-0.018}$       \\
$< 100s$ & $\varpi_{inner}$ (degrees)     &      $144.4 ^{+88.2}_{-35.4}$         &      $160.0 ^{+87.4}_{-48.1}$         &      $158.0 ^{+73.6}_{-40.1}$         &      $158.9 ^{+83.1}_{-46.2}$         &      $150.8 ^{+83.5}_{-44.0}$         &      $160.0 ^{+87.4}_{-48.1}$         \\
$< 100s$ & $T_{P,inner}$         &      $4780.3 ^{+0.7}_{-0.3}$       &      $4780.6 ^{+0.7}_{-0.4}$       &      $4780.5 ^{+0.7}_{-0.3}$       &      $4780.5 ^{+0.8}_{-0.4}$       &      $4780.5 ^{+0.8}_{-0.3}$       &      $4780.6 ^{+0.7}_{-0.4}$       \\
$< 100s$ & $P_{outer}$ (days)    &      $448.9 ^{+5.0}_{-3.3}$   &      $449.4 ^{+5.1}_{-4.8}$   &      $450.7 ^{+8.4}_{-7.0}$   &      $452.3 ^{+15.2}_{-10.5}$         &      $449.9 ^{+5.0}_{-4.3}$   &      $449.4 ^{+5.1}_{-4.8}$   \\
$< 100s$ & $e_{outer}$   &      $0.67 ^{+0.11}_{-0.05}$  &      $0.65 ^{+0.08}_{-0.06}$  &      $0.63 ^{+0.11}_{-0.07}$  &      $0.63 ^{+0.12}_{-0.09}$  &      $0.64 ^{+0.06}_{-0.06}$  &      $0.65 ^{+0.08}_{-0.06}$  \\
$< 100s$ & $\varpi_{outer}$ (degrees)     &      $175.4 ^{+2.0}_{-1.8}$   &      $175.4 ^{+1.9}_{-2.3}$   &      $174.9 ^{+2.9}_{-3.8}$   &      $174.6 ^{+4.3}_{-6.0}$   &      $175.3 ^{+1.7}_{-1.9}$   &      $175.4 ^{+1.9}_{-2.3}$   \\
$< 100s$ &   $\eta$    &      $32.2 ^{+36.9}_{-87.2}$  &      $14.4 ^{+51.0}_{-84.1}$  &      $16.0 ^{+41.9}_{-70.3}$  &      $14.4 ^{+48.9}_{-80.3}$  &      $24.3 ^{+45.2}_{-79.6}$  &      $14.4 ^{+51.0}_{-84.1}$  \\
\hline
\\
$< 100s$ & $T_{P,outer}$         &      $4888.9 ^{+1.9}_{-2.0}$       &      $4889.1 ^{+2.2}_{-2.5}$       &      $4888.3 ^{+3.3}_{-4.3}$       &      $4887.7 ^{+5.0}_{-6.7}$       &      $4888.9 ^{+2.4}_{-2.1}$       &      $4889.1 ^{+2.2}_{-2.5}$       \\
$< 100s$ & $T_{T,outer},Bakos$   &      $4867.6 ^{+10.3}_{-6.4}$      &      $4865.0 ^{+8.6}_{-7.2}$       &      $4862.5 ^{+12.0}_{-9.8}$      &      $4861.5 ^{+13.4}_{-12.4}$     &      $4864.7 ^{+6.6}_{-7.5}$       &      $4865.0 ^{+8.6}_{-7.2}$       \\
$< 100s$ & $T_{T,outer},2010$    &      $5314.8 ^{+12.2}_{-6.2}$      &      $5314.3 ^{+9.8}_{-7.2}$       &      $5313.2 ^{+13.4}_{-10.5}$     &      $5313.6 ^{+19.6}_{-14.4}$     &      $5314.0 ^{+9.1}_{-5.8}$       &      $5314.3 ^{+9.8}_{-7.2}$       \\
$< 100s$ & $T_{T,outer},2011$    &      $5765.4 ^{+12.5}_{-10.0}$     &      $5764.3 ^{+12.3}_{-10.9}$     &      $5763.7 ^{+19.2}_{-14.9}$     &      $5765.6 ^{+32.5}_{-21.9}$     &      $5763.8 ^{+11.8}_{-8.7}$      &      $5764.3 ^{+12.3}_{-10.9}$     \\
$< 100s$ & $T_{T,outer},2012$    &      $6215.8 ^{+16.6}_{-15.5}$     &      $6213.7 ^{+16.6}_{-15.4}$     &      $6214.4 ^{+26.8}_{-20.7}$     &      $6217.4 ^{+46.1}_{-31.4}$     &      $6213.7 ^{+18.4}_{-12.0}$     &      $6213.7 ^{+16.6}_{-15.4}$     \\
\hline
\end{tabular}
\label{TAB:PARAMETERS2}
We strove to make our data comparable to the discovery paper where at all possible: as such, it should be noted that the parameter fits given for $P$, $e$, $\varpi$ and $T_{peri}$ are based on a coplanar MCMC analysis, while the transit time calculations for the outer planet are made using the assumption that the outer planet is inclined at $90^{\circ}$ to the plane of the sky.\\
$^{\rm a}$ All transit and pericenter passage times are in Julian days - 2,450,000\\
$^{\rm b}$ All jitter constraints are in $m\,s^{-1}$\\
$^{\rm c}$ All pericenter alignments are in degrees
%
\end{table}
%

\subsection{Summary of MCMC Results}\label{MCMCSummary}
Our re-analysis of the radial velocity data and subsequent introduction of a coupled TTV analysis leads to the following conclusions:

Comparing our results with those of the HAT-P-13 discovery paper B09 (where $\sigma_j = 3.0\,m\,s^{-1}$ was used), we find that a rather larger range of fits to the RV data are possible. In particular, we find (for the S1 data set used in B09) that even a rather modest levels of jitter ($2.0 < \sigma_j \sim 4.5\,m\,s^{-1}$) will allow 
$e_{inner} = 0.017^{+0.013}_{-0.009}$,
$e_{outer} = 0.73^{+0.12}_{-0.06}$,
$P_{outer} = 429.7^{+4.4}_{-3.7}$ days,  
and relative pericenter alignment, $\eta = 15.9^{+50.6}_{-83.7}$. 

Similarly, when we look at the S2 data set as used in W10, a level of jitter $2.0 < \sigma_j \sim 4.5\,m\,s^{-1}$ will allow 
$e_{inner} = 0.038 ^{+0.022}_{-0.018}$,
$e_{outer} = 0.65^{+0.10}_{-0.06}$,
$P_{outer} = 449.4^{+5.0}_{-4.8}$ days,  
and relative pericenter alignment, $\eta= -12.4^{+52.2}_{-83.9}$. 

Table \ref{TAB:PARAMETERS} contains a summary of all the parameters and the best fits resulting from our analysis of S1, while Table \ref{TAB:PARAMETERS2} contains a summary of all the parameters and the best fits resulting from our 2-planet + slope analysis of S2.  We note that within the range of our uncertainties, all of our results are consistent with those of the discovery paper B09, but we do in general find a much broader spread of allowed values for all parameters. Similarly for the comparison of the S2 analysis with the results of W10, our results are consistent with those of W10 (except for the inner eccentricity) but the error bars on our fitted parameters are much broader. To ``make the eccentricities consistent'', one has to appeal to even lower jitter values $2.0 < \sigma_j \sim 3	.5\,m\,s^{-1}$, by which point the lower 1-sigma limits from our analysis start to overlap with the upper limits from the W10 analysis.

We should point out that while our prior on the jitter (\S 2.1) does penalize high jitter values, it may be that an even more punitive prior is warranted. This is due to the observed fact that stellar activity (and therefore stellar jitter) declines with age \citep{Isaacson10}. As HAT-P-13 is old ($\sim 5Gyr$, it is possible that it has a jitter level that is drawn from a population with lower values than that quoted from the general population in \citet{Wright05}. As such, it is possible that a more limited range of jitters should be allowed. However, we note that (i) our jitter allows for contributions over and above pure ``stellar jitter'', and (ii) we list results for a number of different jitter ranges in Tables \ref{TAB:PARAMETERS} \& \ref{TAB:PARAMETERS2} from which we see that  even if one only wishes to accept a limited ranges for the jitter values resulting from the MCMC analysis (e.g. $2.0 < \sigma_j \sim 4.5\,m\,s^{-1}$), the associated errors on the fitted planetary parameters are still much broader than those presented in the results of either B09 or W10.

Given that any ``jitter'' values arising from our analysis will intrinsically be ``system'' jitter (the combined contribution of jitter from stellar sources, instrumental noise, undetected planets, etc) rather than pure stellar jitter, we are hesitant to produce detailed figures regarding any ``best-fit'' jitter parameters arising from our analysis. However, for the same of completion, we note that the analyses corresponding to the three columns of Fig 1 (S1, S2 assuming 2 planets $+$ slope, and S3 assuming 3 planets) find overall system jitter levels of $\sigma_j= X^{+Y}_{-Z}$, $\sigma_j= X^{+Y}_{-Z}$ and $\sigma_j= X^{+Y}_{-Z}$ respectively.

Subsequent to the pure MCMC analysis, if we then introduce an additional TTV analysis step (based upon the output of the MCMC routine), we can somewhat constrain the eccentricity of the outer planet, finding that now $0.6 < e_{outer} < 0.85$. However, the remainder of the other elements are essentially unconstrained by this additional analysis. Tables \ref{TAB:PARAMETERS} \& \ref{TAB:PARAMETERS2} also contain a summary of all the parameters and the best fits resulting from our MCMC + TTV analysis. 

We thus conclude this section of our investigation by suggesting that previous methodologies of excluding jitter from an MCMC analysis significantly overestimates the precision with which other system parameters (e.g. planetary orbital elements) can be estimated from RV data. We thus recommend that jitter be included as a model parameter in future MCMC analyses of RV data, allowing a true estimation of system parameters and their uncertainties to be determined.

\section{Further Investigation of Transit Timing Variations}\label{FURTHER_TTVS}
In the coplanar MCMC + TTV analysis of section \ref{MCMC}, it was suggested that the RMS TTV amplitude could be of some diagnostic power in constraining certain system parameters (E.g. the eccentricity of the outer planet, $e_{outer}$). In \citet{Nesvorny09} and \citet{Payne10a}, it has been found that planets which are significantly inclined relative to one another can give rather different signal profiles and amplitudes as compared to the coplanar case. This inclination dependence, combined with the differing planetary masses that would be required in an inclined system to satisfy the RV constraints, suggests that a more detailed investigation of the inclination dependence of the TTV signal in the HAT-P-13 case is warranted. 

Our ultimate aim is to find whether it might then be possible to use the TTV characteristics to break the degeneracy of the RV $M_{outer} \sin{i_{outer}}=15.2M_J$ relation.

\subsection{Examples of Transit Timing Variation Signals for HAT-P-13}\label{Profiles}
%
\begin{figure}
\centering
\begin{tabular}{c}
\includegraphics[trim = 0mm 0mm 0mm 130mm, clip, angle=-90, width=0.75\textwidth]{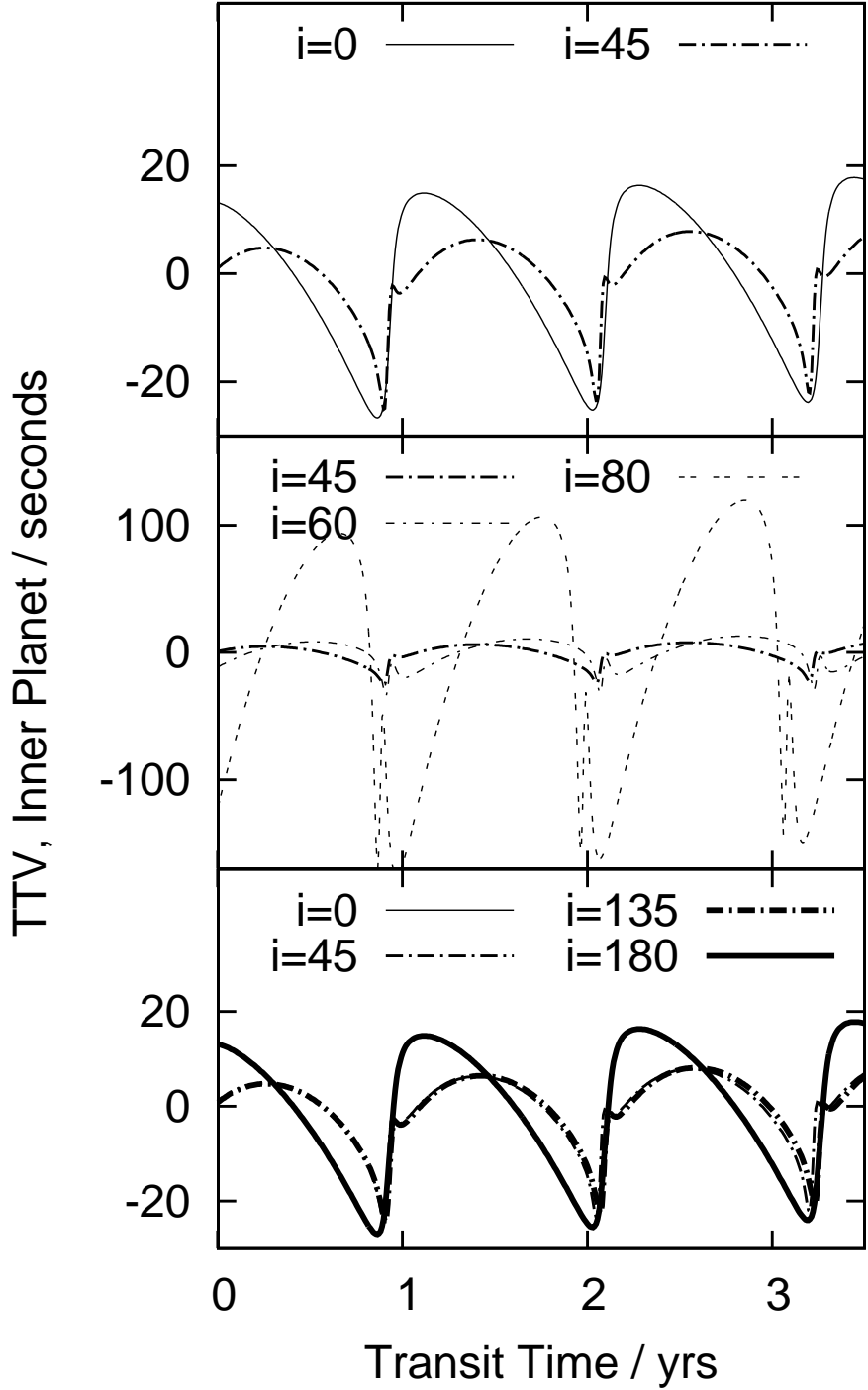}
\end{tabular}
\caption{Variation of the relative planetary inclination in the HAT-P-13 system. The system parameters are as observed, such that $a_{inner}=0.043\,AU,\,e_{inner}=0.02,\,m_{inner}=0.85M_J,\,a_{outer}=1.188,\,e_{outer}=0.691,\,m_{outer}\sin{i_{outer}}=15.2M_J$. While the pattern of behaviour is non-trivial, we can make certain statements:
(i) For the most likely prograde cases $i\sim 0\,^{\circ}$ and $i\sim 45\,^{\circ}$, there is a significant difference in the predicted TTV profile, with much sharper features being expected for the high inclination cases;
(ii) The TTV amplitude for these cases is $\sim 20$ seconds, well within the current observational constraints;
(iii) Systems with planets close to perpendicular are likely to give highly distinctive, high amplitude TTV signals;
(iv) The retrograde systems can give rather similar amplitudes and overall profiles to those of systems in prograde orbits (extra care may be needed to analyses and distinguish these cases).
It should be noted that the origin of the time axis on this plot is somewhat arbitrary, given that it is a simulation, but the translation to ``real-life'' can easily be done by noting that the deepest troughs in the TTV plots occur as the outer planet approaches pericenter, e.g. April 16th, 2010. 
}
  \label{FIG:INC_HAT_1}
\end{figure}
%
\begin{figure}
\centering
\begin{tabular}{c}
\includegraphics[trim = 0mm 0mm 0mm 0mm, clip, angle=-90, width=\textwidth]{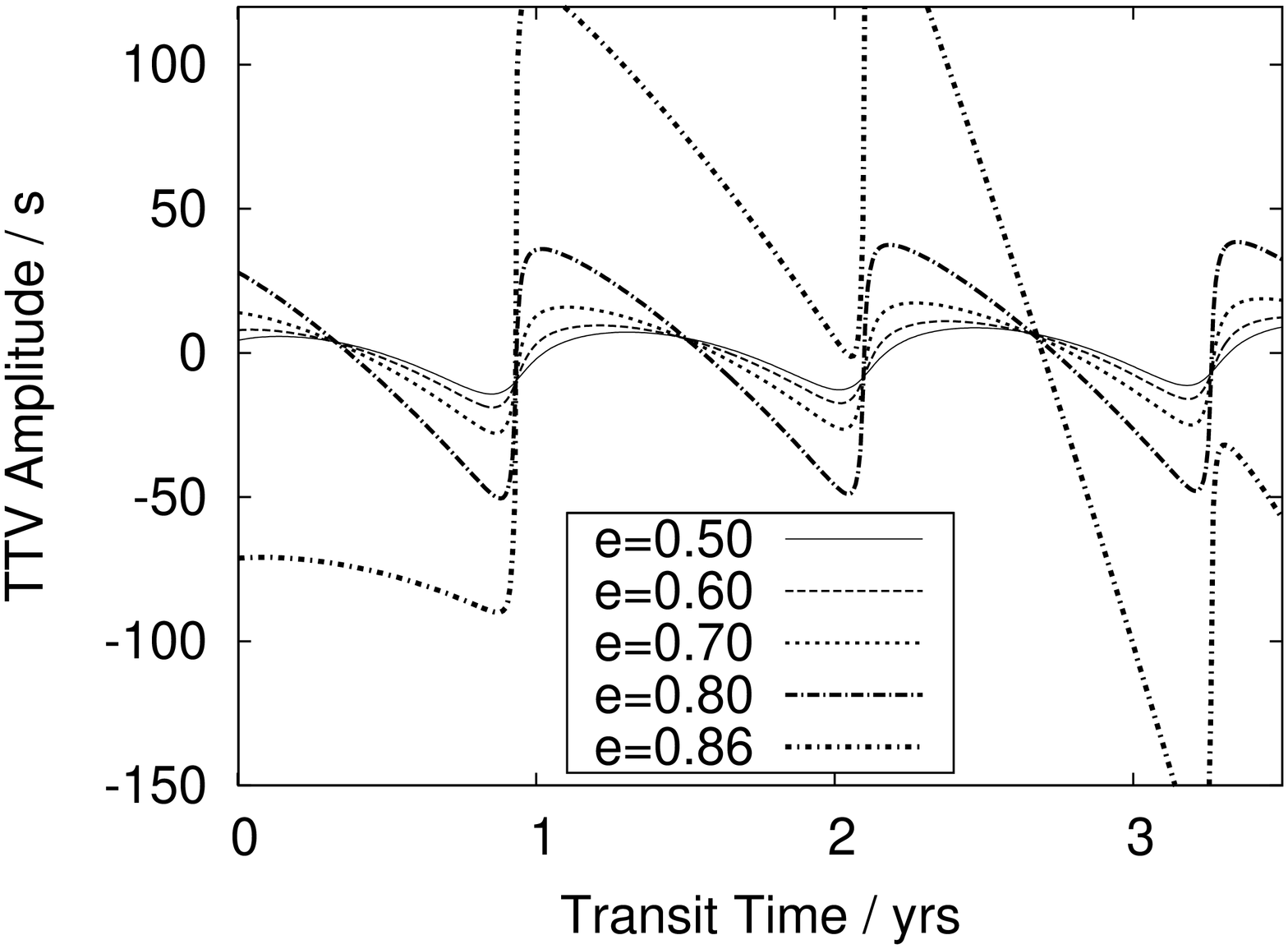}
\end{tabular}
\caption{TTVs as a function of time for various values of the outer planet eccentricity, $e_{outer}$, in the prograde co-planar HAT 13 b system. We find from this particular example that increasing $e_{outer}$ from 0.5 to 0.8 causes only a minor increase in the TTV amplitude, but that going to higher eccentricities ($e_{outer} \sim 0.85$) leads to the significant amplitude variations seen in \Fig{FIG:TTV1}, potentially associated with unstable planetary orbits.
As in \Fig{FIG:INC_HAT_1}, it should be noted that the deepest troughs in the TTV plots occur at the time of pericenter passage of the outer planet.
}
  \label{FIG:INC_HAT_2}
\end{figure}
%
We commence this more detailed investigation of the inclination dependence of the TTV signal variation in the HAT-P-13 system by plotting in \Fig{FIG:INC_HAT_1} some sample results which illustrate the effects of varying the orbital inclination of the outer planet. Note that we set $a_{inner}=0.043\,AU,\,e_{inner}=0.02,\,m_{inner}=0.85M_J,\,a_{outer}=1.188,\,e_{outer}=0.691,\,\&\,m_{outer}\sin{i_{outer}}=15.2M_J$ in line with the standard values from B09. N.B., as we vary the inclination of the outer planet, $i_{outer}$, we also simultaneously vary the mass of the outer planet, $m_{outer}$.

Given the work of \citet{Mardling10} discussed in \S \ref{INTRO}, we concentrate our analysis on a few key relative inclinations: For the prograde cases we concentrate on $i_{rel} \sim 0^{\circ}$ and $i_{rel} \sim 45^{\circ}$, as these are thought to be the most likely states at late times. We also examine the symmetric retrograde cases, $i_{rel} \sim 180^{\circ}$ and $i_{rel} \sim 135^{\circ}$. Finally, for completeness we also look at a couple of example plots from the ``forbidden'' region of parameter space $55^{\circ} < i_{rel} < 125^{\circ}$ from which it is thought the planets are likely to be excluded due to the combined effects of Kozai forcing and tidal dissipation.

We can immediately see from the $i_{rel} = 0^{\circ}$ and $i_{rel} = 45^{\circ}$ plots in \Fig{FIG:INC_HAT_1}, that (a) the expected TTV amplitudes are well below the current observational limits, so either inclination is plausible in that sense, but (b) perhaps more importantly, the profiles have rather different shapes, with the $i_{rel} = 45^{\circ}$ plot exhibiting prominent spikes in TTVs over a relatively short period of time, in addition to having an overall reduced amplitude. We discuss further in section \ref{Projections} a plausible observational strategy to extract this information.

We can also see that the completely retrograde case, $i_{rel} = 180^{\circ}$, is very similar in both shape and amplitude to the $i_{rel} = 0^{\circ}$ prograde case, suggesting that they would be rather difficult to distinguish observationally. 

For the plots in the ``forbidden'' $55^{\circ} < i_{rel} < 125^{\circ}$ region, it seems that the systems in which the planets are close to perpendicular will have extremely high TTV amplitudes (due, no doubt, to the significantly increased $m_{outer}$ required in order to satisfy the $m_{outer}\sin{i}$ constraint). These highly distinctive profiles mean that if for some reason a planet \emph{were} able to occupy this region of parameter space, it would be readily identifiable (and indeed, certain particularly high inclinations may already be ruled out by the TTV constraint - see section \ref{Contours} below for further discussion).

Finally, we note in passing that in section \ref{MCMC} it was found that eccentricities for the outer planet were constrained to lie below $e_{outer} \sim 0.85$. We show in \Fig{FIG:INC_HAT_2} some details of the TTV plots which result from increasing the eccentricity of the outer planet. Keeping the semimajor axes and masses of the planets fixed at the best-fit values from the discovery paper (and with $i_{rel} = 0^{\circ}$), we increase $e_{outer}$ from 0.5 to 0.85. We find that for values of the eccentricity in the range $ 0.5 < e_{outer} < 0.8$, increases in eccentricity result in only marginal increases in the TTV amplitude. However, at very high eccentricities ($e_{outer}\sim 0.85$), the systems start to exhibit ``kicks / step-changes'' in their behavior as the outer planet starts to significantly perturb the inner orbit (as the outer planet passes through pericenter). We note that whilst such systems \emph{may} be stable against collisions, the relatively large perturbations experienced during close pericentre encounters can still lead to significant orbital evolution and hence large TTV amplitude variation (see \citet{Payne10a} and \cite{Veras10} for more detailed discussions).


\subsection{Transit Timing Variation Contour Maps}\label{Contours}
%
\begin{figure}
  \centering
  \begin{tabular}{ccc}
    \multicolumn{1}{c}{\psfig{figure=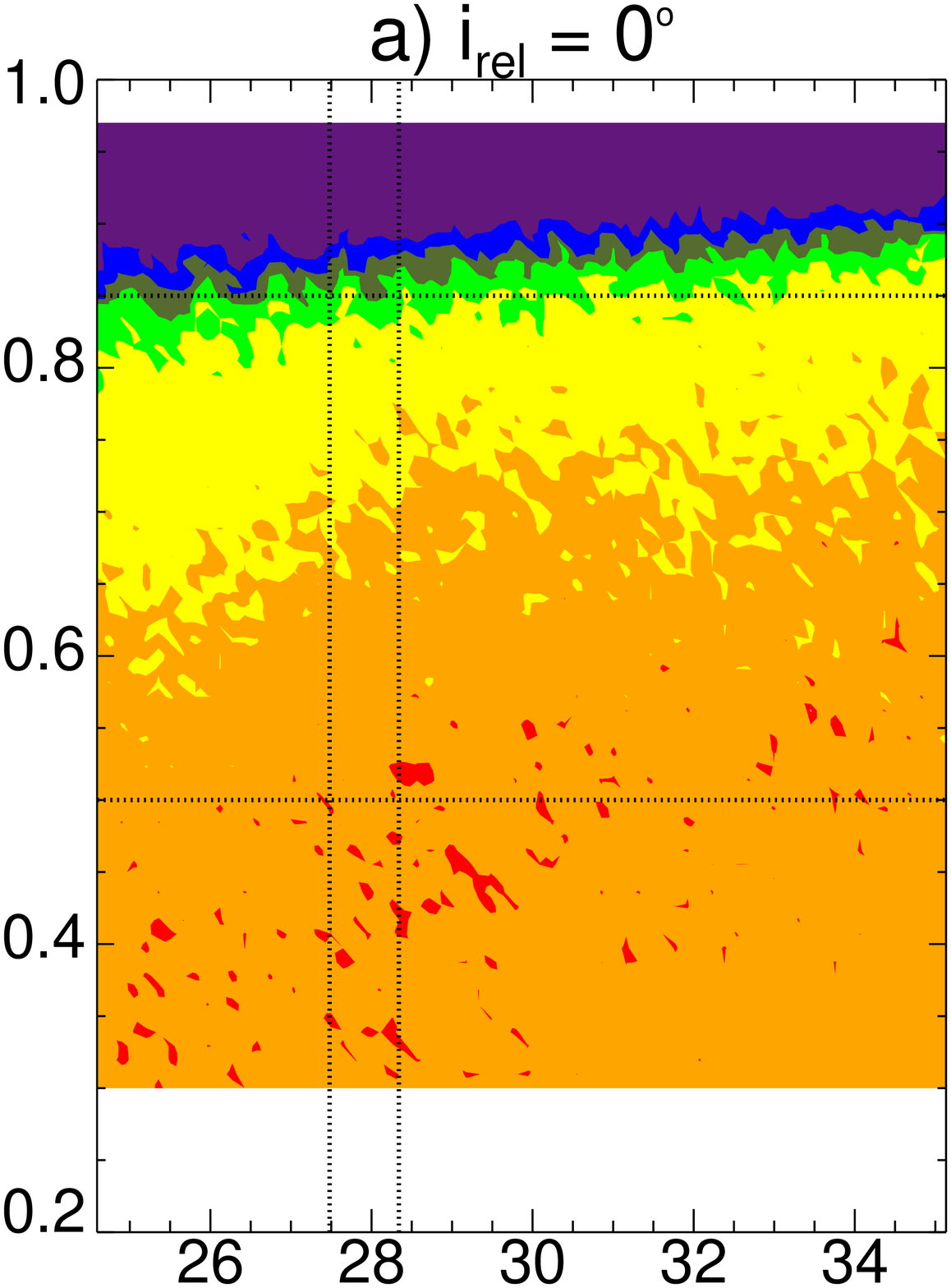,width=0.3\textwidth}}&
    \multicolumn{1}{c}{\psfig{figure=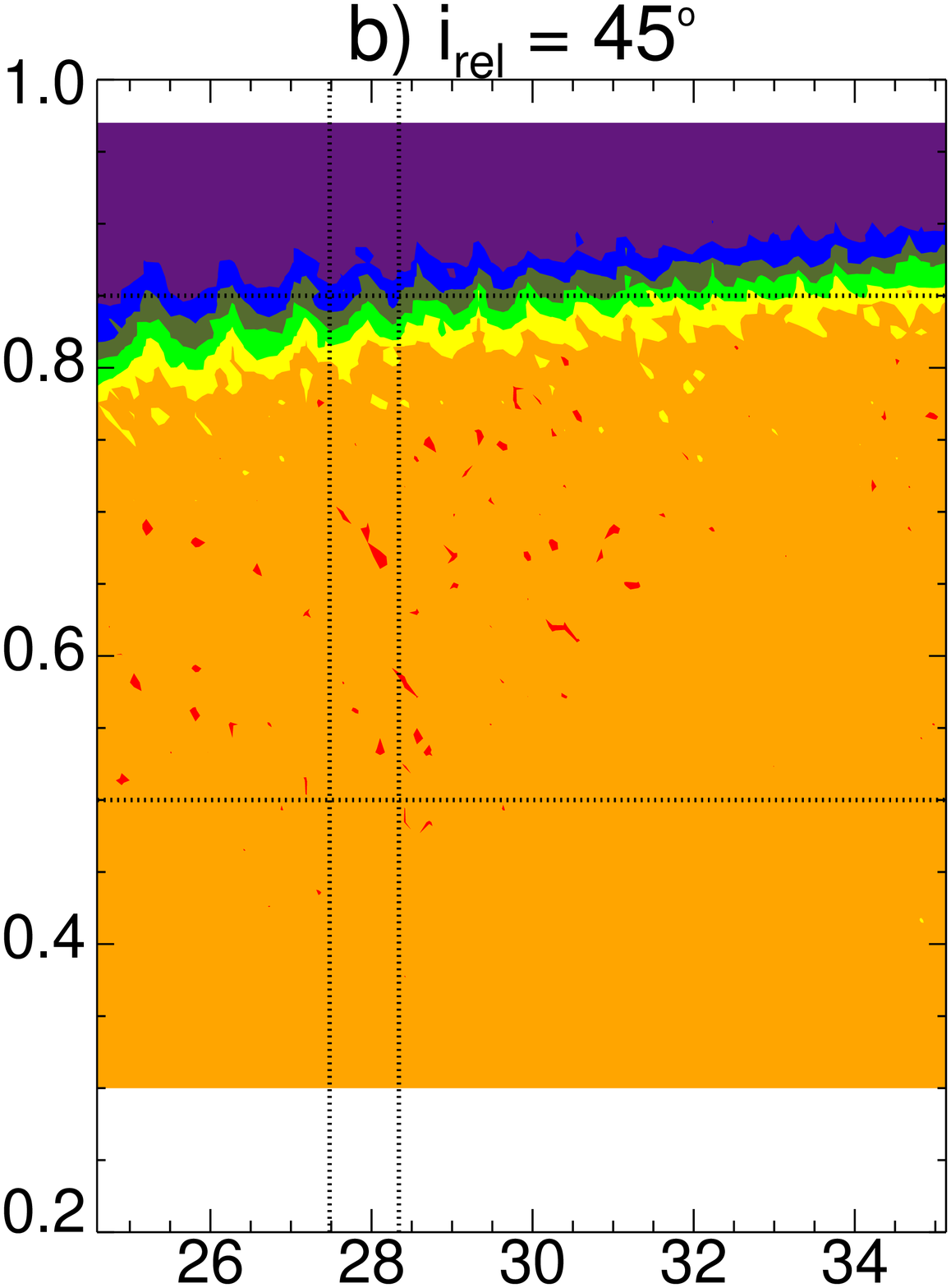,width=0.3\textwidth}}&
    \multicolumn{1}{c}{\psfig{figure=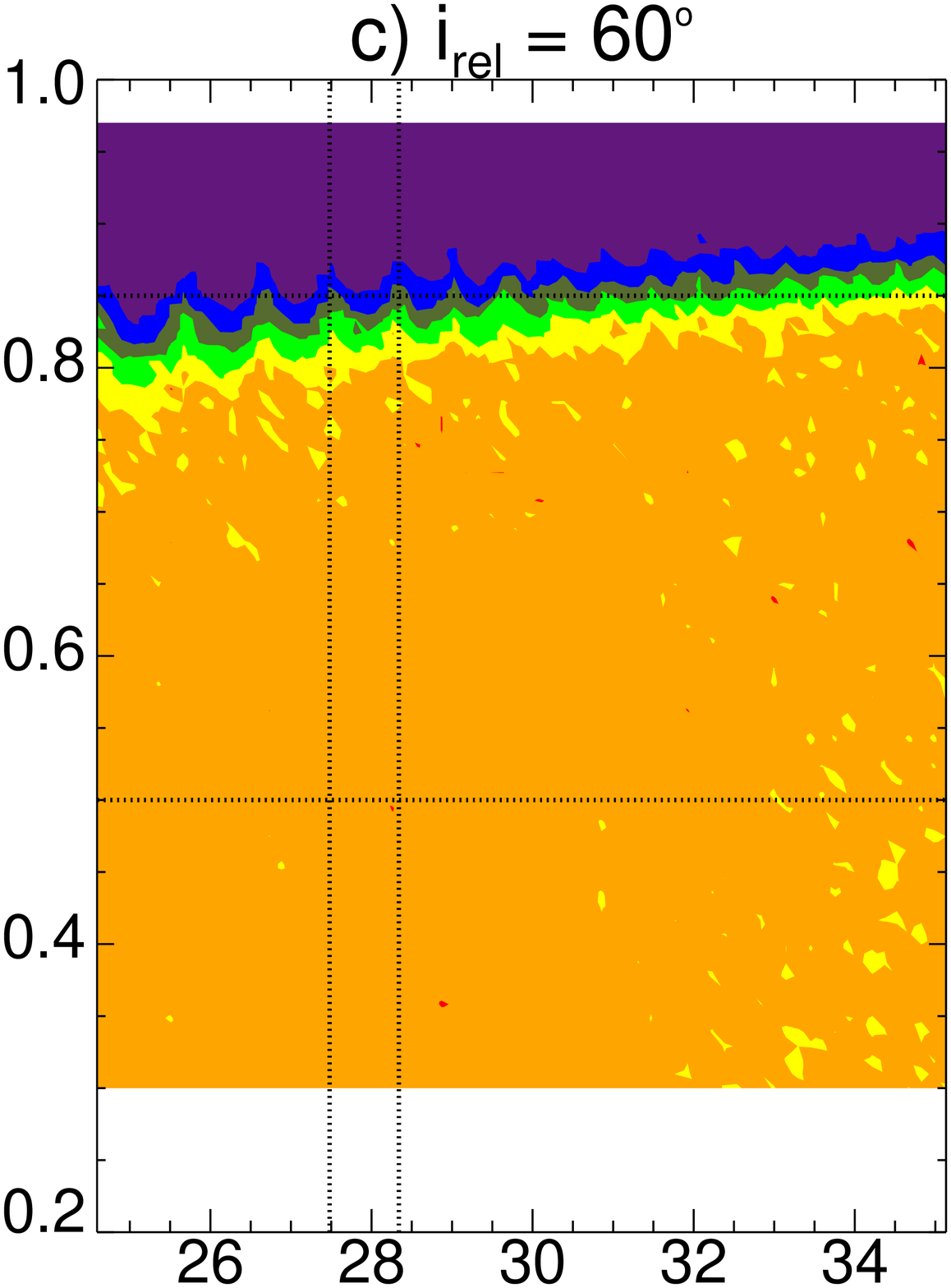,width=0.3\textwidth}}\\
    \multicolumn{1}{c}{\psfig{figure=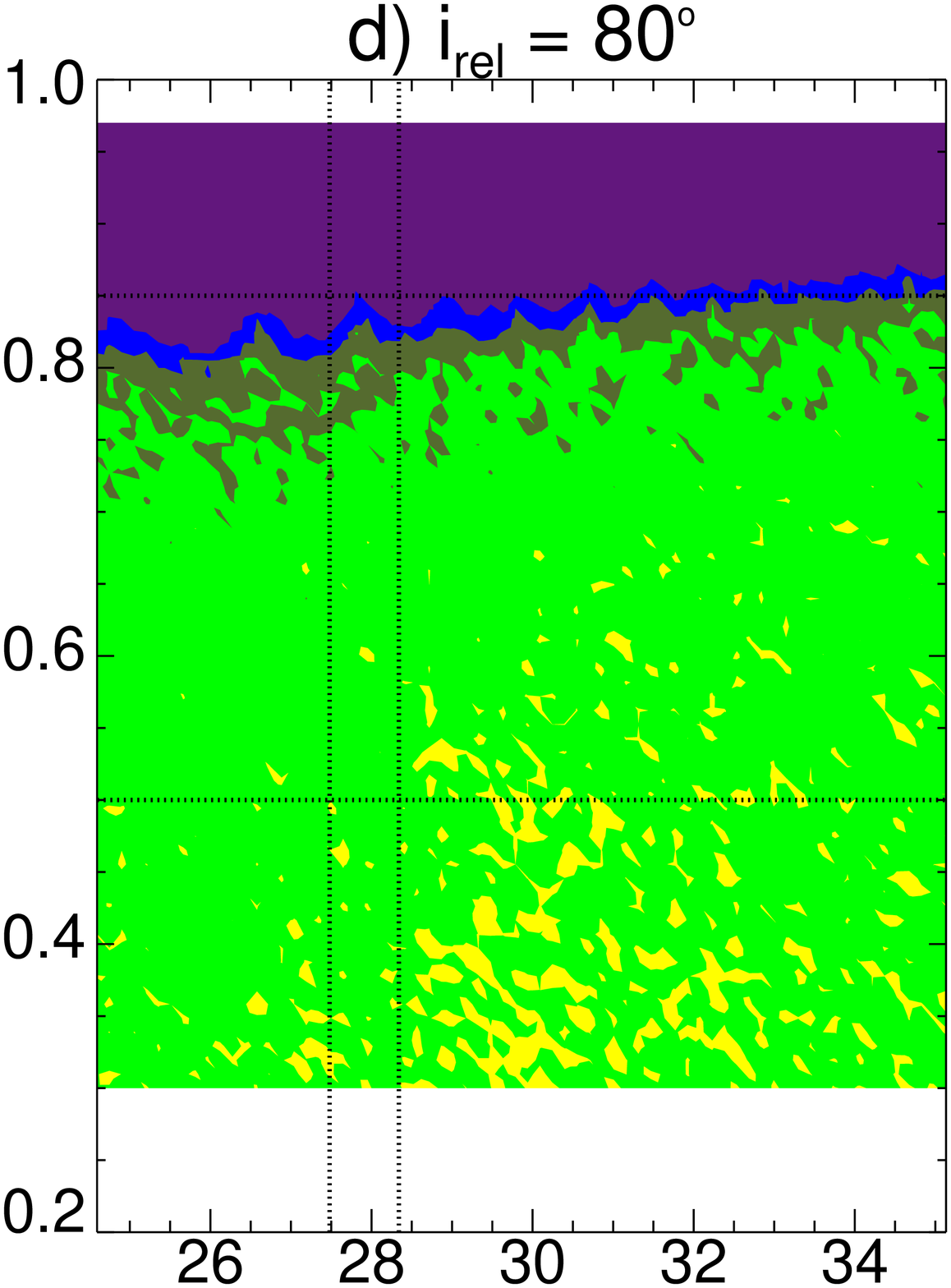,width=0.3\textwidth}}&
    \multicolumn{1}{c}{\psfig{figure=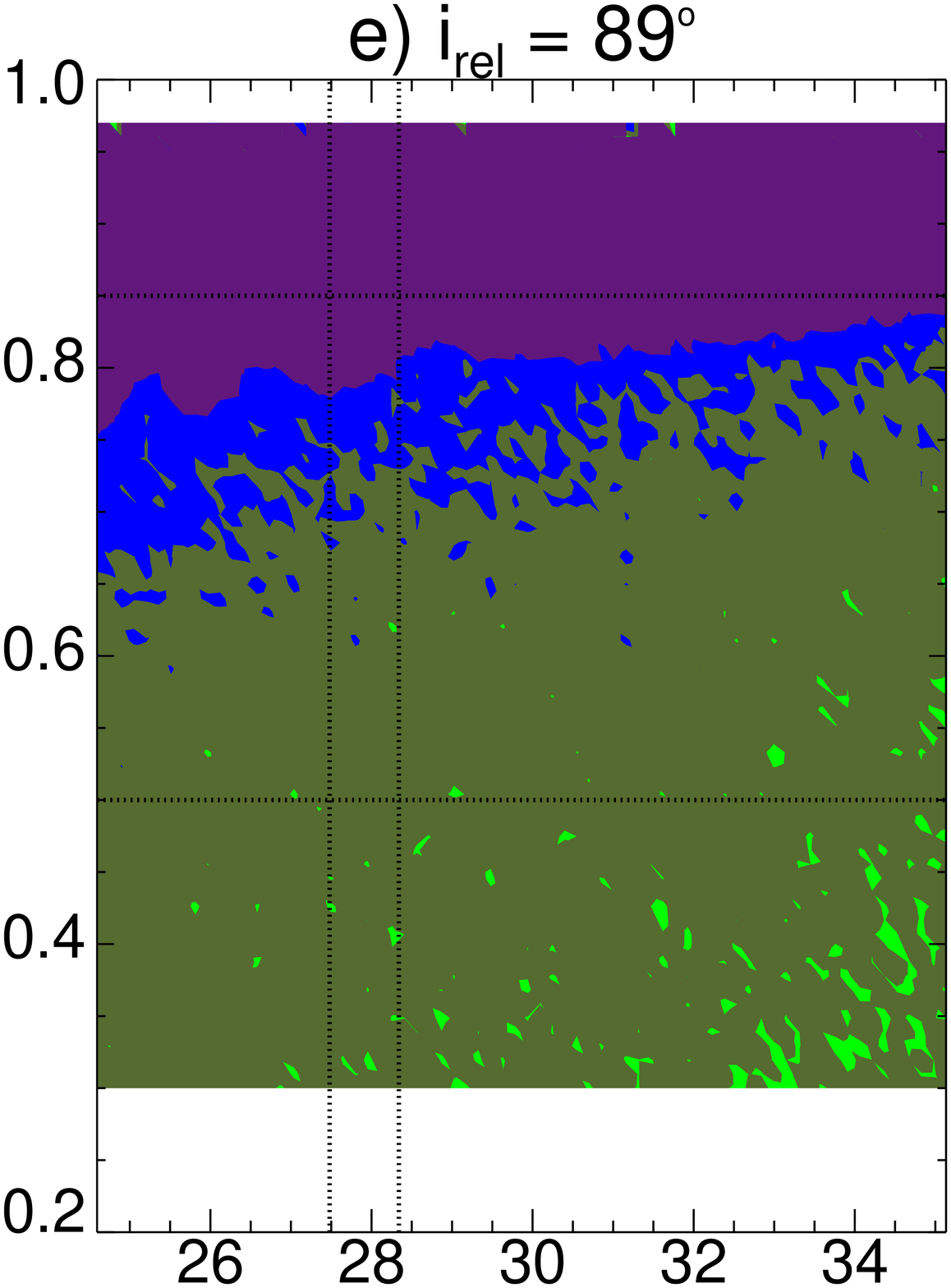,width=0.3\textwidth}}&
    \psfig{figure=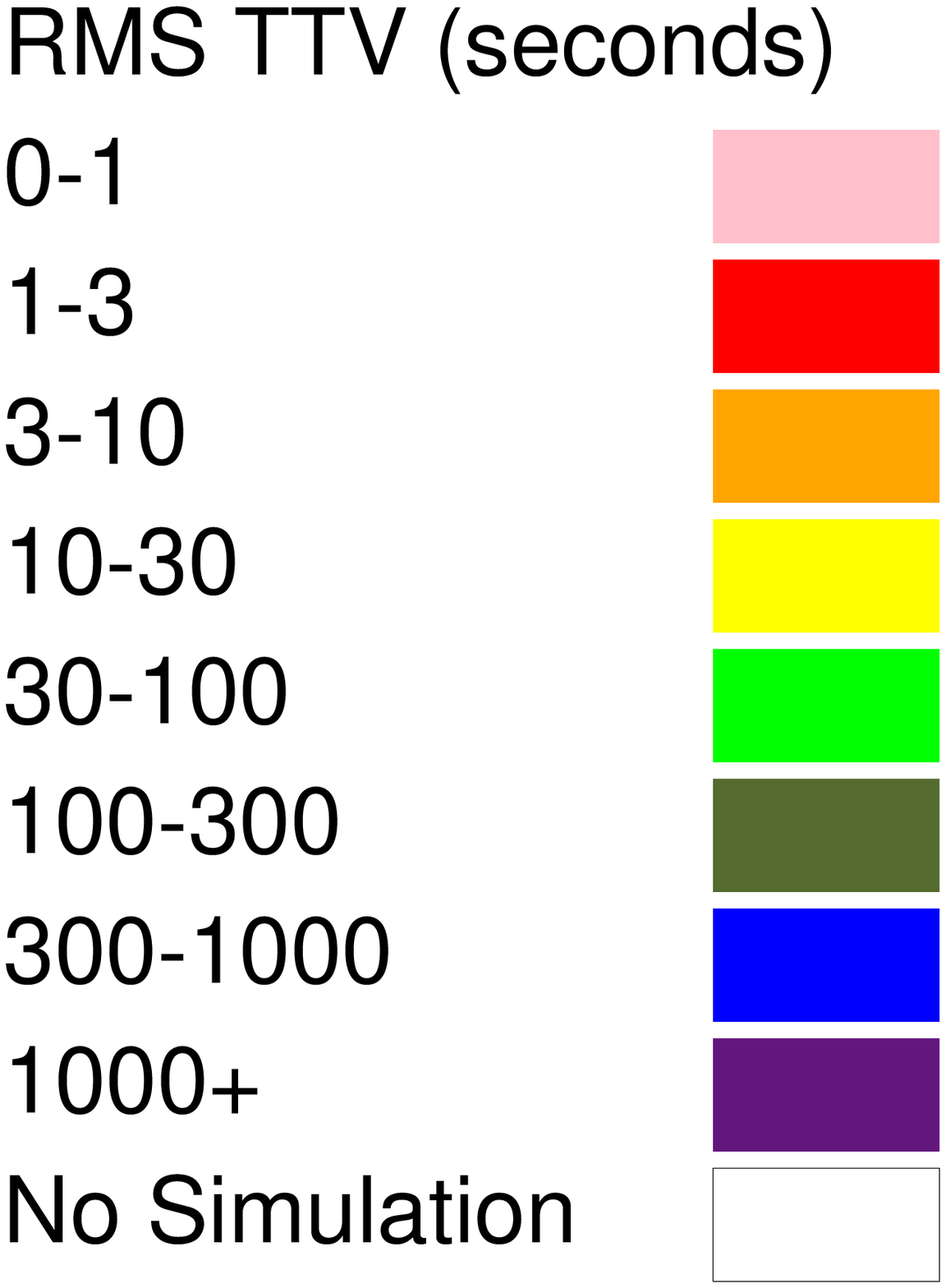,width=0.3\textwidth}\\
    \multicolumn{1}{c}{\psfig{figure=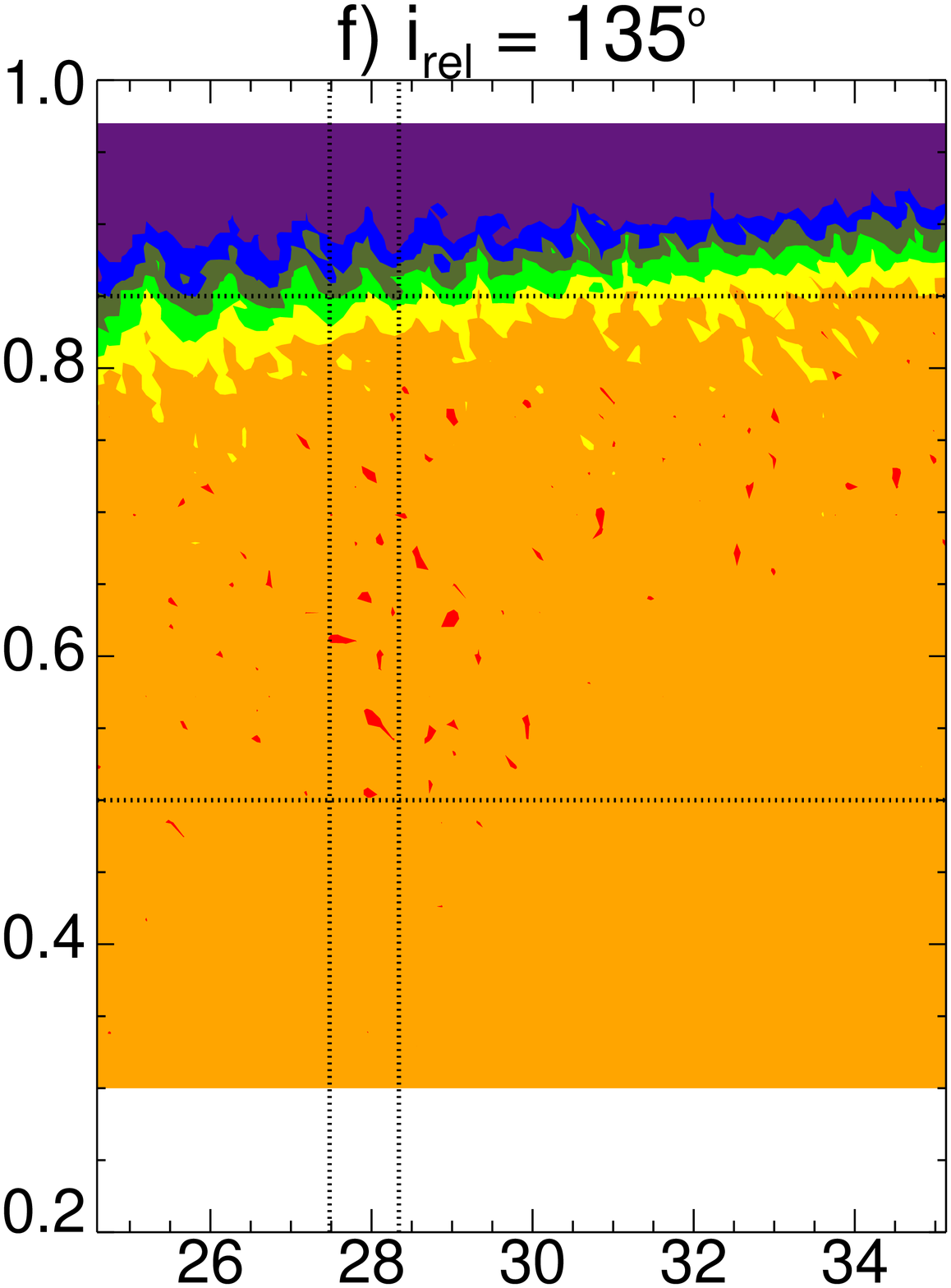,width=0.3\textwidth}}&
    \multicolumn{1}{c}{\psfig{figure=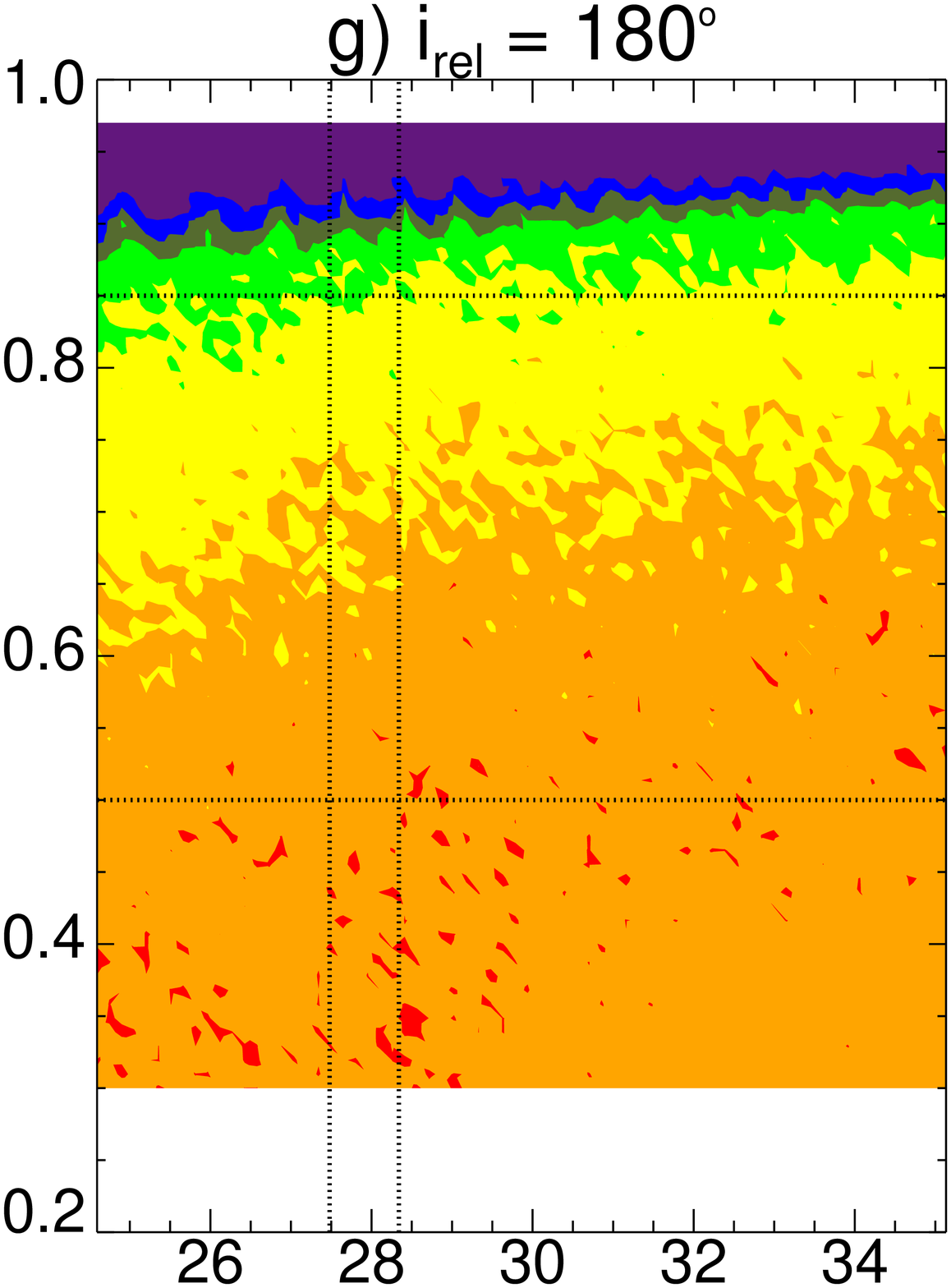,width=0.3\textwidth}}&
  \end{tabular}
  \put(-495,-70){{\begin{sideways} \centering {\bf \large Outer Eccentricity, $e_{outer}$} \end{sideways}}}
  \put(-330,-310){\centering \bf \large Semimajor Axis Ratio $(a_{outer}/a_{inner})$}
  \caption{Contour plots in the $(a_{outer}/a_{inner},e_{outer})$ plane, illustrating the contours of the predicted RMS TTV amplitude at different relative inclinations. The horizontal and vertical lines give the approximate observational constraints from \Fig{FIG:TTV2} ($0.5<e_{outer}<0.85$, $420<P_{outer}<440$ days). The plots make clear that the TTVs can always be very large towards the top of the allowed eccentricity range, but that there is little variation with semi-major axis. Very high inclinations result in TTVs that are significantly greater than the observational constraints from B09 across the entirety of the parameter space.}
  \label{FIG:FLAMES:35MEDIAN}
\end{figure}
%
\begin{figure}
  \centering
  \begin{tabular}{cc}
    \multicolumn{1}{c}{\psfig{figure=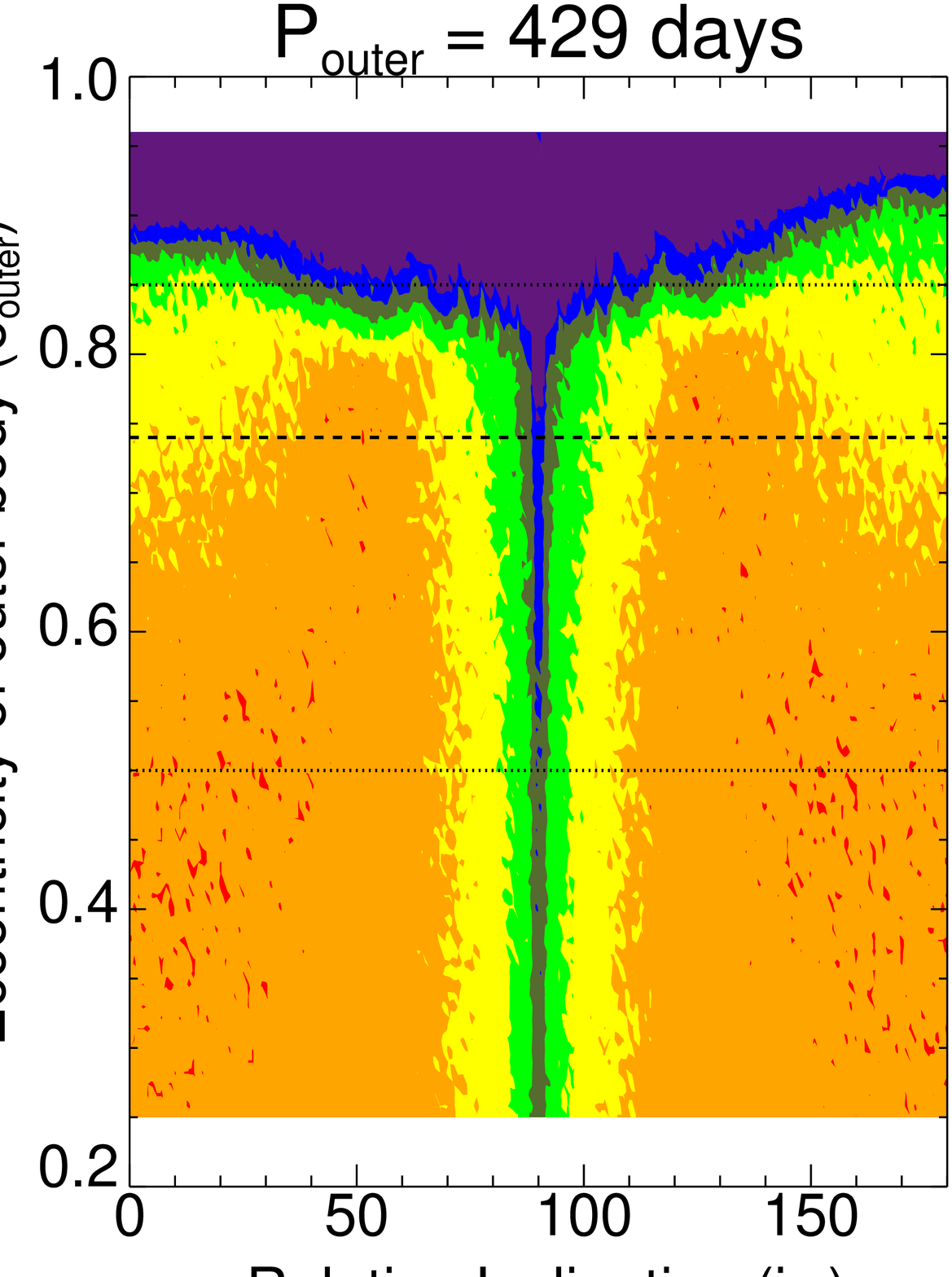,width=0.8\columnwidth}}
  \end{tabular}
\vspace{1cm}
  \caption{Contour plots in the $(i_{rel},e_{outer})$ plane, illustrating the contours of the predicted RMS TTV amplitude (The contours used are the same as those in \Fig{FIG:FLAMES:35MEDIAN}). The horizontal lines indicate the same upper and lower bounds as given in Fig. 6, as well as the best fit value for $e_{outer}$ from the right-hand column of Table 2, W10.
  We can clearly see that, irrespective of the eccentricity of the outer planet, inclinations $ 88^{\circ} < i_{rel} < 92^{\circ}$ will produce TTVs larger than the RMS constraints from \citet{Bakos09}. Moreover, it is clear that if these observational constraints can be tightened-up, the allowed range of $(i_{rel},e_{outer})$ will be significantly reduced, allowing them to be of significant more practical use.}
  \label{FIG:FLAMES:35MEDIANINCSCAN}
\end{figure}
%
To gain a further insight into the range of TTV signatures possible for HAT-P-13, we keep the mass and orbit of the inner planet fixed and vary the orbit of the more massive outer planet. For each set of configurations - i.e. semi-major axis, $a$, eccentricity, $e$, \& inclination, $i$ of the non-transiting planet - we typically conduct 5 simulations randomising the angular orbital elements (argument of pericentre, $\omega$, longitude of ascending node, $\Omega$, \& initial mean anomaly $M$). We assume that the known transiting planet has the mass, semi-major axis and eccentricity of HAT-P-13b ($M_{inner} = 0.85\,M_J$, $a_{inner}=0.043\,AU$ and $e_{inner}=0.02$) and constrain the outer planet to have $M_{outer} \sin{i_{outer}} = 15.2\,M_J$).

In \Fig{FIG:FLAMES:35MEDIAN} we calculate the RMS of $\delta t(i)$ for each configuration and present contour plots of the median RMS TTV amplitude in the $(\frac{a_{outer}}{a_{inner}},e_{outer})$ plane for multiple inclinations.  Note that these plots use simulations conducted for 3.5 year simulations, but similar simulations conducted over 10 year periods are essentially identical. Moreover, given that the plots showing the \emph{maximum} RMS TTV amplitude and those showing the \emph{median} RMS TTV amplitude have qualitatively and quantitatively similar values across the parameter space, we use the median, 3.5 year plots for all figures below.

Looking at the $i_{rel}=0\,^{\circ}$ prograde case in \Fig{FIG:FLAMES:35MEDIAN}, we find that the values of the outer planet's semi-major axis and eccentricity as given in the B09 or W10 papers($a_{outer} \sim 1.2\,AU$ \& $e_{outer} \sim 0.6 - 0.7$) would be expected to give an RMS TTV amplitude of $\sim 10s$. Furthermore, we find that there is a rather large range of parameter space which gives similarly low TTVs, meaning that the current TTV observations cannot help to constrain the outer orbit, other than saying that $e_{outer} < 0.8$ ($a_{outer}$ is essentially unconstrained by the TTVs in the range $1 < a_{outer} < 1.5\,AU$ which was examined).

As the relative inclination of the orbits is increased, we find that there is a slight \emph{decrease} in TTV signal amplitude in the $e_{outer} \sim 0.7$ region as $i\rightarrow 45\,^{\circ}$. 

As the relative inclination of the orbits increases further towards $i_{rel} = 90\,^{\circ}$, there is an increase in RMS TTV signal amplitude across the entirety of the parameter space, such that at $i_{rel} = 85\,^{\circ}$ almost all regions have amplitudes $>30$ seconds. For $i \equiv 90\,^{\circ}$, the mass of the outer planet would become formally infinite, so we cannot investigate this case, but we do note that in the $i_{rel} = 89\,^{\circ}$ case, almost the entirety of parameter space has TTVs higher than the observed limits. 

As the orbits become retrograde, we find that the TTV amplitudes decrease again, with the results at $i_{rel} = 135\,^{\circ}$ ($i_{rel} = 180\,^{\circ}$) being very similar to those seen at $i_{rel} = 45\,^{\circ}$ ($i_{rel} = 0\,^{\circ}$). I.e. there is some kind of approximate symmetry about $i_{rel} = 90\,^{\circ}$.

The clear difference in TTV amplitude as $i \rightarrow 90\,^{\circ}$ is a rather different behaviour to that observed by \citet{Payne10a} in their general investigation of inclinations effects on TTVs in hot-Jupiter + hot-Earth systems. There they found that TTV amplitudes were very generally highest at $i_{rel}=0\,^{\circ}$, with an approximately monotonic decline as $i \rightarrow 180\,^{\circ}$, with no special behaviour seen at $i_{rel} = 90\,^{\circ}$. However, it is difficult to make a precise comparison between their results and our results here, as in \citet{Payne10a} the planetary masses were kept constant as the inclination was increased, whereas in this investigation we are varying the outer mass as $M_{outer}  = 15.2\,M_J/\sin{i_{outer}}$.

For completeness, we include in \Fig{FIG:FLAMES:35MEDIANINCSCAN} a plot in the $(i_{rel},e_{outer})$ plane, again plotting the median RMS TTV values from 3.5 year simulations. In these simulations we fix the semi-major axis of the outer planet and then vary the inclination and eccentricity of the outer planet. We reproduce here the plot for $a_{outer}/a_{inner}=27.2$, i.e. an outer planet of $P_{outer} = 429$ days.  As seen from \Fig{FIG:FLAMES:35MEDIANINCSCAN}, the upper allowed eccentricity \emph{decreases} at larger semi-major axes. There is a critical range of inclinations above which the TTV amplitude becomes too great to fit observational constraints: For the current 100 second limits, the excluded range is very narrow $\sim 90^{\circ}\pm 2^{\circ}$. Clearly this is of little benefit, as such regions are already excluded in any practical sense, as such a highly inclined object would be so massive as to be a star and hence likely be readily visible, but if future observations constrain the TTV amplitude to be less than 10s (for example), then the inclination restrictions become much more severe, and all inclinations $\sim 90^{\circ}\pm 20^{\circ}$ could be ruled out. Moreover, we see that as the inclination increases from $0^{\circ}- 45^{\circ}$, the ability to constrain the eccentricity using TTV limits varies significantly: E.g. the yellow region mapping out the $10-30s$ TTV contour narrows rapidly.

We also made similar plots at different semimajor axes, but we find that there is little difference between the results at different semimajor axes, as might be anticipated from \Fig{FIG:FLAMES:35MEDIAN}.

\section{Future Transit Observations}\label{Projections}
\subsection{Inner Planet: Transit Timing Scheduling}\label{inner}
\begin{figure}
  \centering
  \begin{tabular}{c}
    \multicolumn{1}{c}{\psfig{figure=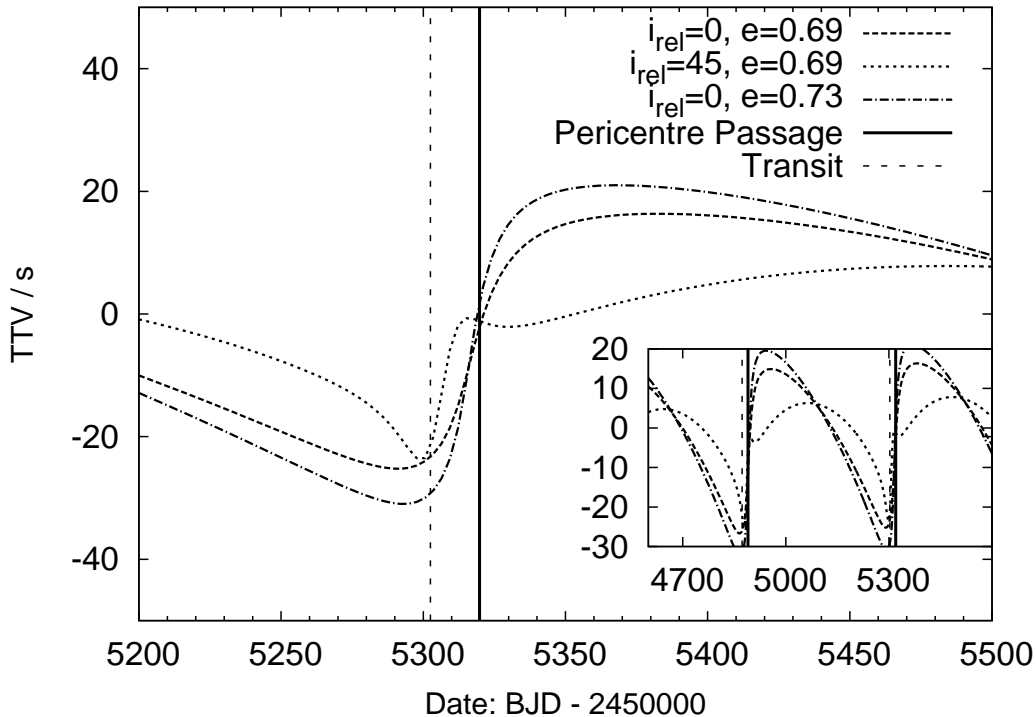,angle=-90,width=0.8\columnwidth}}
  \end{tabular}
  \caption{Using transits of the inner planet to determine system inclination. We show examples of TTVs for the cases of $i_{rel}=0^{\circ}$ and $i_{rel}=45^{\circ}$ favored by the \citet{Mardling10} analysis. In the inset we provide TTV profiles over approximately 3 years, but in the main plot we focus on the region close to pericenter passage, illustrating that over the course of $\sim 50$ days ($\sim 17$ transits), the variation chages from $\sim -30$ to $\sim 15$. We suggest that sufficiently precise observations over this period should be able to distinguish the different inclination states. 
We also plot an example of an $i_{rel}=0^{\circ}$ with a different eccentricity (0.73), illustrating that sufficiently precise observations can also distinguish eccentricity states. We note that these illustrative simulations were chosen to have their time of pericenter passage at BJD = 2455302.2, the best fit date from Table \ref{TAB:PARAMETERS} for $2.0 < \sigma_j \lesssim 4.5\,m\,s^{-1}$ and TTV amplitude less than 100 seconds: given the uncertainty in our fits for $T_{P,outer}$ (see section \ref{outer}), the absolute dates may shift, but the general message regarding the relative timing of the large changes in signal amplitude will not. All plots have $a_{inner}=0.043\,AU,\,e_{inner}=0.02,\,m_{inner}=0.85M_J,\,a_{outer}=1.188,\,\&\,m_{outer}\sin{i_{outer}}=15.2M_J$.)   
 }
  \label{FIG:HATP13INNERTRANSITS}
\end{figure}

As noted in section \ref{FURTHER_TTVS} from \Fig{FIG:INC_HAT_1} and \Fig{FIG:INC_HAT_2}, we have a potential means of distinguishing between different relative inclination and/or (outer) eccentricity states of the system. To do so will require a detailed investigation of the inner transit times to an accuracy greater than $\sim 5$ seconds. In this section we focus on the time-frame close to the pericenter-approach of the outer planet. It is at this point that the TTVs induced on the inner planet are predicted to go through a sharp change in values in only a small number of orbits. In \Fig{FIG:HATP13INNERTRANSITS} we display the TTVs expected for three different system configurations: (i) $i_{rel}=0^{\circ}$ and $e_{outer}=0.69$; (ii) $i_{rel}=45^{\circ}$ and $e_{outer}=0.69$; (i) $i_{rel}=0^{\circ}$ and $e_{outer}=0.73$, in which we have fixed the longitude of ascending node to be $=0$, and chosen the other system parameters to be those of the best fit value from the far right-hand column of Table 1. 

We see that in all three cases, the TTV signal changes sharply during the time between the predicted transit of the outer body (see section \ref{outer} below) and the predicted date of pericenter passage for the outer body. From Julian date $\sim 2455300$, over the next $\sim 50$ days (i.e. over $\sim 17$ transits of the inner planet), the signal would be expected to switch from being $\sim -30s$ to being $\sim +15s$. A similar pattern would be expected to occur at potential transit dates in 2011 \& 2012 at or around the dates given in Table 2. If enough observations can be made of high enough precision over this period, it should be possible to (i) At the very least confirm that a TTV variation is seen, (ii) give some significantly improved constraints on the magnitude of the TTVs in the system and hence point towards a better understanding of the system parameters, as well as (iii) pointing the way towards a subsequent, longer term observational campaign to further track the TTV profile over the course of an entire orbit of the outer planet.
%
%

We would thus suggest that an observational campaign targeting the transits of the inner planet could give informative results about the relative inclination of the planets in the system. We suggest that observations extending over the period of pericenter passage (for the outer planet) will initially be the most useful in determining the nature and amplitude of the TTVs, but that longer term observations can also give important information that allow different inclination states to be distinguished. In addition, it should be emphasized that such observations would need to be of a precision similar to (or better than) that demonstrated in the study of WASP-10 by \citet{Johnson09} and \citet{Johnson10}, where the mid-transit times were determined to a precision of $\sim 7$ seconds. If such a level of precision could be achieved over the course of 10 - 15 transits of the inner planet, then it is plausible that significant inclination information could be extracted.

We caution that the precise date(s) of the pericenter passage for the outer planet are, of course, highly dependent on the values of the fitted orbital parameters $a_{outer}$, $e_{outer}$, etc. Given that we have demonstrated in \S 3 that these fitted parameters are rather uncertain, it should be clear that the precise time of pericenter passage is similarly uncertain, and so any scheduled observations should take these uncertainties into account. Moreover, if/when further additional RV observations become available, it would be prudent to repeat the analysis of \S 3 to allow the uncertainties in the planetary parameters to be improved, and hence allow any future observational strategy to be refined. We also caution that in the examples shown in \Fig{FIG:HATP13INNERTRANSITS}, we have made specific assumptions regarding the longitude of ascending node of the planets (fixed $=0$). If such assumptions are relaxed, then TTV profiles can be generated that differ in the degree and extent to which their amplitude changes over the course of the pericenter passage. We emphasize that in fitting / modeling any putative future observations, such degeneracies would need to be accounted for.

\subsection{Outer Planet: Transit Timing Projections}\label{outer}
It is clear from sections \ref{FURTHER_TTVS} and \ref{inner} that further observations of the precise timings of the transit of the inner planet will be the key to understanding the TTVs, and subsequently the relative inclinations, of the planets around HAT-P-13. However, we feel that is is worth emphasizing that a (fortuitous) transit observation of the \emph{outer} planet would help to pin down the inclination, period and most other parameters of the planetary system much more precisely (to say nothing of giving the first ground-based observation of a system with two transiting planets!). While the a-priori probability of HAT-P-13c transiting is relatively low if all inclination angles are considered equally likely ($\sim 1\%$, see \citet{Seagroves03,Kane09}), this probability can be significantly increased (to $\sim 8\%$) for certain particular inclination and observation angles \citep{Beatty10}.

If the outer planet does indeed transit, then there remains the issue of knowing when to observe the star to detect the transit. If we use our MCMC data to predict the times of the next few transits, assuming that the outer planet is inclined at $90^{\circ}$ to the plane of the sky (as was also assumed in B09), then using the method outlined in \citet{Kane07,Kane09}, we find that there is a significant uncertainty in the predicted timing of any potential transit. We note at this point that our potential transit date for the year 2009 ($T_{T,outer, Bakos}$, Table 1 \& 2) is typically a few days later than that derived by B09 (but with a large range of uncertainties) and that our fitted outer period is slightly longer (see Table \ref{TAB:PARAMETERS}). 

\begin{figure}
  \centering
  \begin{tabular}{c}
    \multicolumn{1}{c}{\psfig{figure=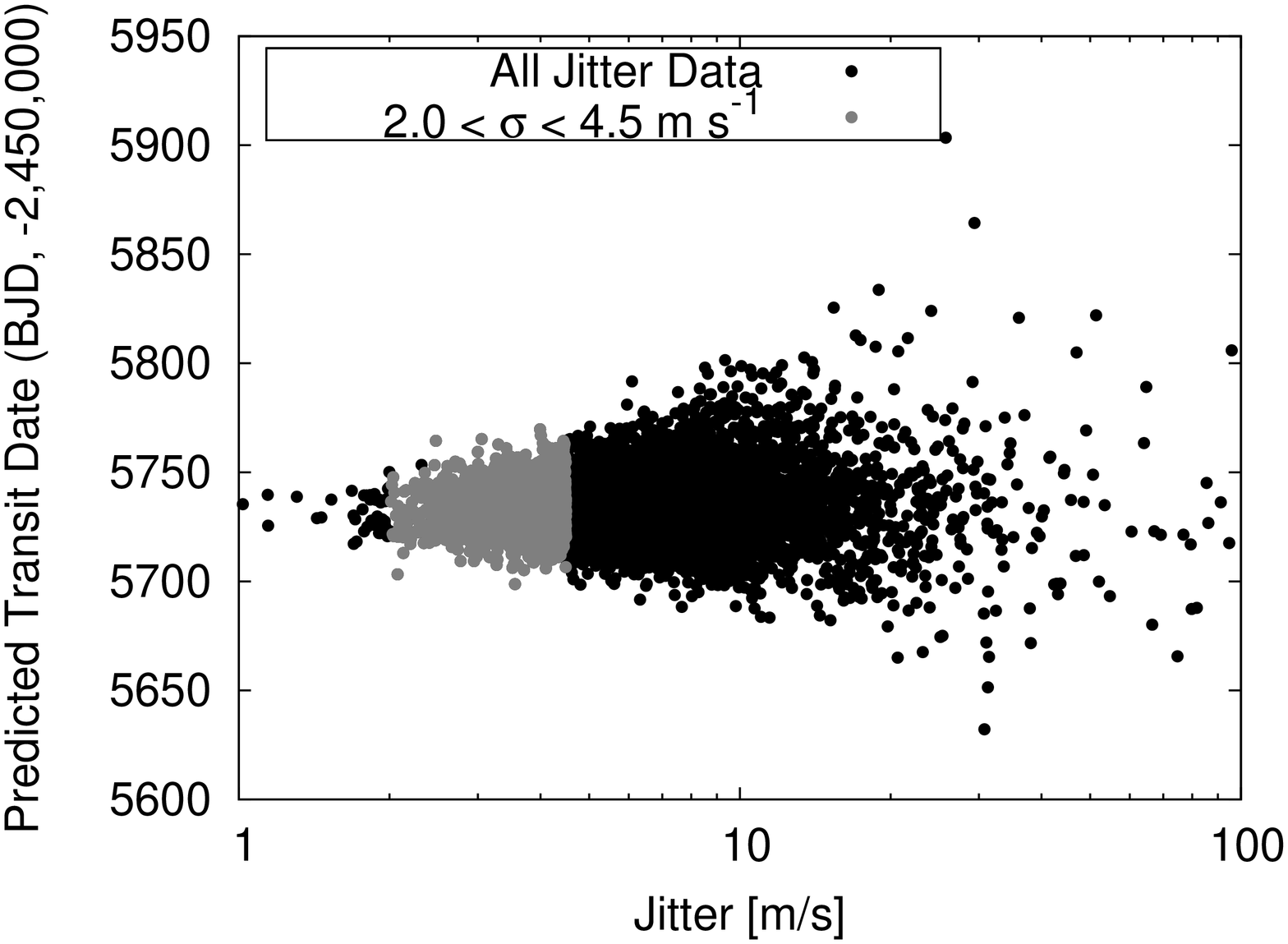,angle=-90,width=0.5\textwidth}}\\	
    \multicolumn{1}{c}{\psfig{figure=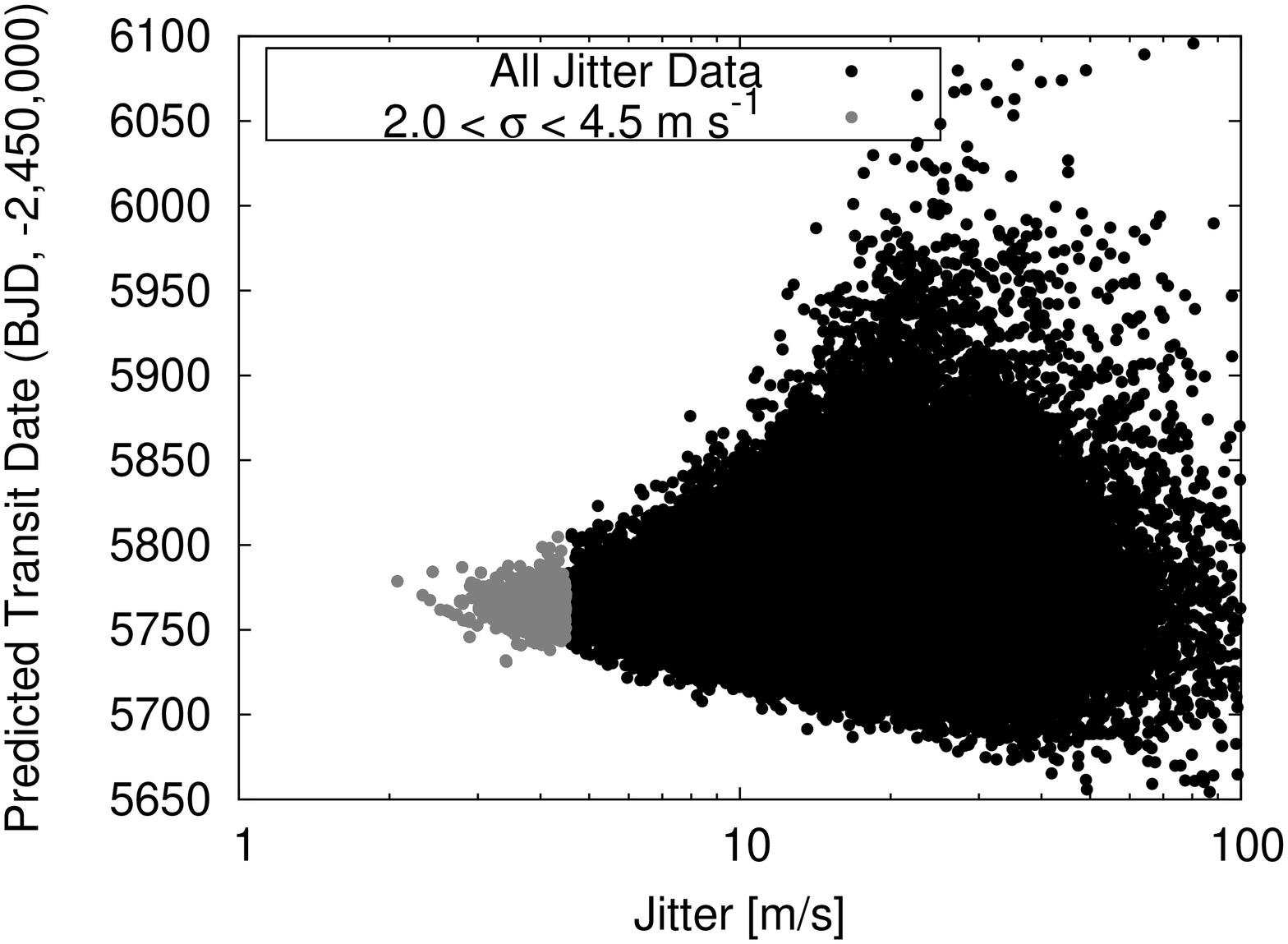,angle=-90,width=0.5\textwidth}}\\	
    \multicolumn{1}{c}{\psfig{figure=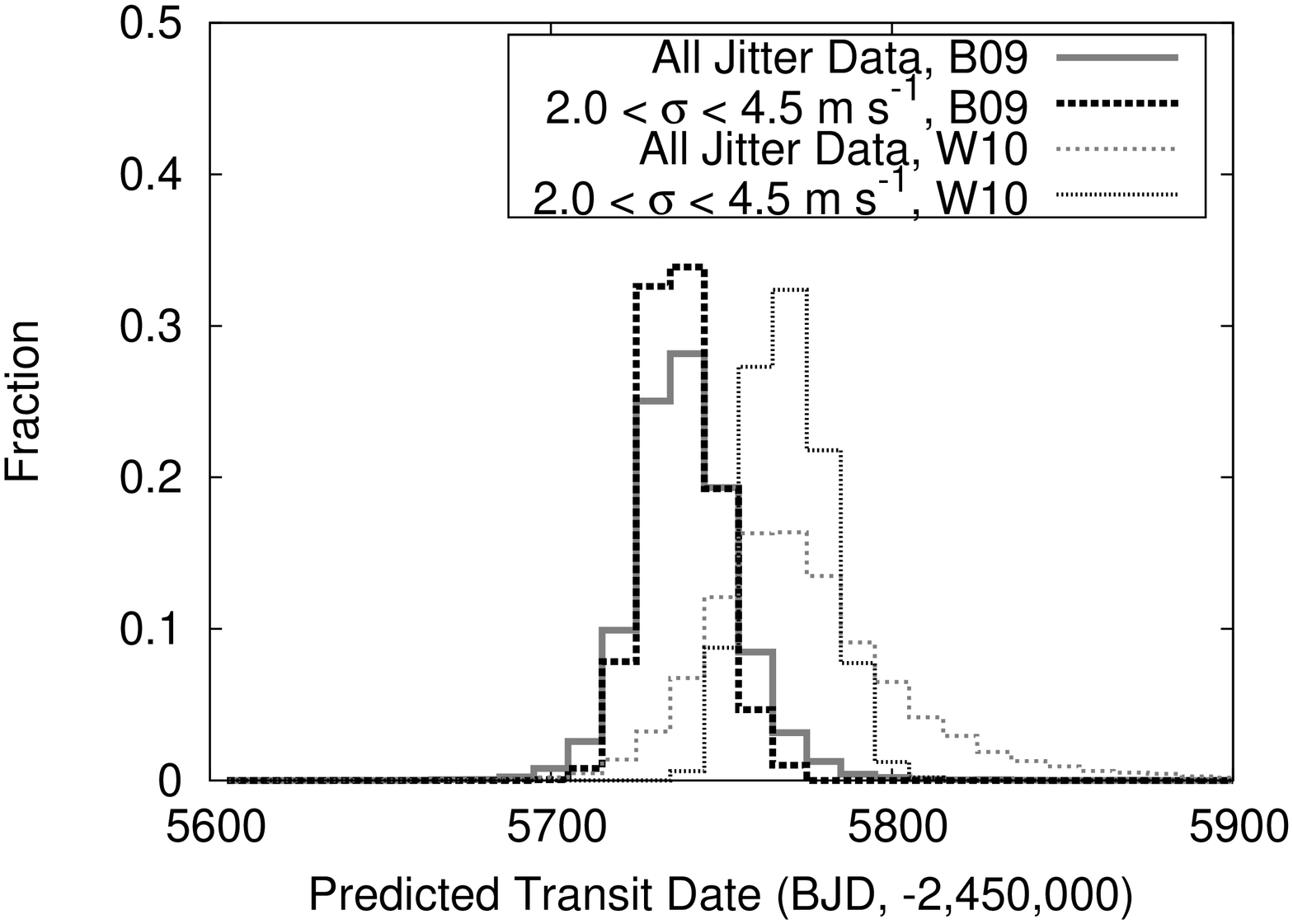,angle=-90,width=0.5\textwidth}}
  \end{tabular}
  \caption{Predicted transit dates for HAT-P-13c using the MCMC sample fits from B09 (top) and W10 (middle). We assume an inclination for the orbit of $90^{\circ}$ to the plane of the sky and restrict our attention here to only those systems which have TTVs less than 100 seconds. The dates are Julian dates -2450000. 
  {\bf Top \& Middle} It is clear that the transit time distribution is of the same form as that of the period distribution in \Fig{FIG:Jitter}. We see that there is a significantly increased scatter in the predicted date for the W10 data set, and that moreover the distribution is asymmetrical. 
  {\bf Bottom} The histograms comparing the restricted jitter sample ($2.0 < \sigma_j \sim 4.5\,m\,s^{-1}$) and the full data sample (but both have TTVs $<100$ seconds) for the B09 (S1) \& W10 (S2) data sets. These make it clear that the distributions for the different jitter levels are similar within a data set, but that the different data sets can have the center of the distributions shifted relative to one another due to the preference for a larger $P_{outer}$ in the S2 analysis}
  \label{FIG:HATP13TRANSITS}
\end{figure}

Specific predictions require us to make some assumptions about the stellar jitter, so as an example we consider a low stellar jitter case and select the results in which the \emph{overall} jitter lies in the range $2.0 < \sigma_j < 4.5\,m\,s^{-1}$ (and take the cases with TTV amplitude less than 100 seconds). In this case, we find that the predicted transit dates for 2010 - 2012 are $T_{T,outer, 2010}=2455314.3^{+9.8}_{-7.2}$, $T_{T,outer, 2011}=2455764.3^{+12.3}_{-10.9}$ and $T_{T,outer, 2012}=2456213.7^{+16.6}_{-15.4}$ respectively, which translates into April 27th 2010 $^{+9.8}_{-7.2}$ days, July 21st 2011 $^{+12.3}_{-10.9}$ days and Oct 13th 2012 $^{+16.6}_{-15.4}$ days. We display in \Fig{FIG:HATP13TRANSITS} the scatter in the predicted transit times for 2011 as a function of the assumed jitter and an accompanying histogram to illustrate the spread in predicted transit times, as well as the difference that results from taking a restricted range of jitter values. We do this for both the B09 S1 and W10 S2 data sets. It is clear from the scatter plot that within a given data set (e.g. B09) the distribution is essentially the same as that of the outer period plotted in \Fig{FIG:Jitter}. The histograms suggest that the systems with jitter in the range $2.0 < \sigma_j \sim 4.5\,m\,s^{-1}$ tend to have predicted transits which are more sharply peaked towards slightly lower values than the distribution which includes all jitter values.

However, when one compares the S1 results to those of S2, we see that there is an offset in the predictions, due to the preference for longer periods for the outer planet in S2, as well as the asymmetry in the parameter space. 

We again emphasize that different assumptions regarding the level of jitter in the system act to shift the predicted date (as well as the error bars) of any potential transit. We also stress that the uncertainties on our figures cover only two-thirds of the data: more extreme excursions exist. It would thus seem prudent for any future campaign of transit observations to be undertaken significantly before and after the mid-point of any predicted transit date, to allow for the uncertainties: e.g. for the 2011 potential transit, don't just observe on July 21st, 2011, but rather take observations over a large range of nights (perhaps starting around July 9th and continuing for $\sim 3$ weeks) to give the highest probability of observing any transit that might occur. 

On a practical point, we note from the object visibility calculator at ``http://catserver.ing.iac.es/staralt/'' that HAT-P-13 will be visible for many hours per night in the northern hemisphere during October 2012, but that observations around July 2011 would be rather more challenging, as the star is likely to only be visible for $\sim 1$ hour per night.  


\subsection{Additional Thoughts on Variations in $e_{inner}$ and $\eta = \varpi_{inner} - \varpi_{outer}$}
We note from the study by \citet{Mardling10} of the likely dynamics of the HAT-P-13 system, that the secular timescale for $e_{inner}$ and $\eta$ to vary around the limit cycle is of the order a few thousand years (see her Fig. 3b). We can also see that the amplitudes of the oscillations are $\sim 50^{\circ}$ for $\eta$ and $\sim 5\times 10^{-3}$ for $e_{inner}$. This would suggest that over the course of a 10 year observational campaign, the change in these quantities would be $\Delta\eta(10yrs) \sim 0.2^{\circ}$ and $\Delta e_{inner}(10yrs) \sim 10^{-5}$, both very small quantities indeed.

\section{Conclusion}\label{Conclusion}
We have performed an investigation of the HAT-P-13 system, concentrating our efforts on  re-evaluating the RV analysis, including jitter as an intrinsic part of the MCMC analysis, and then combining this evaluation with an analysis of the transit timing variations (TTVs) in the system. 

We find that:
\begin{itemize}
\item If the methodology for including jitter within an RV analysis is changed and jitter is included as a model parameter within the MCMC analysis, a significantly larger range of parameter space opens up, i.e. the masses and orbital parameters for the system are significantly less well-defined. As an example, based on the extended RV data set analyzed in Winn et al. 2010, if the overall system jitter is in the range $2.0 < \sigma_j < 4.5\,m\,s^{-1}$ rather than $\sigma_j \sim 3.4\,m\,s^{-1}$, the eccentricity of the inner planet has a one-sigma best fit value of $e_{inner} = 0.038^{+0.022}_{-0.018}$, the eccentricity of the outer planet has best fit value of $0.65^{+0.10}_{-0.06}$, while the period of the outer planet would have best fit values of $P_{outer} = 449.4^{+5.0}_{-4.8}$ and the relative pericenter alignment of the two planets becomes essentially unconstrained, with a best fit value of $\eta = 12.4^{+52.2}_{-83.9}$. With the current data set, even higher jitter values are plausible, and this would act to further increase the uncertainty in the observations of system parameters (see Section \ref{Jitter}).
\item If we include the current weak TTV constraints ($<100$ seconds) in addition to the RV analysis, then we find that we can exclude eccentricities for the outer planet larger than $\sim 0.85$. If future observations can pin-down the RMS TTV amplitude to a much narrower range ($< 10$ seconds for example), then this would strongly constrain the eccentricity of the outer planet ( $\lsim 0.6$ for the case of TTVs less than 10 seconds.). See Section \ref{MCMC_PLUS_TTV} for a discussion of the implications of the TTV  amplitude.
\item The current TTV constraints already suggest that the planets in the system do \emph{not} have relative inclinations in the range $88^{\circ} < i_{rel} < 92^{\circ}$, but any future tightening of the TTV constraint would act to exclude a much wider range of inclinations centered on $ i_{rel} = 90^{\circ}$ ($70^{\circ} < i_{rel} < 110^{\circ}$ would be excluded \emph{if} TTVs amplitudes were less than 10 seconds - See Section \ref{Contours})
\item The TTV profile can in many circumstances act as an efficient diagnostic tool in determining the relative inclination between the two planets. In particular, it is easy to distinguish between systems with $i\sim 0^{\circ}$ and $i\sim 45^{\circ}$ (the approximate angles found to be the most likely in the analysis of \citet{Mardling10}) through suitably timed transit observations (see Section \ref{Profiles}). We thus suggest that an observational campaign targeting the transits of the inner planet could give informative results about the relative inclination of the planets in the system. We suggest that observations extending over the period of pericenter passage (for the outer planet) will initially be the most useful in determining the nature and amplitude of the TTVs, but that longer term observations can also give important information that allow different inclination states to be distinguished.
\item We note that a transit by the outer planet is rendered \emph{more} likely following the exclusion of regions of inclination space by the analysis of \citet{Mardling10}, supporting the idea of an observational campaign, but the large spread in allowed orbital periods resulting from our jitter analysis would suggest that any observations should be conducted over a $\sim 3$ week period centered approximately on either July 21st 2011, or October 13th 2012. (Observations in 2011 will be significantly more difficult than in 2012 - See Section \ref{Projections}.)
\item Given the value of transit timing observations near the time of pericenter of the outer planet, the long orbital period of the outer planet, and the poor observability during pericenter passages in 2011, we suggest that observers should consider observing whenever it would be possible to observe full ingress or full egress, rather than only observing full transits.  Measuring a large fraction of transits/ingress/egress times during this critical time will require observations from a network of telescopes at multiple longitudes.  
\end{itemize}

\section*{Acknowledgements}
This material is based upon work supported by the National Science Foundation under Grant No. 0707203 and is also based on work supported by NASA Origins of
Solar Systems grant NNX09AB35G. MJP is grateful to Ben Nelson for helpful discussions regarding the ``Systemic Console'', to Karen Kinemuchi for discussions regarding observations and to Phil Armitage and CU for their hospitality during part of the writing of this manuscript. EBF appreciates conversations with Rosemary Mardling and thanks the Isaac Newton Institute for Mathematical Sciences for their hospitality during the 2009 program on the Dynamics of Disks and Planets.
Finally, the authors would like to thank the two anonymous referees for their suggestions and comments which lead to a significant improvement in the quality of the work presented.



\appendix
For the sake of completeness, we return to consider \Fig{FIG:TTV1} in which we demonstrated that the fitted values for the overall RV offset, as well as the RV Amplitude, $K_{outer}$ , longitude of pericenter, $\varpi_{outer}$, and the eccentricity, $e_{outer}$, of the outer planet all showed a correlation with the RMS TTV amplitude. In Fig. 10 we plot $P_{outer}$ for which there is \emph{no} obvious correlation between the RMS TTV amplitude and the fitted system parameter. 

\begin{figure*}
\centering
\begin{tabular}{cc}
     	\psfig{figure=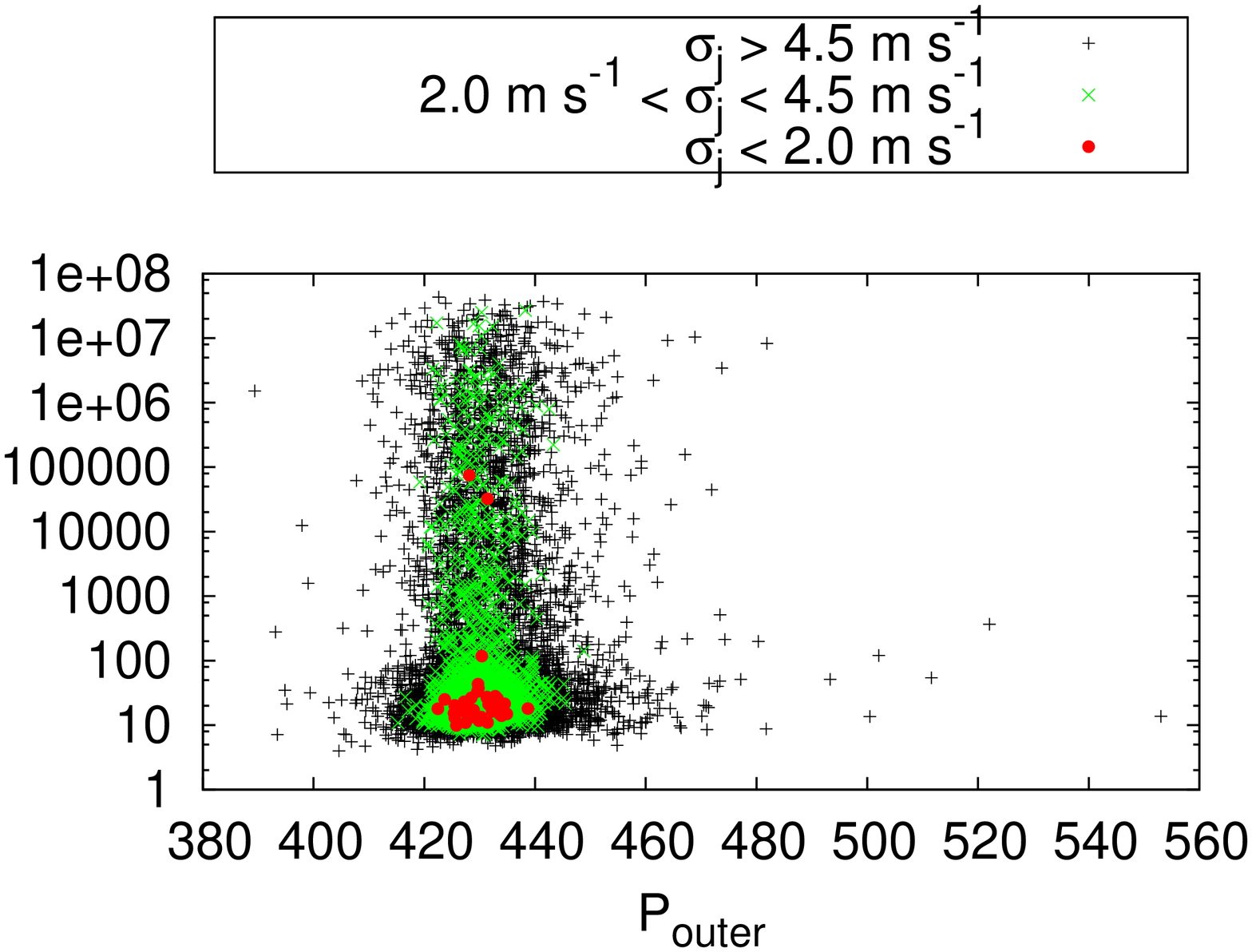,angle=-90,width=0.43\textwidth}&
	     	\psfig{figure=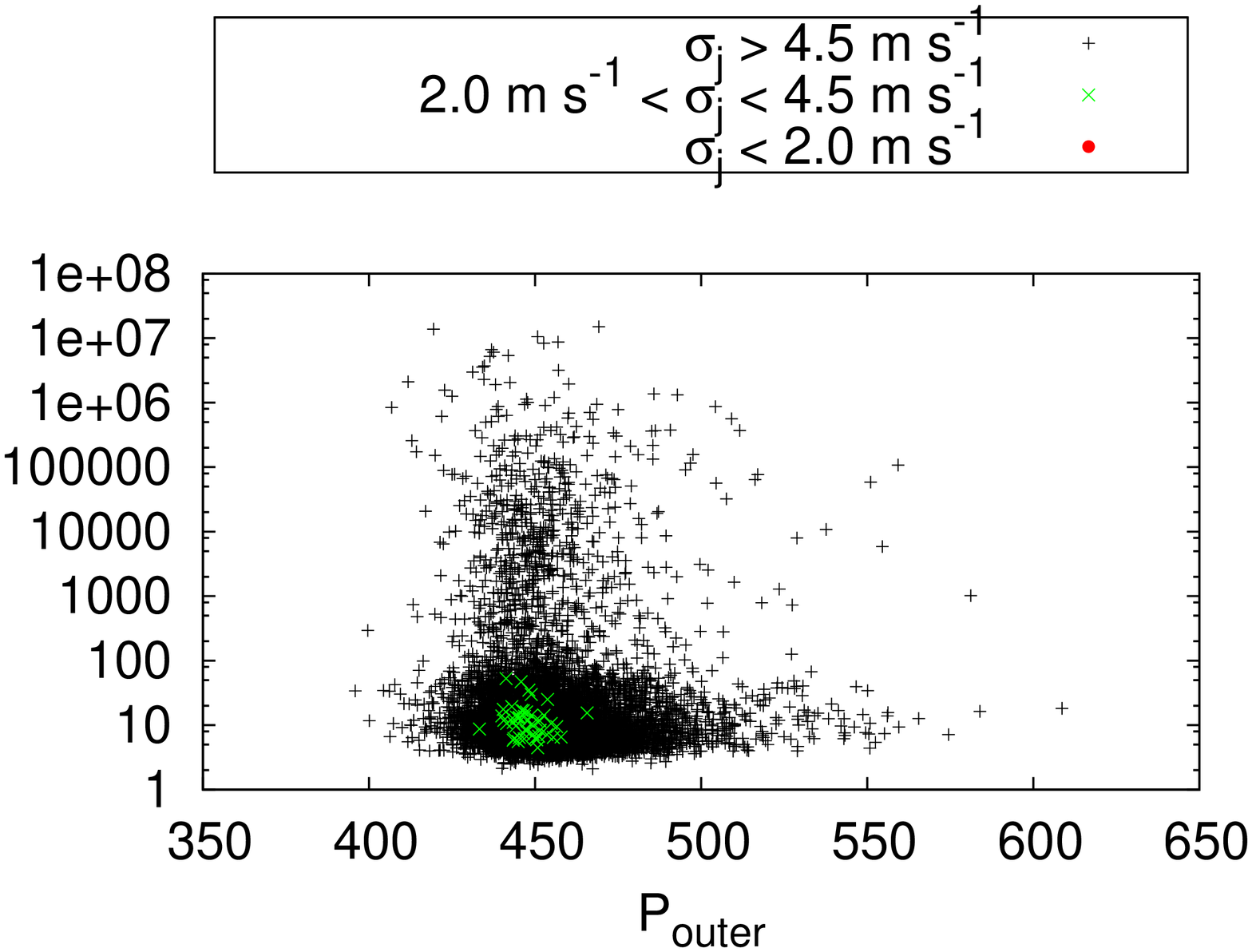,angle=-90,width=0.43\textwidth}\\
\end{tabular}
  \put(-450,-75){{\begin{sideways} \centering {\bf \large RMS TTV Amplitude, [s]} \end{sideways}}}
\caption{TTV Amplitude vs. System Parameters for parameters in which there is no obvious correlation between amplitude and parameter value.
In the various plots above we show the RMS TTV amplitude as a function of various system parameters (planetary eccentricity, etc). As is Fig. 1, systems with $\sigma_j< 2.0\,m\,s^{-1}$ are plotted using a red circle (gray in the print version), those with $2.0 < \sigma_j< 4.5\,m\,s^{-1}$ are plotted using a green cross (gray in the print version), and those with $\sigma_j > 4.5\,m\,s^{-1}$ are plotted using black addition symbols. 
As in Fig. 1, in the left-hand column we present results using only the subset of data known at the time of publication by B09, assuming a 2-planet fit. In the center and right-hand columns we present results obtained using the full data set of W10, with the central column assuming 2-planets + a linear trend, while the right-hand column assumes a 3-planet fit.
We plot results for the period of the outer planet. Unlike the results plotted in \Fig{FIG:TTV1}, we see that there is no obvious correlation between the RMS TTV amplitude and the system parameter value.
}
\label{FIG:TTV_AMP}
\end{figure*}


\end{document}